\documentclass[
 reprint,
 amsmath,amssymb,
 aps,
]{revtex4-2}

\usepackage{graphicx}
\usepackage{dcolumn}
\usepackage{bm}
\usepackage{url}
\usepackage[breaklinks]{hyperref}
\usepackage[dvipsnames]{xcolor}
\usepackage{placeins}
\usepackage[mathlines]{lineno}
\usepackage{siunitx}
\usepackage{braket}
\usepackage{tikz}

\hypersetup{colorlinks=true,breaklinks=true,urlcolor=blue,linkcolor=blue,citecolor=blue}
\DeclareSIUnit\gauss{G}

\renewcommand\vec{\mathbf}
\newcommand{\eps}{\varepsilon}

\begin{document}

\preprint{APS/123-QED}

\title{The moving Fermi polaron}

\author{Johanna Hennebichler}
\thanks{These two authors contributed equally to this work.}
\affiliation{Institute for Quantum Optics and Quantum Information (IQOQI) \\Technikerstra{\ss}e 21a, 6020 Innsbruck, Austria}
\affiliation{Institute for Experimental Physics, University of Innsbruck \\Technikerstra{\ss}e 25, 6020 Innsbruck, Austria}

\author{Ruben Erlenstedt}
\thanks{These two authors contributed equally to this work.}
\affiliation{Institute for Quantum Optics and Quantum Information (IQOQI) \\Technikerstra{\ss}e 21a, 6020 Innsbruck, Austria}
\affiliation{Institute for Experimental Physics, University of Innsbruck \\Technikerstra{\ss}e 25, 6020 Innsbruck, Austria}

\author{Matteo Caldara}
\thanks{Contact author: matteo.caldara@monash.edu}
\affiliation{International School for Advanced Studies (SISSA) \\Via Bonomea 265, 34136 Trieste, Italy}
\affiliation{School of Physics and Astronomy, Monash University \\Victoria 3800, Australia}
\author{Erich Dobler}
\affiliation{Institute for Quantum Optics and Quantum Information (IQOQI) \\Technikerstra{\ss}e 21a, 6020 Innsbruck, Austria}
\affiliation{Institute for Experimental Physics, University of Innsbruck \\Technikerstra{\ss}e 25, 6020 Innsbruck, Austria}

\author{Cosetta Baroni}
\thanks{Contact author: cosetta.baroni@uibk.ac.at}
\affiliation{Institute for Quantum Optics and Quantum Information (IQOQI) \\Technikerstra{\ss}e 21a, 6020 Innsbruck, Austria}
\affiliation{Institute for Experimental Physics, University of Innsbruck \\Technikerstra{\ss}e 25,
6020 Innsbruck, Austria}

\author{Georg M. Bruun}
\affiliation{Center for Complex Quantum Systems, Department of Physics and Astronomy, Aarhus University\\Ny Munkegade 120, DK-8000 Aarhus C, Denmark}
\author{Pietro Massignan}
\affiliation{Departament de Física, Universitat Politècnica de Catalunya \\Carrer de Jordi Girona 1-3, 08034 Barcelona, Spain}
\author{Rudolf Grimm}
\affiliation{Institute for Quantum Optics and Quantum Information (IQOQI) \\Technikerstra{\ss}e 21a, 6020 Innsbruck, Austria}
\affiliation{Institute for Experimental Physics, University of Innsbruck \\Technikerstra{\ss}e 25,
6020 Innsbruck, Austria}

\date{\today} 

\begin{abstract}

The Fermi polaron, formed by a mobile impurity interacting with a surrounding Fermi sea, exemplifies the canonical quasiparticle concept as a cornerstone in our description of quantum many-body systems across a wide range of energy scales. 
Experiments on atomic quantum gases have provided profound insights into the universal nature of the Fermi polaron.  
While most previous studies have focused on the case of zero impurity momentum, finite-momentum properties have remained largely uncharted. 
Here, we investigate the moving Fermi polaron by combining a novel Raman acceleration scheme with high-precision radio-frequency spectroscopy, thus exploring the quasiparticle’s dispersion relation over a wide range of momenta.
We compare our measurements of energy shifts and spectral linewidths, recorded for both attractive and repulsive interactions, with a microscopic theory and reach quantitative agreement for all momentum regimes.
For low momenta, we find the energy of the moving polaron to be fully consistent with the usual Fermi liquid picture of a dressed particle with a constant effective mass.
At high momenta, the polaron approaches the behavior of a weakly interacting bare particle, featuring small energy shifts and weak broadening.
In the intermediate momentum regime, broadening is generally larger and, most strikingly, the polaron shows a markedly different behavior for attractive and repulsive interaction.
While the repulsive polaron exhibits a smooth connection between both regimes along with a monotonic change of the energy shift, the attractive case shows a peculiar non-monotonic behavior. With increasing momentum, the attractive polaron enters a new regime, where its energy deviates from the constant effective mass expression and broadening suddenly increases.
By comparing these observations with theory, we show that the rather abrupt behavior coincides with the attractive polaron entering a molecule-hole continuum, where it is no longer the ground state. We interpret these observations as a motion-induced polaron-molecule transition.

\end{abstract}

\maketitle

\section{\label{sec:introduction}Introduction}

The concept of quasiparticles is of paramount importance in  quantum many-body systems, where it provides a powerful framework to capture the emergent behavior of particles moving in a complex medium.
Polarons, in particular, were originally introduced by Landau and Pekar to describe an electron moving through an ionic crystal, where its motion is accompanied by lattice distortions~\cite{Pekar1948}. Since then, the description in terms of polarons has become a widespread tool to model the behavior of mobile particles across a wide range of physical systems including semiconductors, strongly correlated materials, and quantum fluids~\cite{BaymPethick1991book}. Depending on the nature and strength of the interactions with  the surrounding medium, polarons can exhibit different properties, ranging from weakly dressed quasiparticles to strongly bound composite objects, and 
understanding their behavior remains a central problem in many areas of physics.

A mobile impurity particle interacting with a Fermi gas becomes dressed by excitations of the medium, forming a paradigmatic realization of a quasiparticle known as the Fermi polaron.
Because of its conceptual simplicity and the fact that it can be realized under pristine conditions in ultracold atomic gases with widely tunable interactions, the Fermi polaron has been extensively studied in both experiment and theory~\cite{Massignan2014pdm, Schmidt2018umb, Varenna2022book, Scazza2022rfa, Parish2025review, Massignan2026pia}. 
This has greatly advanced our understanding of the polaron's fundamental properties based on microscopic interactions. 
Previous studies have revealed the existence of attractive and repulsive polaron branches and established the system as a benchmark problem in many-body physics~\cite{Schirotzek2009oof, Nascimbene2009coo, Kohstall2012mac,Koschorreck2012aar, Scazza2017rfp, Yan2019bau, Baroni2024mib}.
Bose polarons, i.e., impurities embedded in a weakly interacting ensemble of bosons, share many similarities with Fermi polarons but also feature some critical differences, and have also been studied in great detail over the past few years~\cite{Hu2016bpi,Jorgensen2016ooa,Penaardila2019aab,Yan2020bpn,Skou2021neq,Skou2022lad,Morgen2025qbs,Etrych2025uqd,Henke2025ror,Grusdt2025iap,Massignan2026pia}. 
In the broader context of condensed matter, it has turned out that an exciton interacting with a surrounding electron gas in two-dimensional semiconductors can form a Fermi polaron quite similar to its counterpart in an ultracold atomic gas. This allows for essentially the same theoretical descriptions and illustrates the universal validity of the Fermi polaron concept~\cite{Massignan2026pia}. 

Most experiments on polarons in ultracold quantum gases have focused on quasiparticle properties at zero momentum. The role of finite momentum, having received much less attention so far~\cite{Mathy2012qfo, Hu2022rso, vonMilczewski2024, Shi2025vtd, Horvath2026odf, Andrade2026cbp}, is highly intriguing from a fundamental point of view because of its intimate relation to the dynamic response of the medium. Moreover, it is highly relevant for applications, since moving quasiparticles underpin the transport coefficients of materials and can introduce qualitatively new physics.
To lowest order in momentum, well below the characteristic momentum scale set by the medium, the essential physics of the moving quasiparticle can be captured by an effective mass and described by a simple parabolic dispersion relation.
Experimental observations of the effective mass in the low-momentum regime have been reported for the Fermi polaron~\cite{Nascimbene2009coo, Navon2010teo, Koschorreck2012aar, Scazza2017rfp, Yan2019bau} and, very recently, also for the Bose polaron~\cite{Henke2025ror}. 

For higher momenta, the character of interaction changes,  
the dressing cloud can no longer follow the motion of the bare particle, and the simple effective-mass description becomes inadequate.
In a quantum gas with near-resonant interactions, for example, 
the scattering amplitude is naturally limited at high momenta by unitarity or effective-range effects~\cite{Chin2010fri} and the rapidly moving quasiparticle will inevitably undress and develop into a bare particle. This behavior will be reflected in a richer, non-parabolic dispersion relation.
Exploring this relation and the connection between regimes of low and high momenta on a microscopic level is essential for our understanding of the quasiparticle concept and its validity, but it has so far remained uncharted. 

In our combined experimental and theoretical work, we investigate the properties of the moving Fermi polaron over a wide range of momenta. Our experimental model system~\cite{Kohstall2012mac, Fritsche2021sab, Baroni2024mib, Grimm2025fqm} is an optically trapped sample of $^{41}$K impurity atoms immersed in a large Fermi sea of $^6$Li atoms, featuring interaction tuning via an interspecies Feshbach resonance~\cite{Chin2010fri}. A newly developed, species-selective Raman acceleration scheme allows us to impart a quasi-continuous momentum on the impurity~\cite{Dobler2026PhD}. Combined with high-precision spectroscopic measurements in an ultracold Fermi gas, we systematically map out the polaron energy shift and spectral linewidth as functions of momentum, thus uncovering the complete dispersion relation.  Our results are compared with a diagrammatic T-matrix calculation, showing excellent agreement, and they reveal a nontrivial evolution of the polaron properties. In the low-momentum regime, the energy of the attractive polaron is given by a simple Fermi liquid expression where the bare mass is replaced by an effective mass. In this regime, weak broadening of the polaron spectral width indicates a relatively long lifetime. For higher momenta, the simple description in terms of a constant effective mass becomes inadequate and the broadening increases rapidly, which signals a transition to a new regime as the polaron enters a continuum of molecule-hole excitations. We interpret this as a motion-induced polaron-molecule transition. Finally, for high momenta the properties of the attractive polaron approach those of a weakly interacting bare particle as the scattering becomes unitarity limited. The repulsive polaron is found to behave in a different and much simpler way, as it smoothly evolves from a quasiparticle at low momenta to a weakly interacting bare particle at high momenta. 

The paper proceeds as follows: 
In Sec.~\ref{sec:shift}, we start with introducing the basic concepts to understand motion-induced energy shifts of the polaron from low to high momenta, and we outline our theoretical approach that captures the full range of momenta.
In Sec.~\ref{sec:experiment}, we introduce our experimental platform with its essential features enabling precise spectroscopic investigations of polarons after controlled momentum transfer.
In Sec.~\ref{sec:resultsEnergy}, we then present our measurements of the energy shift and the linewidth of the polaron peak and compare them with predictions from our theoretical model. 
In Sec.~\ref{sec:pol_mol_transition} we discuss the momentum-induced polaron-molecule transition as a special feature of interest. 
We finally conclude in Sec.~\ref{sec:conclusion} and present details on various theoretical aspects and experimental methods in a set of Appendices~\ref{app:theory}-\ref{app:dataanal}.

\section{Momentum-dependent energy shift}
\label{sec:shift}

We consider an impurity atom with mass $m_\downarrow$ moving in a sea of degenerate fermions with mass $m_\uparrow$ and density $n$, with Fermi wavenumber $k_F=(6\pi^2 n)^{1/3}$, Fermi momentum $p_F = \hbar k_F$, and Fermi energy $E_F = p_F^2/2 m_\uparrow$. To understand the effect of impurity momentum $p$, we first consider the limiting cases of low and high momenta, and later on we discuss the connection between these regimes.

{\em Low-momentum regime ($p \ll p_F$):}  
The situation of a polaron at rest ($p=0$) has been the subject of numerous theoretical and experimental investigations~\cite{Massignan2014pdm,Varenna2022book,Parish2025review,Massignan2026pia}. This elementary situation is now well understood and constitutes a cornerstone in the description of the many-body problem. 
Depending on the sign of the interaction, the impurity attracts or repels atoms from the medium, leading to the formation of the `attractive polaron' ($\eps<0$) or the `repulsive polaron' ($\eps>0$); here $\eps$ represents the interaction energy at zero impurity momentum.
The two-body interaction between an impurity atom and an atom in the Fermi sea is characterized by the $s$-wave scattering length $a$, which in the experiment can be tuned via a magnetically controlled Feshbach resonance~\cite{Chin2010fri, Kokkelmans2014fri, Lous2018pti}. For the theoretical description (see App.~\ref{app:theory} for full details), we introduce the dimensionless interaction parameter $X \equiv - 1/(k_Fa)$, where $-1\lesssim X\lesssim 1$ corresponds to the strongly interacting many-body regime. 

For a slowly moving polaron, the energy can be conveniently expressed within a Taylor expansion~\cite{Massignan2026pia}, which results in the simple parabolic dispersion relation
\begin{equation}
    E = \eps + \frac{p^2}{2 m_\downarrow^*} \, .
    \label{eq:Epol}
\end{equation}
The first term represents the energy $\eps$ at zero momentum, and the
the second contribution describes the momentum dependence in terms of the constant {effective mass} $m_\downarrow^*$, which incorporates the effect of the dressing by excitations of the surrounding medium. The effective mass is usually larger than the bare mass $m_\downarrow$, since the dressing tends to increase the inertia of the quasiparticle.

The approximation of Eq.~\eqref{eq:Epol}, where the effective mass is treated as a constant, holds if the impurity momentum $p$ remains well below a characteristic momentum determined by the medium. 
For our situation, the Fermi momentum $p_F$ represents the only relevant momentum scale set by the medium. Consequently, the characteristic momentum that limits the effective-mass description will be naturally given by $p_F$ with a dimensionless factor. In the most general case, this factor depends on three dimensionless quantities: the interaction strength $X$, the mass ratio $m_\downarrow/m_\uparrow$ and, for `narrow' Feshbach resonances~\cite{Chin2010fri}, the parameter $k_FR^*$, with $R^*$ describing the effective range of the specific resonance~\cite{Petrov2004tbp}.

\begin{figure}[t]
    \centering
   \includegraphics[trim=50 0 0 0,clip,width=0.85\columnwidth]{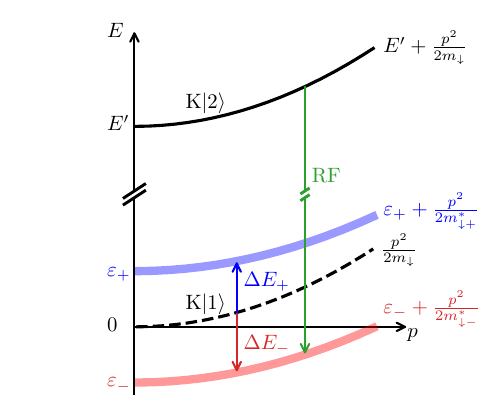}
   \vspace{-0.3cm}
    \caption{Energies of impurity states in the low-momentum regime, which is described by Eq.~\eqref{fig:EvsP} under the assumption of constant effective masses. In the internal state K$|1\rangle$, the impurity features widely tunable interaction with the medium. This allows us to realize polaronic regimes (red and blue line) in direct comparison with the non-interacting case of a bare particle (dashed line). In this way, we test our theoretical predictions for the interaction-induced energy shifts $\Delta E_{\mp}$ and the predicted parameter values $\eps_\mp$ for the zero-momentum energy shift and $m^*_{\downarrow\mp}$ for the effective mass. Here the subscripts $-$ and $+$ refer to the attractive/repulsive polaron. For clarity in the illustration, we have set $|\eps_-| = |\eps_+|$ and $m^*_{\downarrow-} = m^*_{\downarrow+}$, but note that these are generally different quantities. The internal state K$|2\rangle$ serves us as a non-interacting reference to probe the energy shifts $\Delta E_\mp$ by RF injection spectroscopy (green arrow).
    }
    \label{fig:EvsP}
\end{figure} 

To characterize the effect of interaction, we now consider the energy of the moving impurity relative to the energy $p^2/2m_\downarrow$ of the bare particle (see Fig.~\ref{fig:EvsP}). The corresponding energy shift as a function of $p$, given by
\begin{equation}
    \Delta E = \eps - \frac{p^2}{2 m_\downarrow} \left( 1 - \frac{m_\downarrow}{m_\downarrow^*} \right) \, ,
    \label{eq:shift}
\end{equation} 
represents the difference between the dispersion relations of the polaron and the bare particle and contains all the relevant interaction physics.
In a real experiment, the shift can be probed spectroscopically~\cite{Vale2021spo} by coupling the polaronic state to a non-interacting reference state. This can be implemented by means of a radio-frequency (RF) field (green arrow in Fig.~\ref{fig:EvsP}), which is the basic idea of RF spectroscopy~\cite{Liu2020rfr, Torma2016pou,Kohstall2012mac,Massignan2014pdm}.

\begin{figure}[t]
    \centering
    \includegraphics[width=1\columnwidth]{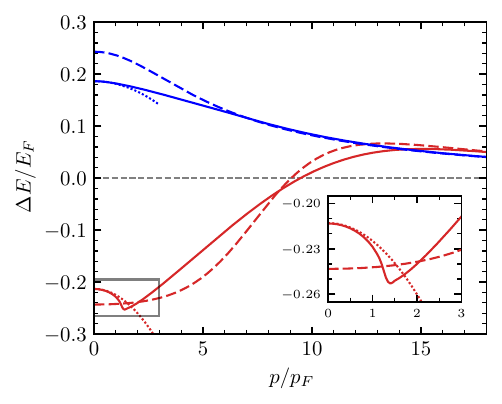}
    \vspace{-8mm}
    \caption{Momentum-dependent energy shift for the attractive (red) and the repulsive (blue) polaronic branch as predicted by the different theoretical models discussed in the text.
    The solid curves represent the results of the diagrammatic T-matrix theory (applicable at any momentum). 
    The dashed curves result from the simple two-body scattering approach discussed in App.~\ref{app:pscatt} (which is strictly justified only for sufficiently large momenta).      
    The dotted lines represent parabolic fits based on Eq.~\eqref{eq:shift} to the predicted low-momentum behavior, assuming a constant effective mass.
    The inset zooms into the low-momentum region $p/p_F \le 3$.
    The parameter values are chosen for our Li-K mixture ($m_\downarrow/m_\uparrow = 41/6$) and typical experimental conditions ($X=\pm1$, $k_FR^*=0.75$); see Sec.~\ref{sec:experiment}.
    For this specific example, we extract values for the coefficients entering Eq.~\eqref{eq:shift}: For the attractive polaron at $X=+1$ (repulsive polaron at $X= -1$), we obtain 
     $m^*_{\downarrow}/m = 1.091$ ($1.034$) and
     $\eps/E_F = -0.213$ ($0.187$).
    }
    \label{fig:theories}
\end{figure}

For the attractive polaron at small momentum, Eq.~\eqref{eq:shift} predicts an increase of the energy shift in magnitude (red arrow in Fig.~\ref{fig:EvsP}, labeled $\Delta E_-$). In this regime, a moderate motion of the impurity will {increase} the overall strength of the attractive interaction effect.
This surprising behavior is characteristic of the attractive polaron in the approximation of Eq.~\eqref{eq:Epol}.  It stands in contrast to the behavior of the repulsive polaron, where the impurity motion will decrease the magnitude of the repulsive interaction effect (blue arrow in Fig.~\ref{fig:EvsP}, labeled $\Delta E_+$). This strikingly different behavior of the attractive and the repulsive polaron is also demonstrated by the theoretical curves in Fig.~\ref{fig:theories} in the low-momentum regime.

{\em High-momentum regime ($p\gg p_F$):} For rapidly moving impurities, the effect  of near-resonant scattering with atoms in the medium is limited by unitarity, which simplifies the theoretical description. For $p \gg p_F$, even a system that is strongly interacting in the low-momentum regime becomes weakly interacting and thus loses its many-body character. 
The energy shift $\Delta E$ can then be described within a straightforward approach based on two-body scattering processes alone, which gives 
\begin{equation}
    \Delta E = - \frac{2 \pi \hbar^2 n}{m_r} \textrm{Re}[f_v(k_p)],
    \label{eq:DeltaE_fastpolarons}
\end{equation}
where $f_v(k_p)$ is the two-body scattering amplitude in vacuum, evaluated at the on-shell collision wavenumber $k_p$ (for details, see App.~\ref{app:pscatt}).
Note that this simplification is analogous to the case of weak interactions ($|a| \ll 1/k_F$ or $|X| \gg 1$), which allows for a treatment in the standard mean-field approximation. 

The predictions based on Eq.~\eqref{eq:DeltaE_fastpolarons} for the momentum-dependent energy shifts are shown in Fig.~\ref{fig:theories} (dashed lines) for both the attractive (red) and the repulsive (blue) polaron.
Remarkably, clearly beyond the validity of the two-body approach, we find the simple model to provide reasonable estimates even for the energy shifts in the zero-momentum limit. However, in this approach, the low-momentum behavior does not show the initial parabolic momentum dependence expected as a signature of the effective mass according to Eq.~\eqref{eq:shift}. This is not surprising as the effective mass is a many-body concept beyond two-body scattering physics.

In the high-momentum limit, where the simple model is well justified, we expect an energy shift $\Delta E \propto 1/p^2$, with positive sign for both the attractive and the repulsive polaron. While the latter keeps its positive sign at any momentum, the former undergoes a sign reversal from negative to positive with increasing momentum. This behavior is explained by the momentum-dependent resonance shift~\cite{Horvath2017ats}, which occurs for narrow Feshbach resonances as a consequence of the finite range of the two-body scattering potential (see App.~\ref{app:pscatt} for more details). 

{\em Connecting high- and low-momentum regimes:}
The combination of the above cases suggests an interesting characteristic behavior. For the attractive polaron, the energy shift will exhibit a {\em non-monotonic} dependence on momentum. With increasing momentum, the magnitude of the negative shift will first increase as long as the approximation of Eq.~\eqref{eq:Epol} holds, which is based on the assumption of constant effective mass (dotted red curve in Fig.~\ref{fig:theories}). To connect with the high-momentum regime (dashed red curve), the magnitude will then have to decrease with further increasing momentum. Here, the competition between the two different regimes will result in non-monotonic behavior. 
For the repulsive polaron, in contrast, both effects will act together with the same sign, leading to a simple monotonic decrease of the energy shift with increasing momentum without any further features.

{\em Theory at any momentum:} 
For a quantitative description of the energy shift within the full momentum range, we have adopted a diagrammatic T-matrix approach, presented in detail in App.~\ref{app:theory}. 
In essence, we extract the fundamental quasiparticle properties of polarons from the impurity self-energy at zero temperature, which we calculate as
\begin{align}
    \Sigma(\mathbf{p}, E) =& \sum_{\textbf{q}}^{q< p_F} \mathcal{T}(\mathbf{p}+\mathbf{q}, E + \epsilon_{\mathbf{q} \uparrow}-E_F), 
    \label{eq:self-energy_main}
\end{align}
with the T-matrix $\mathcal{T}(\mathbf{q},E) $ describing the two-body scattering processes at energy $E$ between the impurity and particles with momentum $\mathbf{q}$ inside the Fermi sea (see App.~\ref{app:theory} for its explicit expression), and $\epsilon_{\mathbf{q} \uparrow} = q^2/2m_\uparrow$ is the dispersion of a majority particle. 

The predictions from our T-matrix theory on the momentum-dependent energy shifts are shown by the solid lines in Fig.~\ref{fig:theories}. The full theory captures the complete behavior discussed for low and high momenta.
In particular, for both the attractive and the repulsive polaron, it shows the initial parabolic behavior (dotted lines) characteristic for a constant effective mass. As expected, this regime is restricted to low momenta ($p/p_F \lesssim 1$).

Equipped with the diagrammatic T-matrix approach as a powerful tool, we analyze our high-precision measurements taken in an ultracold quantum mixture of K impurities moving in a Fermi sea of Li atoms. The combined approach of theory and experiments allows us to address the open questions in our understanding of the dispersion relation and the underlying physics: 
(i) To what extent can a finite-momentum impurity be accurately described as a quasiparticle with a constant effective mass? (ii) What underlying physical mechanisms govern the breakdown of this effective-mass description? (iii) What is the fate of the quasiparticle beyond this regime?

\section{\label{sec:experiment}Experimental platform and procedures}

Our experiments are based on an optically trapped mixture of $^{41}$K (impurity atoms with mass $m_\downarrow$) and $^6$Li (atoms with mass  $m_\uparrow$ forming the Fermi sea). Our choice of the specific K isotope is due to practical reasons (natural abundance, laser cooling properties, transitions for optical imaging) and the availability of an interspecies Feshbach resonance well suited for interaction tuning.
All experiments reported here are carried out at low impurity concentration close to the single-impurity limit, where effects of mediated interactions~\cite{Baroni2024mib} remain very small. As a consequence, the quantum statistics of the impurities is largely irrelevant for what is discussed here.

In Sec.~\ref{sec:experiment_A}, we summarize the main experimental procedures developed in previous work~\cite{Kohstall2012mac, Cetina2015doi, Cetina2016umb, Fritsche2021sab, Baroni2024mib}. In Sec.~\ref{sec:experiment_B}, we introduce a new feature of our set-up, which plays a central role in the present experiments: An `impurity accelerator'~\cite{Dobler2026PhD} allows us to give well-defined photon momentum kicks to the impurity atoms via optical Raman transitions.

\subsection{Sample preparation and spectroscopy}\label{sec:experiment_A}

Following an all-optical approach~\cite{Spiegelhalder2010aop}, we prepare our mixed-species system in a crossed-beam optical dipole trap, operated with 1064-nm near-infrared light. Based on the evaporation and spin-preparation scheme described in detail in previous work~\cite{Lous2018pti, Lous2018PhD, Fritsche2021sab}, we obtain a mixture of fermionic lithium atoms in the lowest hyperfine spin state Li$|1\rangle$ ($F=1/2$, $m_F=1/2)$ and potassium atoms in the second- or third-lowest state K$|2\rangle$ ($F=1$, $m_F=0)$ or K$|3\rangle$ ($F=1$, $m_F= -1)$, respectively (see App.~\ref{app:Ramanaccel}). The optical trap generates an elongated (aspect ratio $\sim 7.3$) species-specific trapping potential with axial trap frequencies
$\omega_z^{\text{K}}/2\pi = 39\;\text{Hz}$,   
$\omega_z^{\text{Li}}/2\pi = 65\;\text{Hz}$
and radial frequencies
$\omega_r^{\text{K}}/2\pi = 283\;\text{Hz}$,   
$\omega_r^{\text{Li}}/2\pi = 475\;\text{Hz}$.
Under typical experimental conditions, the sample contains $N_{\rm Li} = 4\times 10^5$ lithium atoms and $N_{\rm K} = 7\times 10^3$ potassium atoms. The sample is fully thermalized at a typical temperature of $T = 160\,$nK, which is about 25\% above the critical temperature for Bose-Einstein condensation of the K component; see Ref.~\cite{Lous2017toa} for accurate Fermi gas thermometry based on bosonic impurities. For the Li Fermi sea, this corresponds to a reduced temperature of $T/T_{F,\,{\rm trap}} \approx 0.1$, with $T_{F,\,{\rm trap}} = (\hbar / k_B) (6N_{\rm Li} \omega_r^2 \omega_z)^{1/3} \approx 1.7\,\mu$K representing the globally defined Fermi temperature in the three-dimensional harmonic trap.

The impurity atoms reside close to the center of the trap, where they experience a nearly homogeneous environment, with local density variations of the fermionic medium remaining very small. This benefit is due to the much smaller size of the K cloud as compared with the Li Fermi sea, resulting from the tighter optical confinement of the K atoms in combination with the strong Fermi pressure of the light Li atoms~\cite{Lous2017toa}.  
We account for the residual inhomogeneity by introducing effective quantities for the Fermi energy and the number densities, defined as averages over the spatial distribution of the K atoms~\cite{Fritsche2021sab}. Typical values of the effective Fermi energy are around $E_F = k_B \times 1.6\,\mu{\rm K} = h \times 33\,{\rm kHz}$, which is less than 3\% below the peak values in the center of the trap. The residual variation of the Fermi energy as experienced by the impurity atoms in different regions of the trap can be quantified by a relative standard deviation~\cite{Cetina2016umb}, which under our typical conditions is as low as 5\%. This justifies our treatment as a quasi-homogeneous system. 
For the number densities, we calculate effective values
$n_{\rm Li} = 4.1\times 10^{12}\,{\rm cm^{-3}}$
(all atoms in the lowest spin state) and $n_{\rm K} = 7.2\times 10^{11}\,{\rm cm^{-3}}$ 
(in one of the non-interacting states K$|2\rangle$ or K$|3\rangle$). 
Experimental values for the number densities, the temperature, and the Fermi energy are extracted from independent measurements of atom number, temperature, and trap frequencies with estimated relative uncertainties of typically 15\%.

We employ a magnetically tuned Feshbach resonance~\cite{Chin2010fri} to control the $s$-wave interaction between the impurities and the fermionic medium. Our resonance~\cite{Lous2018pti, Fritsche2021sab} 
is located near $335\,$G and couples the energetically lowest spin states of the two species, K$\,|1\rangle$ ($F=1$, $m_F=1$) and Li$|1\rangle$. 
The resonance enables precise tuning of the interspecies $s$-wave scattering length according to $a = a_{\rm bg} (1 - \Delta/\delta B)$, with the background scattering length $a_{\rm bg} = 60.865\,a_0$ (Bohr's radius $a_0$), the resonance width $\Delta = 948.7\,$mG, and the variable magnetic detuning $\delta B = B-B_0$. The exact center of the resonance is subject to an upshift of typically $50$\,mG, which is caused by the trap light itself~\cite{Lous2018pti}. Under the conditions of our present experiments, $B_0 = 335.107(2)\,$G.

In view of the expected universal behavior, we will express energies and momenta in units of the Fermi energy $E_F$ and momentum $p_F$, as defined at the beginning of Sec.~\ref{sec:shift}.
The situation is further characterized by two dimensionless parameters:  $X = -1/(k_F a)$ describes the interaction strength in the zero-momentum limit, whereas $k_F R^*$ quantifies a momentum-dependent resonance shift originating from the effective range in two-body scattering~\cite{Petrov2004tbp}. The latter can be neglected in the limit of broad Feshbach resonances~\cite{Chin2010fri}, but needs to be taken into account for narrow resonances as in our case. With $R^* = 2241(7)\,a_0$~\cite{Lous2018pti} and $k_F  \approx (3040\,a_0)^{-1}$, we typically obtain $k_F R^* \approx 0.74$.

We probe our system using radio-frequency (RF) {\em injection} spectroscopy~\cite{Kohstall2012mac, Liu2020tor}. The method, as applied to the same system in our previous work~\cite{Fritsche2021sab, Baroni2024mib}, is based on the transfer of impurity atoms from an initial state, for which the interaction with the Fermi sea is negligibly weak (here K$|2\rangle$), to a resonant state (here K$|1\rangle$), see Fig.~\ref{fig:EvsP}.  
Spectroscopy is performed by varying the frequency of the RF pulse that transfers impurity atoms from K$\,|2\rangle$ to K$\,|1\rangle$, after a well-defined momentum has been imparted to the impurities, as outlined in Sec.~\ref{sec:experiment_B} and described in more detail in App.~\ref{app:Ramanaccel}. 
After applying the RF pulse (duration 500\,$\mu$s or 250\,$\mu$s, depending on the momentum transferred), we release the atoms into free space and, after an expansion time of 0.05 and 1.05\,ms, measure the number of atoms in state K$\,|1\rangle$ and K$\,|2\rangle$, respectively, by state-selective absorption imaging. 
Our spectroscopic signal is defined as the fraction of atoms transferred, 
measured as a function of the pulse detuning $\Delta \nu = \nu_0-\nu$, where $\nu$ is the central frequency of the RF pulse and $\nu_0$ represents the frequency of the bare transition between K$|1\rangle$ and K$|2\rangle$  at the chosen magnetic field ($\sim57.9$\,MHz). We use Blackman-shaped pulses to avoid side lobes in the spectra, adjusting their power such that $\pi$-pulses are realized in the absence of interactions with the fermionic medium.

Our central observable is the narrow peak in the RF spectrum, which for the polaron reflects the coherent part of the quasiparticle wavefunction~\cite{Massignan2026pia}.
This peak is generally accompanied by a broader incoherent background (width $\sim$$E_F/h$)~\cite{Schirotzek2009oof, Kohstall2012mac, Cetina2016umb}. The present measurements of the peak position require high precision better than $0.01 \, E_F/h$ ($\sim300\,\mathrm{Hz}$). 
The reduction of statistical uncertainties to such a level requires many repeated measurements over several hours. During that time, the magnetic field has to be kept stable on the milligauss level. To meet these stringent requirements~\cite{Baroni2024mib}, we have developed a toolbox of experimental and data analysis methods, as described in App.~\ref{app:dataanal}.

\subsection{Raman acceleration scheme} \label{sec:experiment_B}

\begin{figure}[t]
    \centering
    \includegraphics[trim=2 0 10 5,clip,width=1\columnwidth]{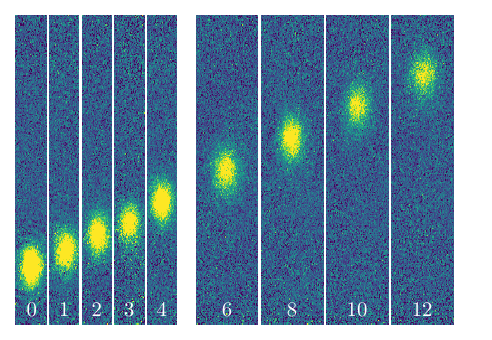}
    \vspace{-5mm}
    \caption{Controlled photon momentum transfer on the $^{41}$K impurity atoms. Absorption images of the cloud, containing 7000 atoms, taken after 4 ms time of flight for transferred momenta $\Delta p$ between 0 and 12\,$\hbar k$. The image crop measures 140\,$\mu$m $\times$ 960\,$\mu$m for the five images on the left and 280\,$\mu$m $\times$ 960\,$\mu$m for the four images on the right. The numbers at the bottom indicate the number of transferred photon momenta. The distance traveled by the cloud increases linearly with the imparted momentum, demonstrating precise control of the center-of-mass motion of the impurity cloud. In particular, the low momentum range up to $3\,\hbar k$, realized with the new double-pulse scheme, connects well to the established scheme for larger momenta. The atoms are imaged in the non-interacting state K$|2\rangle$ and each image is averaged over three experimental runs to reduce background noise.
    }
    \label{fig:ramanImages}
\end{figure}

Right before application of the RF spectroscopy pulse, we accelerate the impurity cloud by short momentum kicks (few~$\mu$s). The controlled momentum transfer, directed along the weak axis of the trap, is based on two-photon Raman transitions induced by a pair of counterpropagating laser beams~\cite{Kasevich1991avs}. Our Raman beams are red-detuned by about 25\,GHz from the D$_2$ line of K (wavelength $\lambda = 766.7\,$nm) and, with the proper choice of polarizations~\cite{Dobler2026PhD}, couple the states K$|2\rangle$ and K$|3\rangle$. In these two spin states, we can safely neglect the weak background interaction with the medium.
A single Raman $\pi$-pulse in the counterpropagating beam configuration (typical duration $2\,\mu$s) changes the atomic momentum by $2\,\hbar k = 2\,h/\lambda$, which corresponds to a relatively large momentum change of about $2.6\,p_F$. 

The application of multiple Raman pulses in short succession, a scheme well established in matter-wave optics (e.g.\ Refs.~\cite{Weitz1994aib, Jaffe2018eas, Mcguirk2000lal}), allows us to realize larger momentum changes, as shown for $\Delta p = 4, 6, 8, 10$, and $12\,\hbar k$ in Fig.~\ref{fig:ramanImages}. This is essential to explore the high-momentum regime ($p\gg p_F$), but the discrete steps of photon momentum transfer are too large to resolve the particularly interesting behavior related to the effective mass in the low-momentum range $p\lesssim p_F$. 
To overcome this limitation, we have developed an alternative scheme~\cite{Dobler2026PhD}, which is based on two separate Raman pulses in combination with the near-harmonic sloshing oscillation along the weak axis of the trap. This scheme allows us to apply quasi-continuous momentum changes of up to $4\,\hbar k$ to the impurity cloud. The absorption images for $\Delta p = 1, 2,$ and $3\,\hbar k$ in Fig.~\ref{fig:ramanImages} show the impurity cloud after applying this double-pulse scheme. The two Raman acceleration schemes employed in this work are described in detail in App.~\ref{app:Ramanaccel}.

\begin{figure}[t]
    \centering
   \includegraphics[trim=5 8 0 0,clip,width=1\columnwidth]{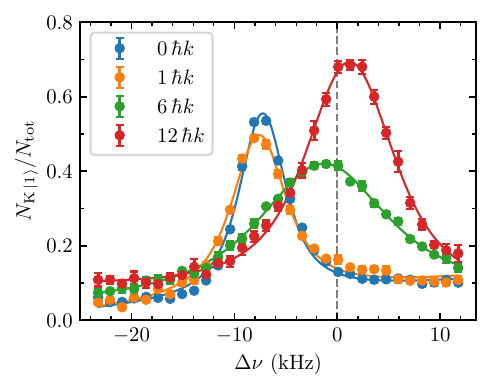}
    \caption{Effect of photon momentum transfer on the peak in the RF  spectrum of the attractive polaron ($X=1.03$, $E_F/h = 32.7\,$kHz). The experimental data points are shown with standard errors, derived from typically 10 repetitions. The solid lines represent fits based on the heuristic line-shape model described in App.~\ref{app:dataanal}. Note that the spectra for $\Delta p =  6\,\hbar k$ and $12\,\hbar k$ are taken with a shorter pulse than the spectra for $\Delta p =  0\,\hbar k$ and $1\,\hbar k$: While this can affect the amplitudes of the peak, it does not affect its position (see App. \ref{app:Ramanaccel}).}
    \label{fig:polaronpeaks}
\end{figure} 

\begin{figure}[t]
    \centering
    \includegraphics[trim=5 5 0 0,clip,width=1\columnwidth]{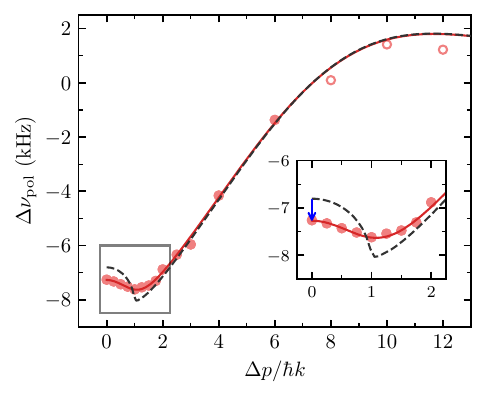}
    \caption{Polaron peak shift $\Delta \nu_{\rm pol}$ (in frequency units) versus momentum change $\Delta p$ (in units of photon momentum). The dots represent the experimental measurements for the attractive polaron with $X = 1.03$. Our fit analysis is based on the data points with $\Delta p \le 6\,\hbar k$~\cite{note6hk}, represented by the filled symbols. The solid curve illustrates the resulting best fit (for parameter values see Table~\ref{tab:my_label}).
    The dashed curve indicates the zero-temperature behavior predicted by theory.
    The inset zooms into the low-momentum region ($\Delta p\le 2 \,\hbar k$), exhibiting the characteristic behavior related to the effective mass. The blue arrow indicates the thermal shift for $\Delta p = 0$.
    The data points represent averages over typically 10 independent measurements. Error bars are omitted as they are smaller than the symbol size.
    }
    \label{fig:shiftvskicks}
\end{figure}

\section{\label{sec:resultsEnergy}Experimental results}

Here we present the main results of our measurements of the energy shift (Sec.~\ref{sec:energyshift}) and the linewidth (Sec.~\ref{sec:resultsDecay}) of the polaron peak for both the attractive  and the repulsive case. Our measurements are compared with the theoretical predictions from Sec.~\ref{sec:shift}. In particular, we confirm the non-monotonic energy shift expected for the attractive polaron with increasing momentum.

\subsection{Energy shift}\label{sec:energyshift}
In Fig.~\ref{fig:polaronpeaks}, we present four selected spectra that were recorded for the attractive polaron with $X = 1.03$. The main quantity of interest is the energy shift of the peak, $\Delta E_{\rm pol} = h \Delta \nu_{\rm pol}$, measured as a function of the momentum $\Delta p$ transferred to the impurity atoms. This dependence allows us to study the expected effects of impurity momentum, as qualitatively introduced in Sec.~\ref{sec:shift}.

Without any photon momentum transfer ($\Delta p = 0$, blue data points in Fig.~\ref{fig:polaronpeaks}),  
the interaction with the medium shifts the peak position to $\Delta \nu_{\rm pol} = -7.3$\,kHz (corresponding to about 22\% of the Fermi energy). 
The transfer of one photon momentum ($\Delta p = 1\,\hbar k$, orange) leads to a slightly larger shift of the peak to about $-7.6\,$kHz. 
A six times larger momentum change ($\Delta p = 6\,\hbar k$, green) then reverses the initial trend and causes a much smaller shift of the peak position to $\Delta \nu_{\rm pol} = -1.35$\,kHz. 
A further increase in the momentum transfer by a factor of two ($\Delta p = 12\,\hbar k$, red) leads to a sign reversal and shifts the center of the peak to $\Delta \nu_{\rm pol} = +1.2$\,kHz.

In Fig.~\ref{fig:shiftvskicks}, we show a typical set of measurements, taken at $X = 1.03$. The experimental data points were extracted from spectra recorded at 16 different values of $\Delta p$, including the four spectra shown in Fig.~\ref{fig:polaronpeaks}. The observed dependence reveals a distinct feature at low momentum, consisting of an initial increase in the magnitude of the frequency shift (downshift of $\Delta\nu_{\rm pol}$) followed by a decrease towards zero shift.
This non-monotonic behavior qualitatively fits our expectations (Sec.~\ref{sec:shift}) for the connection of a many-body `effective-mass' regime to a regime dominated by two-body scattering.
For a complete quantitative analysis of the results, we apply our diagrammatic T-matrix approach (App.~\ref{app:theory}) together with thermal averaging over the impurity motion (App.~\ref{app:thermav}). 
The solid line in Fig.~\ref{fig:shiftvskicks} shows a corresponding fit of our theory to the experimental data points, which demonstrates excellent agreement.

In all our fits, we treat $E_F$ and $X$ as free parameters. Reasonable estimates for their values can be obtained from independent measurements, which do not rely on any polaron theory, but such model-independent measurements suffer from relatively large uncertainties (typically $E_F$ with a relative uncertainty of 15\% and $X$ with an absolute uncertainty of $0.1$) and the fits can deliver more accurate results. 
We can also take $T/T_F$ as a third free parameter, but we find that corresponding three-parameter fits do not always converge well, depending on the number and quality of data points in a given set of measurements. In such cases, we fix the reduced temperature to the independently determined value $T/T_F = 0.1$, determined within 20\% uncertainty, and carry out two-parameter fits. The temperature uncertainty turns out to have only minor effects on the fit results. 

The mean thermal momentum of the impurity cloud without any momentum transfer ($\Delta p = 0$) is given by $\bar{p} = 
\sqrt{8 m_\downarrow k_B T/\pi}$.
For our typical conditions ($T= 160$\,nK), this corresponds to $\bar{p} \approx 0.93 \, p_F$,
which tells us that thermal effects are potentially important.
We therefore consider thermal averages of our observables, which we calculate as described in App.~\ref{app:thermav}. 
The effect of thermal averaging on the momentum-dependent energy shift is shown in Fig.~\ref{fig:shiftvskicks} by the solid curve compared to the zero-temperature dashed curve.  
While there is little effect on the slowly varying energy shift at higher momenta, thermal averaging leads to a clearly visible smoothing effect in the low-momentum regime (inset in Fig.~\ref{fig:shiftvskicks}). The zero-temperature curve exhibits a kink near $\Delta p \approx 1.0\,\hbar k \approx 1.3\,p_F$ (see also Fig.~\ref{fig:theories}), which is caused by the polaron 
entering a continuum of molecule-hole states, as we discuss in detail in Sec.~\ref{sec:pol_mol_transition}.
At finite temperature, the kink loses its sharpness and the whole low-momentum feature is partially washed out. Nevertheless, at our temperature, the non-monotonic feature of interest survives with reduced contrast, as can be clearly seen in the experimental data.

Our measurements at zero momentum transfer ($\Delta p = 0$) demonstrate the pure effect of thermal averaging. We observe a downshift of $-460$\,Hz (corresponding to $-0.014\,E_F$), as indicated by the blue arrow in the inset of Fig.~\ref{fig:shiftvskicks}.
The assumption of a constant effective mass (see Sec.~\ref{sec:shift}) greatly simplifies the calculation of the thermal averages based on Eq.~\eqref{eq:shift} (see App.~\ref{app:thermav} for more details) and, within this approximation, we arrive at a linear relation between the thermal shift $\langle \Delta E \rangle_{\rm th}$ and the temperature, with a proportionality constant essentially determined by the effective mass:
\begin{equation}
    \frac{\langle \Delta E \rangle_{\rm th}}{E_F} = - \frac{3 \, T}{2 \, T_F} 
    \left(1-\frac{m_\downarrow}{m_\downarrow^*} \right) \, .
    \label{eq:thshift}
\end{equation}
With $m_\downarrow^* = 1.080\,m_\downarrow$ calculated within our T-matrix approach for the given experimental parameters, we obtain a shift of $-350\,$Hz, which is about 30\% below the above experimental value. This small deviation is due to fact that in the relevant momentum range ($\bar{p} \approx 0.93\,p_F$), the effective mass already tends to increase with momentum, as we discuss in detail in App.~\ref{sec:effective_mass}.
In view of universality, it is interesting to note that the thermal polaron peak shift that was measured for a fermionic spin mixture of $^6$Li  in Ref.~\cite{Yan2019bau} follows Eq.~\eqref{eq:thshift} remarkably well.

\begin{table}
    \caption{Parameter values obtained by fitting the theoretical T-matrix model to the observed momentum-dependent energy shifts of the polaron. We list the results from five different  data sets, recorded at different magnetic detunings for the attractive (A-C) and the repulsive polaron (D, E). Note that the range parameter $k_F R^*$ is not an independent quantity, but it directly follows from $E_F$ (since $R^*$ is known).}
    \begin{ruledtabular}
    \begin{tabular}{ccccc}
         data set & $X$     & $E_F/h$ (kHz) & $T/T_F$ & $k_FR^*$\\
\hline  
         A   & 1.30(5) & 30.9(7)       & 0.1$^a$     & 0.718(8) \\
         B   & 1.03(2) &  32.7(3)      & 0.10(2) & 0.739(3) \\
         C   & 0.87(3) &  33.7(5)      & 0.07(3) & 0.750(6) \\
         D   & -0.93(3)&  33.0(6)      & 0.1$^a$     & 0.742(7) \\
         E   & -1.47(9)&  34.7(1.5)    & 0.1$^a$     & 0.761(16)\\
    \end{tabular}
\end{ruledtabular}
$^a$ Parameter value fixed.
    \label{tab:my_label}
\end{table}

We have recorded in total three data sets for the attractive polaron  (labeled A, B, C) with the interaction parameter set to different values in the regime of main interest ($X = 1.30, 1.03, 0.87$). Data set B is the one discussed above. We have also recorded two data sets (D and E) for the repulsive polaron (corresponding to $X = -0.93, -1.47$, respectively). Besides the controlled variation of $X$ by setting different magnetic detunings, all other parameters are kept fixed, but they are subject to small variations caused by fluctuations and long-term drifts in our apparatus. All data sets have been analyzed and fitted by the T-matrix theory in the way described above with set B serving us as an example.
The resulting fit parameter values are summarized in Table~\ref{tab:my_label}. 

\begin{figure}[t]
    \centering
    \includegraphics[trim=5 5 0 0,clip,width=1\columnwidth]{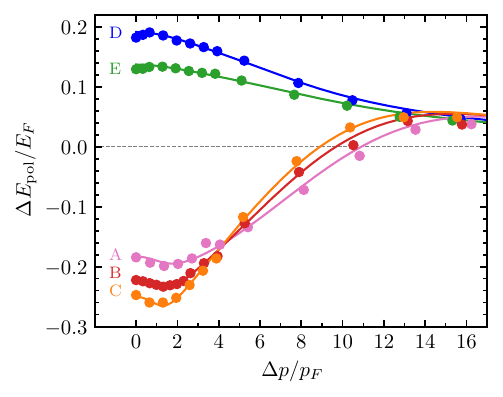}
    \vspace{-7mm}
    \caption{
    Momentum-dependent energy shifts for the attractive (A-C) and the repulsive  polaron (D, E) in a universal representation using reduced units. The solid lines represent the fits to the experimental data, based on the  T-matrix theory described in App.~\ref{app:theory} augmented by the thermal average presented in App.~\ref{app:thermav}. 
    The relevant parameter values are summarized in Table~\ref{tab:my_label}.
    The data points represent averages over typically 10 independent measurements. Error bars are smaller than the symbol size.}
    \label{fig:theory_and_exp}
\end{figure}

With precise values for $E_F$ and $p_F$ obtained from the fits, we can now introduce reduced units for energy and momentum and cast our results in a universal representation, independent of experimental details and the specific choice of the system.   
In Fig.~\ref{fig:theory_and_exp}, we show all our five experimental data sets together with the corresponding theory fits. 
The comparison highlights the remarkable agreement between experiment and theory in both polaronic branches in the full momentum range explored. 
In the regime of high momenta $\Delta p\gg p_F$ we see how both branches approach each other and become independent of $X$. Here, all curves finally exhibit the same positive shift, as predicted within the two-body scattering model (App.~\ref{app:pscatt}). For the attractive branch, this also implies a momentum-induced sign reversal, with attraction turning into an effective repulsion. The corresponding zero crossing appears at a position ($\sim$$10\,p_F$ in Fig.~\ref{fig:theory_and_exp}), in agreement with Eq.~\eqref{eq:zerocrossing}.

All three data sets and the corresponding fits (A-C in Fig.~\ref{fig:theory_and_exp}) exhibit the low-momentum behavior characteristic for the polaron's effective mass. 
We observe the maximum energy shift (minimum of $\Delta E_{\rm pol}$) at typically $1.5\,p_F$, which provides a reasonable estimate for the validity range of the assumption of a constant effective mass.
This also agrees with the momentum where the effective-mass behavior can be expected to connect with simple two-body scattering behavior (see inset in Fig.~\ref{fig:theories}).
Concerning the dependence on $X$, our three data sets indicate a general trend:  With decreasing $X$, the amplitude of the feature increases, while the momentum range over which it is  observed becomes narrower. This behavior is fully consistent with the results of our theoretical T-matrix description.

For the repulsive polaron (D and E in Fig.~\ref{fig:theory_and_exp}), the experimental data, recorded at $X = -0.93$ and $-1.47$, display the expected overall monotonic decrease of the energy shift with increasing momentum (see Sec.~\ref{sec:shift}). This simple behavior is in good agreement with the theoretical T-matrix model, and can be understood by the fact that both the emergence of the effective mass and the limitations of the scattering by unitarity act together to reduce the energy shift. 

We note that, for very low momenta $\Delta p \lesssim 0.5\,p_F$, we observe a slight downward deviation of the measured energy shift from the theoretical prediction by just a few percent. This surprising feature, which we observe only for the repulsive polaron, appears in a much narrower momentum range than the broader effective-mass feature. This deviation is attributed to a mechanism beyond our model. Not to be misled by this effect of unknown origin, we have excluded the first two data points in data sets D and E from our fit analysis. We speculate that the feature may be caused by residual effects related to the finite impurity density~\cite{Baroni2024mib}. Such mediated effects are beyond single-impurity physics and would require impurity-medium-impurity interactions, which may be more fragile than the direct impurity-medium interaction. This would explain the narrower momentum range. Another possible effect could be a local depletion of the medium caused by inelastic decay~\cite{Kohstall2012mac}. Even a rather slow motion would effectively wash out such a local depletion and thus remove the narrow feature. More studies at extremely low impurity densities would be necessary to unambiguously identify or rule out such effects.

\begin{figure}[t]
    \centering
    \includegraphics[trim=5 5 0 0,clip,width=1\columnwidth]{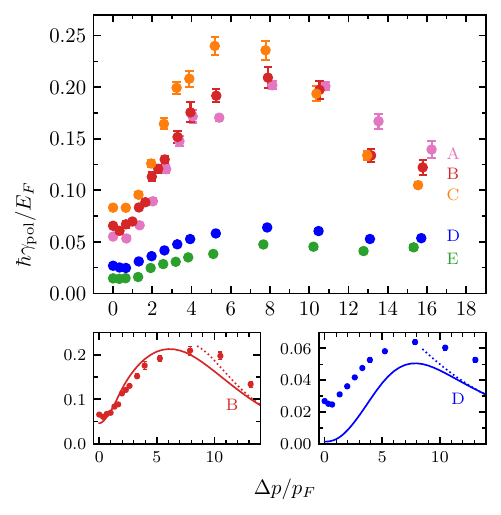}
    \caption{
    Momentum-dependent linewidths of the polaron peak for both the attractive (A-C) and the repulsive (D, E) case. The data have been derived from the same five data sets as the energy shifts reported in Fig.~\ref{fig:theory_and_exp}, with corresponding labels and color codes; for specific parameter values see Table~\ref{tab:my_label}. The large panel at the top presents an overview of all measurements over the full parameter range explored. For comparison with our T-matrix theory, the two lower panels  show  representative data sets for the attractive case (B) and the repulsive case (D). The solid lines result from the full T-matrix theory, including the approximate self-consistency and thermal average, and the dotted lines represent the results from the two-body vacuum scattering model presented in App.~\ref{app:pscatt}. 
    The data points represent averages over typically 10 independent measurements. Most of the error bars are smaller than the symbol size.}
    \label{fig:measured_widths}
\end{figure}

\subsection{\label{sec:resultsDecay}Linewidth of polaron peak}

The linewidth of the polaron peak provides information on the lifetime of the quasiparticle and possible decay mechanisms. In Fig.~\ref{fig:measured_widths}, we present the experimentally determined peak linewidth $\gamma_{\rm pol}$ as a function of the transferred momentum. The linewidths are extracted from the same RF spectra used to determine the energy shifts discussed above (Fig.~\ref{fig:theory_and_exp}); details of the extraction procedure are given in App.~\ref{app:dataanal}.
We note that the residual inhomogeneity sampled by the impurities in the center of the trap (Sec.~\ref{sec:experiment_A}) leads to a slight broadening effect with an estimated linewidth contribution of $\hbar \gamma_{\rm inhom} \sim 0.05 \, \Delta E$, which even for the largest shifts does not exceed $0.01\,E_F$.

In all cases, the experimental data points show a qualitatively similar behavior. At $\Delta p = 0$, the linewidth is relatively small and then starts to increase with momentum until a maximum is reached near $\Delta p/p_F \approx 7$, after which it slowly decreases  at higher momenta. 
The linewidth increases with decreasing interaction parameter $|X|$ (increasing interaction strength), and shows a slight shift of the maximum to lower $\Delta p/p_F$, which corresponds to the shift observed for the zero crossing of the energy in the two-body scattering regime.

As an important quantitative difference, we find that the linewidth of the repulsive polaron is generally much smaller than in the attractive case, comparing both situations for a similar magnitude $|X|$ of the interaction strength. The two selected sets B and D in Fig.~\ref{fig:measured_widths}, 
which correspond to $X\approx 1$ on the attractive branch and $X\approx -1$ on the repulsive branch, respectively, show a linewidth about four times larger for the attractive case. 
This different behavior can be attributed to a combination of finite range effects and the presence of the molecule-hole continuum, as we discuss in the following.

The attractive polaron is the ground state of the system as long as the interaction parameter $X$ stays above a critical value $X_{\rm cr}$ (for our system $X_{\rm cr} \approx 0.3 $~\cite{Massignan2012pad, Baroni2024mib}).

In this case of relatively weak attraction, 
molecular states lie above the polaron, where they form a `molecule-hole continuum'~\cite{Kohstall2012mac}. Here an energy gap separates this continuum from the attractive polaron. 
For stronger interaction ($X<X_{\rm cr}$), the gap disappears and a `dressed molecule' becomes the ground state, with the polaron-molecule coupling rendering the polaronic state energetically unstable. In the finite-temperature case, thermal excitations can bridge the gap, coupling the attractive polaron to the molecular continuum, and thus introducing additional decay. This effectively smoothens the polaron-molecule transition, as observed in Ref.~\cite{Ness2020ooa}. 

Our experiments on the attractive polaron have been carried out slightly above the zero-momentum polaron-molecule crossing  (i.e., at $X\gtrsim X_{\rm cr}$) where even small amounts of additional energy (thermal or resulting from the momentum kick) can induce molecule formation and decay. This leads us to a straightforward interpretation of the observed momentum-dependent linewidth. For $\Delta p = 0$, the small but finite linewidth of the attractive polaron peak can be explained as a result of thermal excitations into the molecule-hole continuum with an additional  small  contribution $\propto T^2$ coming from collisional broadening~\cite{Bruun2008cpo}. 
Consistent with this, our data at $\Delta p = 0$ exhibit an increasing linewidth 
when the polaron-molecule crossing is approached from above (sets A$\,\rightarrow \,$C in Fig.~\ref{fig:measured_widths}) and the energy gap between the attractive polaron and the molecule-hole continuum decreases. 
For increasing momentum, the additional energy that results from the controlled momentum transfer allows the attractive polaron to enter the molecule-hole continuum and leads to the rapid increase in the broadening seen in Fig.~\ref{fig:measured_widths}.
We further analyze this `motion-induced polaron-molecule transition' in Sec.~\ref{sec:pol_mol_transition}.
Finally, for large momenta the system leaves the many-body regime and the behavior can be understood in terms of elastic two-body collisions. Indeed, in the high-momentum case, where the width is observed to decrease, we find that $\gamma_{\rm pol}$ essentially reflects the unitarity-limited collisional broadening of an impurity atom scattering with the medium (dashed lines in Fig.~\ref{fig:measured_widths}).

The repulsive polaron is a metastable state, which in principle always has open decay channels. In the strongly interacting regime (vicinity of $X \approx -1$), decay can be attributed to two mechanisms~\cite{Massignan2011rpa,Kohstall2012mac, Scazza2017rfp, Scazza2022rfa}: In a two-body process, the repulsive polaron can decay into an attractive one, which then rapidly disintegrates within the molecular continuum. Apart from this indirect process, a three-body process can directly produce molecular excitations. 
Figure \ref{fig:measured_widths} shows that, in spite of the existence of these decay channels, the repulsive polaron linewidth is smaller than for the attractive polaron. The reason for this is two-fold: The energy of the repulsive polaron is above the molecule-hole continuum so that decay to molecules takes place via the slower three-body process. Moreover, the relatively large effective range associated with the resonant Li-K scattering substantially reduces collisional broadening for a positive scattering length, as discussed in App.~\ref{app:pscatt}.

Let us now compare the experimental results with predictions from our diagrammatic T-matrix theory, where the linewidth (half width at half maximum) of the polaron peak is obtained as 
\begin{equation}
\label{gamma_pol}
    \gamma_{\rm pol}=-Z_{\mathbf p}\,\textrm{Im}[\Sigma({\bf p}, E_{\bf p})],
\end{equation}
with $Z_{\mathbf p}$ the polaron residue (see App.~\ref{app:theory_decay} for its definition).

For the theory curves, we fix the parameter values to those extracted from the energy curves and summarized in Table~\ref{tab:my_label}. 
An inherent limitation of our T-matrix theory is the fact that it is not self-consistent, i.e., the self-energy contains only bare impurity propagators, rather than dressed ones. This corresponds to treating the out-going impurity in the scattering processes as non-interacting, which is clearly unphysical. As 
a consequence, the linewidth of the attractive polaron obtained from the non-self-consistent T-matrix theory is strictly zero below a certain threshold (here of order $p_F$).
 It on the other hand follows from Fermi liquid theory that collisions give rise to a broadening
 $\propto p^4$ of a polaron with momentum $p$~\cite{Bruun2008cpo}.
As discussed in detail in App.~\ref{app:theory_decay}, approximate self-consistency may be restored by a simple procedure, which consists in evaluating the self-energy in Eq.~\eqref{gamma_pol} at $E_{\bf p}-\varepsilon$ rather than at $E_{\bf p}$ itself (with $\varepsilon$ being the energy of the attractive polaron at zero momentum). This approach 
is equivalent to approximating the energy of the out-going impurity state in the scattering processes with Eq.~\eqref{eq:Epol} taking $m^*=m$, and it recovers the $p^4$ Fermi liquid collisional broadening.
Finally, as already discussed when presenting our results for the polaron energy, the impurities have a non-negligible mean thermal momentum, and we perform the additional thermal averaging described in App.~\ref{app:thermav} to include this.

The results of our theory are shown as solid lines in the lower panels of Fig.~\ref{fig:measured_widths} together with our corresponding experimental results. 
For the attractive polaron (case B in the lower left panel), the
agreement between theory and experiment is remarkable given the relative simplicity of our theory and the absence of free parameters. For high momenta, the theory recovers the two-body vacuum scattering result for collisional broadening.
For the repulsive polaron, corresponding to case D in the lower right panel, there is again good agreement between theory and experiment. Since the  magnitude of the broadening is much smaller for the repulsive polaron, the relative difference due to residual effects not included in the theory, such as e.g.\ spatial inhomogeneities, is larger. The experimental data for high momenta are again well described by two-body vacuum scattering, which  gives rise to a much smaller collisional broadening compared to the attractive polaron.
This is due to the scattering matrix having a relatively large and positive effective range $R^*$, as 
discussed in App.~\ref{app:pscatt} [see in particular  Eq.~\eqref{eq:MFImSigma}].

\begin{figure*}[t]
  \includegraphics[width=\textwidth]{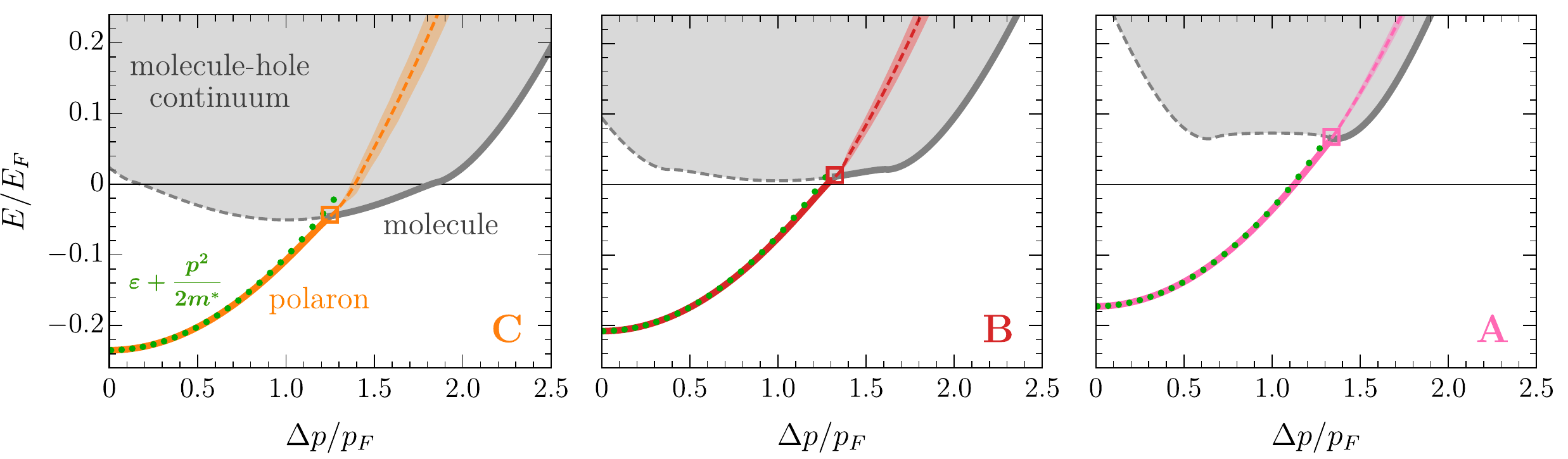}
  \caption{Energy spectrum (not energy shift) as a function of the impurity momentum. From left to right, the three panels are computed using the parameter values corresponding to the three data sets C, B, A. 
   The gray region represents the molecule-hole continuum, whose boundary is given by the Thouless pole condition in Eq.~\eqref{eq:Th_pole}. Beyond the polaron-molecule transition (indicated by empty squares), the polaron is no longer the ground state and it acquires a finite lifetime. The corresponding linewidth given by Eq.~\eqref{gamma_pol} is
  indicated in the spectra.}
  \label{fig:comparison_MHC}
\end{figure*}

\section{\label{sec:pol_mol_transition} Motion-induced polaron-molecule transition}
Our measurements of the energy shift and the linewidth of the attractive polaron peak show two simultaneous features at a critical momentum of the order of $p_F$: A distinct change in the derivative of the energy shift is accompanied by a substantial increase in the peak linewidth. Both of these features can be understood as signatures of the polaron-molecule transition, as already mentioned in the previous Sec.~\ref{sec:resultsDecay}. The transition has attracted considerable interest in theoretical work~\cite{Prokofev2008fpp,Punk2009ptm,Combescot2010ato,Bruun2010dop,Schmidt2011esa,Trefzger2012iia,Cui2020fpr,Peng2021not,Parish2021tso,Shi2025vtd}, but experimental data~\cite{Ness2020ooa} remain scarce. With the ability to vary the impurity momentum in a controlled way, our experiments offer a new degree of freedom to benchmark theoretical predictions.

To characterize the transition,  we make use of our diagrammatic treatment and show in Fig.~\ref{fig:comparison_MHC} the energy spectrum of the impurity problem for the three combinations of interaction parameter and effective range corresponding to the experimental conditions for which the data sets A, B, C were taken. 
The polaron energy is obtained from the T-matrix calculations by solving  Eq.~\eqref{eq:pol_energy}. 
At very small momentum the ground state is a well-defined polaron, whose spectral linewidth only comes from thermal effects as discussed above, and there is an excellent agreement with the effective-mass description in Eq.~\eqref{eq:Epol} (green dotted line).
However, as the impurity momentum increases, the polaron energy  enters the molecule-hole continuum. Its lower rim is given by the so-called molecular Thouless pole, i.e., the energy $E_\mathbf{p}^\text{cont}$, which is a solution of the implicit equation~\cite{Trefzger2012iia}
\begin{equation}    \label{eq:Th_pole}
    {\rm Re}[\mathcal{T}^{-1}((p-p_F)\mathbf{n}_\mathbf{p}, E_\mathbf{p}^\text{cont})] = 0.
\end{equation}
Here, $\mathcal{T}$ is the T-matrix (see App.~\ref{app:theory}) and  $\mathbf{n}_\mathbf{p} = \mathbf{p}/p$ is a unit vector in the direction of $\mathbf{p}$. Physically, Eq.~\eqref{eq:Th_pole} describes an impurity with momentum ${\bf p}$ forming a molecule with a fermion at the Fermi surface with momentum pointing in the opposite direction.
Note that the Thouless pole condition can be equivalently derived from a variational approach which considers the formation of a bare $\uparrow$-$\downarrow$ dimer without particle-hole excitations of the surrounding Fermi sea~\cite{Parish2021tso}. 
Beyond the transition point (identified by an empty square in each panel), the polaron is no longer the ground state in the spectrum. The lowest excitation is now a molecule, which may be formed as a result of the collision between the polaron and a particle from the medium, a process which is faithfully recovered by our T-matrix calculation. This gives rise to a new decay channel for the polaron and a corresponding rapid increase in its spectral linewidth that we observed in Fig.~\ref{fig:measured_widths}. 
\begin{figure}[t]
  \includegraphics[trim=5 5 0 5,clip,width=1\columnwidth]{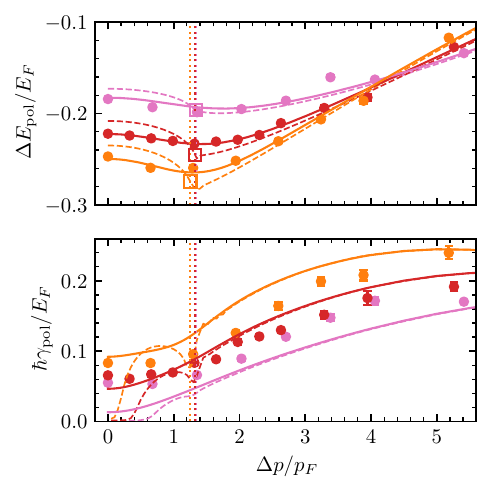}
  \caption{Behavior at small momentum of the energy shift (top panel) and linewidth (lower panel) of the attractive polaron. Data sets and color codes are the same as in Figs.~\ref{fig:theory_and_exp}-\ref{fig:measured_widths} (pink, red, and orange for data sets A, B, and C, respectively). The dashed (solid) lines represent the $T=0$ (thermally averaged) T-matrix result. The empty squares and the vertical lines indicate the location of the polaron-molecule transition, as identified in Fig.~\ref{fig:comparison_MHC}.}
  \label{fig:transition}
\end{figure}
Figure~\ref{fig:transition} further illustrates that the polaron spectrum is significantly affected when it enters the molecule-hole continuum. Indeed, the momentum at which the transition occurs in the $T=0$ theoretical spectra matches the location of the sharp features visible in the $T=0$ theory data (dashed lines in Fig.~\ref{fig:transition}) for both the energy shift and the linewidth. Non-zero temperatures smoothen these sharp features, but nonetheless they remain visible in both the thermally-averaged theory (solid lines) and the experimental measurements (filled circles).

We note here that the polaron-molecule transition is also clearly visible in the discontinuous behavior of the effective mass tensor. 
The latter has longitudinal and transverse components that are larger than the bare mass when the polaron is the ground state. 
However, both components become smaller than the bare mass beyond the transition point, where the polaron becomes an excited state above the molecule. This is discussed in detail in App.~\ref{sec:effective_mass}.

\section{Conclusions and Outlook}\label{sec:conclusion}

We measured the dispersion and spectral linewidth of Fermi polarons over a wide range of impurity momenta, from well below to far above the Fermi momentum. 
We employed an optically trapped mixture of $^{41}$K and $^6$Li with tunable interspecies interaction, where the dilute non-degenerate K component served us as impurity atoms and the Li as the deeply degenerate fermionic medium. Employing a species‑selective optical Raman accelerator, we transferred a well-controlled amount of momentum to the impurity atoms and, by means of high‑precision radio-frequency spectroscopy, we studied the effect on the polaron peak.

Across all momenta, we find our experimental data to be in excellent agreement with a diagrammatic T‑matrix calculation including thermal averaging. We identify a crossover from a low‑momentum many‑body regime, where the polaron obeys a Fermi‑liquid expression with a constant effective mass, to a high‑momentum regime governed by unitarity‑limited two‑body scattering, in which both attractive and repulsive branches approach the behavior of a weakly interacting bare particle with a positive energy shift. Notably, the attractive polaron displays an intriguing non‑monotonic momentum dependence. At low momenta, the interaction energy exhibits a downshift, which is a direct consequence of a constant effective mass, which is larger than the bare mass. An intermediate momentum regime is characterized by an increasing energy upshift and large broadening of the spectral peak, the latter decreasing in the high-momentum regime. This observation is consistent with a motion‑induced polaron–molecule transition, as predicted by our theory. 
In contrast to the rich behavior in the attractive case, the repulsive polaronic branch evolves smoothly from a dressed impurity at low momentum to a weakly interacting bare particle at high momentum, with a monotonic change of the energy shift.

The excellent agreement with our experimental observations represents a benchmark for the general validity of the diagrammatic T-matrix approach. While experiments are naturally limited to specific impurity-bath mass ratios and by the availability of Feshbach resonances of different strengths, the theoretical approach can be applied to any hypothetic system and thus to model specific situations and to identify the most favorable conditions for certain applications of interest. Tunable impurity systems with extreme mass ratios, e.g.\ realized with Li-Cs \cite{Repp2013ooi, Tung2013umo}, Li-Er \cite{Schafer2022fro}, or Li-Dy \cite{Xie2025fso} 
mixtures, represent a promising topic of future research on polaron physics in the heavy-impurity limit \cite{Lippi2024PhD} where Anderson orthogonality arises~\cite{Anderson1967ici, Knap2012tdi},  and of very light ones, which have intriguing connections to few-body physics \cite{Kartavtsev2007let, Greene2017ufb, Naidon2017epa, Christianen2022bpa}. For anisotropic situations with dipolar interactions, the controlled momentum transfer to an impurity in different directions will allow us to explore the theoretically predicted anisotropic polaron dispersion relation \cite{Nishimura2021gso, Kain2014pia}.

Another very interesting future research direction naturally emerging from our results concerns the properties of the moving polaron for a non-zero concentration. Here, effects like Fermi blocking of the polaron scattering states and polaron-polaron interactions will set in, and such experiments will directly explore the fundamental transport properties of Fermi liquids~\cite{Enss2025qti} in a pristine and tunable setting. Another interesting topic is the moving polaron in superfluids and strongly correlated environments~\cite{Astrakharchik2004moa, Castin2015lvc, Koepsell2019imp}. 

Moreover, our species‑selective Raman setup also enables rapid quenches of the impurity–bath interaction, with or without momentum transferred to the impurities, on timescales shorter than the characteristic response time of the fermionic medium. This versatile tool will allow us to probe, in a very flexible way, the initial transient dynamics of both moving and static impurities interacting with the medium, for example by Ramsey interferometry~\cite{Cetina2016umb}. 

Finally, systems with more complex dispersion relations are very common in quantum many-body physics. We expect ultracold gases with spin–orbit coupling~\cite{Lin2011soc, Cheuk2012sis, Wang2012spc, Shi2025vtd}, where the translational motion couples to internal degrees of freedom, to offer many exciting applications and topics for future research, where our toolbox of controlled momentum transfer and spectroscopic probing may provide new insights into the underlying physics.

\section*{Acknowledgments}

We thank Richard Schmidt, Meera Parish and Jesper Levinsen for stimulating discussions.
We also express our gratitude to all members of the ultracold atom groups in Innsbruck for providing a highly collaborative environment and sharing valuable insights into the broader field of quantum-gas mixtures.
The project has received funding from the European Research Council (ERC) under the European Union’s Horizon 2020 research and innovation programme (Grant Agreement No.~101020438 - SuperCoolMix).
P.~M. acknowledges support by the Spanish Ministerio de Ciencia, Innovación y Universidades (grant PID 2023-147469NB-C21, financed by MICIU/AEI/10.13039/501100011033 and FEDER-EU) and by the ICREA {\it Academia} program. G.M.B. acknowledges support from the Novo Nordisk Foundation (grant no. NNF23OC0086599).

\section*{Data availability}

The data that support the findings of this article are openly available~\cite{data-repository}.

\appendix

\section{\label{app:theory}Theoretical model}
In this section, we denote the K$|1\rangle$ atoms as spin  $\downarrow$ impurities with mass $m_\downarrow$, which are immersed in a gas of Li$|1\rangle$ atoms denoted as spin $\uparrow$ majority particles with mass $m_\uparrow$. The density of the majority particles is $n$, corresponding to a Fermi wavevector $k_F = (6 \pi^2 n)^{1/3}$ and Fermi energy $E_F= \hbar^2 k_F^2/2m_\uparrow= p_F^2/2m_\uparrow$.

The retarded Green's function of a $\downarrow$ impurity with momentum $\mathbf{p}$ and energy $\omega$ reads
\begin{equation}
    G_{\downarrow}(\mathbf{p}, \omega) = \frac{1}{\omega+ i0^+ - \epsilon_{\mathbf{p} \downarrow} - \Sigma(\mathbf{p}, \omega)},
    \label{eq:G_imp}
\end{equation}
where $\epsilon_{\mathbf{p} \downarrow}= p^2/2m_\downarrow$ is the dispersion of a bare impurity, and the self-energy $\Sigma(\mathbf{p},\omega)$ describes the effects of the interaction with the surrounding gas of $\uparrow$ particles. Within the ladder approximation, which is known to be remarkably accurate even for strong interactions~\cite{Massignan2026pia,Massignan2014pdm}, the polaron self-energy is given by 
\begin{equation}
    \Sigma(\mathbf{p}, \omega) = \sum_{\textbf{q}} n_F(\xi_{\mathbf{q}\uparrow}) \mathcal{T}(\mathbf{p}+\mathbf{q}, \omega + \xi_{\mathbf{q}\uparrow}), 
    \label{eq:self-energy}
\end{equation}
where $\xi_{\mathbf{q}\uparrow}=\epsilon_{\mathbf{q} \uparrow}-\mu$ with $\epsilon_{\mathbf{q} \uparrow} = q^2/2m_\uparrow$ is the dispersion of the majority particles with chemical potential $\mu$, $n_F(\xi)=1/[\exp(\xi/(k_B T))+1]$ is the Fermi function, and we have set the system volume to unity. 
The self-energy~\eqref{eq:self-energy} includes all the two-body scattering processes between the impurity and majority particles in a many-body environment via the scattering matrix 
\begin{equation}
    \mathcal{T}^{-1}(\mathbf{K}, \omega) = \mathcal{T}_0^{-1}(\mathbf{K}, \omega) - \Pi(\mathbf{K}, \omega),
    \label{eq:LippSchw}
\end{equation}
with $\mathbf{K}$ and $\omega$ the transferred momentum and energy. 
The first term reads~\cite{Bruun2005msa}
\begin{equation}
    \mathcal{T}_0(\mathbf{K}, \omega) = \frac{2 \pi \hbar^2}{m_r} \left[ \frac{1}{a} + R^* k_r^2(\mathbf{K}, \omega) \right]^{-1}
    ,
\label{eq:vacuum_T_mat}
\end{equation} 
where $a$ is the scattering length of the $\uparrow$-$\downarrow$ interaction and 
 $m_r = m_\uparrow m_\downarrow/(m_\uparrow+m_\downarrow)$ is the reduced mass of the two particles. Since the Li-K Feshbach resonance is not broad, finite range effects of the interaction are included through the range parameter $R^*=\hbar^2/(2m_ra_\text{bg} \delta\mu \, \Delta)$~\cite{Petrov2004tbp} and the energy-dependent relative wavenumber
\begin{equation}
    k_r(\mathbf{K}, \omega) = \sqrt{\frac{2m_r}{\hbar^2}\left(\omega - \frac{K^2}{2M}+E_F\right)}.
    \label{kappa_r}
\end{equation}
Here,  $M=m_\uparrow+m_\downarrow$, $\delta\mu$ is the difference between the magnetic moment of the two scattering particles in the open and closed channels, $a_\text{bg}$ is the background scattering length, and $\Delta$ is the width of the Feshbach resonance so that $a=a_\text{bg}[1-\Delta/(B-B_0)]$. 
The second term in Eq.~\eqref{eq:LippSchw} is the pair propagator of an impurity and a majority atom with total center-of-mass momentum $\mathbf{K}$ and energy $\omega$, which is given by 
\begin{align}
    \Pi(\mathbf{K}, \omega) =\sum_{\textbf{k}} \left( \frac{1-n_F(\xi_{\mathbf k\uparrow})}{\omega - \epsilon_{\mathbf{K}-\mathbf{k} \downarrow} - \xi_{\mathbf{k} \uparrow}} + \frac{2m_r}{k^2} \right). 
    \label{eq:pair_prop}
\end{align} 
Interactions between the $\downarrow$ particle and the $\uparrow$ gas give rise to the formation of the polaron with a dispersion $E_{\mathbf p}$ determined from the pole of the impurity Green's function~\eqref{eq:G_imp}, i.e.,
\begin{align}
    E_{\mathbf{p}} = \epsilon_{\mathbf p\downarrow} + {\rm Re}\, \Sigma(\mathbf{p}, E_{\mathbf{p}}).
    \label{eq:pol_energy}
\end{align}
From this, we obtain the energy shift due to interactions as 
$\Delta E_p = E_\mathbf{p}-\epsilon_{\mathbf p\downarrow} = {\rm Re}\, \Sigma(\mathbf{p}, E_{\mathbf{p}})$. 

\subsection{Polaron linewidth}
\label{app:theory_decay}
While the energy obtained from Eq.~\eqref{eq:pol_energy} using the ladder approximation for the self-energy is very accurate when compared with experiments and with Monte-Carlo calculations~\cite{Massignan2026pia,Massignan2014pdm}, it is known  to predict a vanishing collisional broadening for a moving impurity as long as its momentum remains smaller than a critical value $0<p<p_c$ for zero temperature~\cite{Trefzger2013edr}. 
This prediction is unphysical since a moving polaron will always relax towards zero momentum due to scattering with the majority particles, and it can indeed be shown that the collisional broadening coming from momentum relaxation  scales as $p^4$ for $0<p\ll p_F$~\cite{Bruun2008cpo}. To understand the origin of this unphysical prediction of the ladder approximation, we note that collisional broadening gives rise to a  non-zero imaginary part of the self-energy. 
From Eqs.~\eqref{eq:self-energy}, \eqref{eq:LippSchw} and \eqref{eq:pair_prop} it follows that $\text{Im}\Sigma({\mathbf p}, E_\mathbf{p}+i0^+)>0$ requires that the equation
\begin{equation}
E_\mathbf{p}=\epsilon_{\mathbf k\uparrow}-\epsilon_{\mathbf q\uparrow}+\epsilon_{{\mathbf p}+{\mathbf q}-{\mathbf k}\downarrow}
    \label{eq:en_cons}
\end{equation}
has a solution for some values of ${\mathbf k}$ and ${\mathbf q}$. For zero temperature, we have $q<p_F<k$ so that the right hand side of Eq.~\eqref{eq:en_cons} is positive. 
Since $E_\mathbf{p}<0$ for the attractive polaron for a range of momenta, Eq.~\eqref{eq:en_cons} has no solution and the predicted broadening vanishes.  
Equation \eqref{eq:en_cons} expresses energy conservation of a polaron with momentum $\mathbf{p}$ and a majority particle with momentum $\mathbf{q}$ scattering into a \emph{bare} impurity particle with momentum ${\mathbf p}+{\mathbf q}-{\mathbf k}$ and a majority particle with momentum $\mathbf{k}$.
This is of course unphysical since the scattered impurity particle also ends up in a polaron state with energy  $E_{{\mathbf p}+{\mathbf q}-{\mathbf k}}$. When this is taken into account, Eq.~\eqref{eq:en_cons} will have solutions for any non-zero momentum ${\mathbf p}$. 

This deficiency of the ladder approximation can be removed by using full impurity Green's functions instead of bare ones in the self-energy. It is however not clear if such a self-consistent T-matrix approach will give more accurate predictions of the energy as compared to the bare ladder approximation outlined above (especially at low temperature and small impurity concentration~\cite{Hu2018}), and we therefore instead choose a more pragmatic approach. In practice, we simply shift the impurity energy appearing in the pair propagator by the energy $E_{0}<0$ of the static attractive polaron when calculating the collisional broadening,  i.e., we make the replacement $\epsilon_{\mathbf {K-k}\downarrow}\rightarrow \epsilon_{\mathbf {K}-\mathbf {k}\downarrow}+E_0$ in Eq.~\eqref{eq:pair_prop}. 
Physically, this corresponds to neglecting effective mass effects so that the polaron has the energy $E_{\mathbf p}=E_0+p^2/2m_\downarrow$.
This procedure for calculating the broadening of the polaron due to collisions with the majority particles amounts to  evaluating 
\begin{equation}
    \gamma_{\rm pol}
    = - Z_{\mathbf p}\text{Im} \Sigma(\mathbf{p}, E_\mathbf{p}-E_0),
    \label{eq:width_shift}
\end{equation}
where  $Z_{\mathbf p}=[1 - \partial_{\omega}{\rm Re}\Sigma(\mathbf{p},\omega)|_{\omega=E_{\mathbf p}}]^{-1}$ is the quasiparticle residue, which quantifies the overlap between the polaron wave function and the bare impurity state.
It can be shown that this approach gives the correct collisional relaxation rate of the attractive polaron scaling as $p^4$ for small momenta~\cite{Bruun2008cpo}.
The broadening of the Fermi polaron has recently been calculated using a self-consistent theory 
based on the 
functional renormalization group~\cite{vonMilczewski2024}.

\section{Scattering at high momenta}
\label{app:pscatt}
When the momentum of the impurity is much larger than the Fermi momentum, $p\gg p_F$, Fermi blocking in the scattering processes can be ignored, and we can approximate the scattering matrix ${\mathcal T}$  by its vacuum value ${\mathcal T}_v$. This amounts to omitting the Fermi blocking factor (i.e., the Fermi distribution) in the pair propagator Eq.~\eqref{eq:pair_prop}, which can then be computed analytically for zero temperature as
\begin{align}
    \Pi_v(\mathbf{p},\omega) 
    &= - \frac{m_r}{2 \pi \hbar^2} \, i k_r(\mathbf{p}, \omega).
\end{align}
Plugging this result into Eq.~\eqref{eq:LippSchw}, one recovers the standard expression for the low-energy scattering matrix
\begin{equation}
    \mathcal{T}_v (\mathbf{p},\omega) = - \frac{2 \pi \hbar^2}{m_r} f_v(k_r(\mathbf{p},\omega)),
    \label{eq:Tv}
\end{equation}
where the $s$-wave scattering amplitude for a narrow Feshbach resonance reads~\cite{Zaccanti2025mif}
\begin{equation}
f_v(k_r) = - \frac{1}{\frac{1}{a} + R^*k_{r}^2 + i k_{r}} .
\end{equation}
The imaginary part of the scattering amplitude yields the $s$-wave scattering cross section
\begin{equation}
    \sigma = \frac{4\pi}{k_r} {\rm Im} f_v \, .
\end{equation}
For a given collisional momentum $\hbar k_r$, the cross section shows a maximum at a finite negative value of the scattering length $a_{\rm max} = -1/(R^* k_r^2)$, which corresponds to a momentum-dependent resonance shift~\cite{Zaccanti2025mif, Horvath2017ats}. At the resonance point, the cross section acquires its maximum, unitarity-limited value $\sigma_{\rm max} = 4\pi/k_r^2$.

We can now calculate the energy shift of the polaron. For large momenta, the 
scattering energy $p^2/2m_\downarrow$ of the impurity is large compared to $E_F$ and we can approximate 
Eq.~\eqref{eq:self-energy} as 
\begin{equation}
    \Sigma_v(\mathbf{p}) = \sum_{\mathbf{q}} n_F(\xi_{\mathbf{q}\uparrow}) \mathcal{T}_v\left(\mathbf{p},\frac{p^2}{2m_\downarrow}\right) = - \frac{2 \pi \hbar^2 n}{m_r} f_v(k_p),
    \label{eq:sigmaMF}
\end{equation}
where the scattering amplitude is evaluated at the on-shell collision wavenumber 
\begin{equation}
    k_r\left( \mathbf{p},\frac{p^2}{2m_\downarrow} \right) \approx \frac{m_r}{m_\downarrow} \frac{p}{\hbar}\equiv k_p.
    \label{eq:kp}
\end{equation}
Notice that neglecting Fermi blocking effects in the polaron energy is consistent with a \emph{fast-impurity approximation}, where one assumes that the impurity momentum gives the dominant contribution to the collisional wavenumber. 
This situation can be achieved in our experiments when an impurity atom of mass $m_\downarrow= 41\,{\rm u}$ is quickly moving through the fermionic medium composed of atoms of mass $m_\uparrow=6\,{\rm u}$  ($m_r = 5.23\,{\rm u}$). 

The polaron energy shift is readily found taking the real part of Eq.~\eqref{eq:sigmaMF}:
\begin{align}
    \Delta E_{v}(p) 
    &= \frac{2 \pi \hbar^2 n a}{m_r} \frac{1 + a R^* k_p^2}{\left( 1 + a R^* k_p^2 \right)^2 + \left( a k_p \right)^2}.
    \label{eq:MFvsp}
\end{align}
In Fig.~\ref{fig:theories} we have shown example curves for $X=-(k_F a)^{-1}=\pm 1$.
There we have also discussed the behavior of the energy shift and the good agreement with the ${\mathcal T}$-matrix theory in the high-momentum region.

An interesting feature which can be explained in terms of the vacuum scattering  model is the sign reversal that appears for the attractive polaron ($X>0$) at a specific momentum $p_c$. From Eq.~\eqref{eq:MFvsp} it is straightforward to show that the energy shift will vanish at the specific momentum
\begin{equation}
    p_* 
    = \hbar k_F\frac{m_\downarrow}{m_r} \sqrt{ \frac{X}{k_F R^*}  }.
\label{eq:zerocrossing}
\end{equation}
This gives $p_* \approx 9\,p_F$ for our typical experimental conditions, which nicely agrees with our observation.
For higher impurity momenta ($p>p_*$), the initially attractive energy shift becomes repulsive. This is a striking consequence of the momentum-dependent shift~\cite{Zaccanti2025mif} associated with a narrow Feshbach resonance. 
This phenomenon has no counterpart for the repulsive polaron found at $X<0$, whose energy always retains a positive sign.

Finally, it is instructive to consider the large momentum behavior of the polaron energy shift. An expansion of Eq.~\eqref{eq:MFvsp} to leading order in $p \gg p_F$ yields
\begin{equation}
 \Delta E_{v}(p)    
 \simeq \frac{2 E_F}{3 \pi k_F R^*} \frac{(m_\uparrow+m_\downarrow)^3}{m_\uparrow^2 m_\downarrow} \left( \frac{p}{p_F} \right)^{-2},
\end{equation}
which retains information on the narrow resonance (through the range parameter $R^*$), but becomes independent of the interaction parameter $X$. This explains why the two results for the attractive ($X=+1$) and repulsive ($X=-1$) polaron in Fig.~\ref{fig:theories} eventually converge onto the same curve at large impurity momenta. Such behavior is consistent with our experimental data (see Fig.~\ref{fig:theory_and_exp}). 

\begin{figure}[t]
    \centering
    \includegraphics[width=\columnwidth]{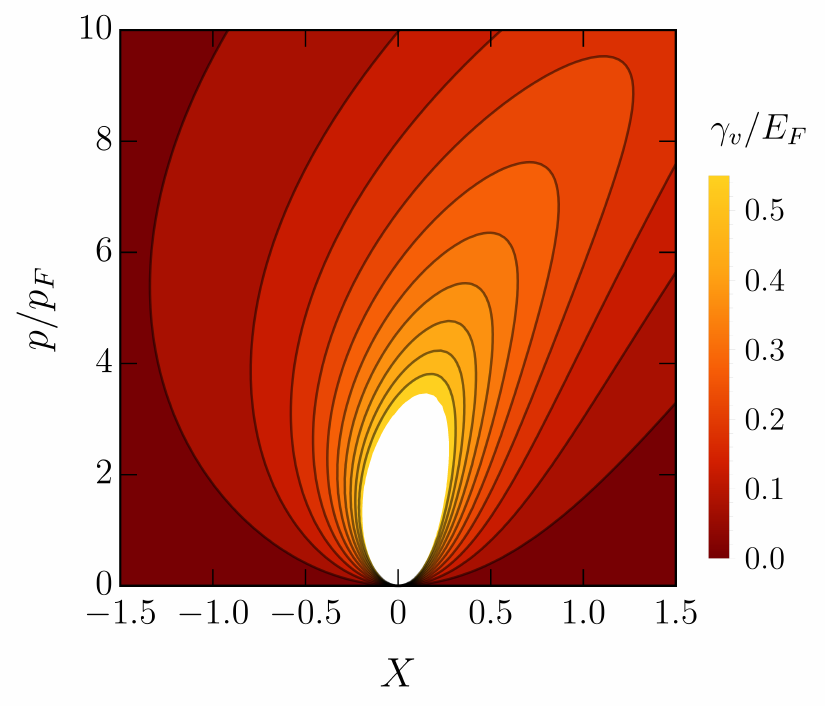}
    \caption{Collisional broadening in the vacuum (two-body) limit as a function of interaction parameter $X$ and momentum $p$, evaluated for $k_FR^*=0.75$.}
\label{fig:gamma_vac}
\end{figure}

For the imaginary part of the vacuum self-energy~\eqref{eq:sigmaMF}, instead, one obtains
\begin{align}
    \gamma_v=&-\text{Im}\Sigma_v(p) 
    = \frac{2 \pi \hbar^2 n a}{m_r} \frac{k_p a}{\left( 1 + k_p^2 a R^*  \right)^2 + \left(k_p a \right)^2} \notag \\
    &\simeq \frac{2 E_F}{3 \pi (k_F R^*)^2} \frac{(m_\downarrow+m_\uparrow)^4}{m_\uparrow^3 m_\downarrow} \left( \frac{p}{p_F} \right)^{-3},
    \label{eq:MFImSigma}
\end{align}
where the second line is the asymptotic behavior at large momenta, which is once more independent of $X$.
A plot of this collisional broadening  in the vacuum approximation is shown in Fig.~\ref{fig:gamma_vac}. Interestingly, already this vacuum (or two-body) approximation shows many of the features visible in the experiment, such as, for example, the fact that the collisional broadening shoots up rapidly for small momenta and then it decreases as a power-law for larger ones. Moreover, when $R^*>0$ the collisional broadening is much larger for positive values of $X$ [i.e., for a given $X_0>0$, we have $\gamma_v(X_0)\gg \gamma_v(-X_0)$]. Notice that the latter feature is specific to a narrow resonance: as can be seen in Eq.~\eqref{eq:MFImSigma}, with $R^*=0$ the collisional broadening rate would be symmetric around resonance.

\section{Polaron effective mass tensor}
\label{sec:effective_mass}

Knowledge of the polaron dispersion gives direct access to the effective mass of the polaron, which describes the dynamical response along direction $i$ due to a momentum kick given along direction $j$. 
The effective mass is therefore a tensor, given by~\cite{Ashcroft1976ssp,Trefzger2013edr}
\begin{align}
    \left[ M^{-1}(\mathbf{p}) \right]_{ij} &= \frac{\partial^2 E_{\mathbf{p}}}{\partial p_i \partial p_j} \notag \\
    &= \left(\frac{p_i p_j}{p^2}\right) \frac{\partial^2 E_\mathbf{p}}{\partial p^2} + \left( \delta_{ij} - \frac{p_i p_j}{p^2} \right) \frac{1}{p} \frac{\partial E_\mathbf{p}}{\partial p},
\end{align}
where the second line holds when the energy dispersion only depends on the magnitude of $\mathbf{p}$, as is the case in an isotropic medium like ours (and we have used that with $p=\sqrt{\sum_i p_i^2}$ one has $\partial p/\partial p_i=p_i/p$). The terms in round parentheses in the latter equation correspond respectively to the projector along the direction of the momentum, and perpendicular to it. This naturally leads to two distinct effective masses, longitudinal and transverse, which account for changes in the dispersion arising respectively from variations in the magnitude and in the direction of the momentum of the quasiparticle:
\begin{subequations}
    \begin{align}
        \label{eq:eff_mass_par}
        m^*_\parallel(p) &= \left( \frac{\partial^2 E_\mathbf{p}}{\partial p^2} \right)^{-1}, \\
        \label{eq:eff_mass_perp}
        m^*_\perp(p) &= 
        \left(\frac{1}{p} \frac{\partial E_\mathbf{p}}{\partial p}\right)^{-1}.
    \end{align}
\end{subequations}
In the $p \to 0$ limit, $E_{\bf p}$ is purely quadratic, and the two quantities converge to the same result, which is the direction-independent $m^*$ introduced in Eq.~\eqref{eq:Epol}.

\begin{figure}[t]
    \centering
    \includegraphics[width=\columnwidth]{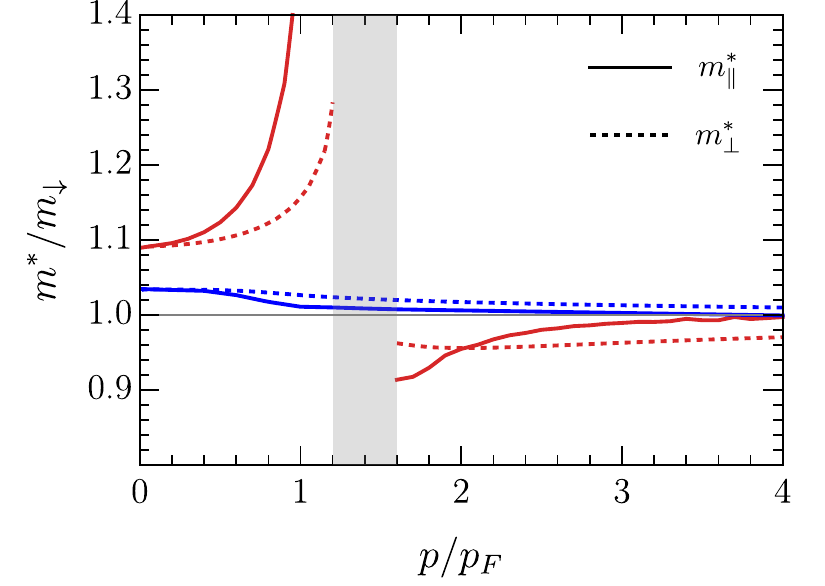}
    \caption{Longitudinal (solid lines) and transverse  (dashed lines) effective masses of the attractive (repulsive) polaron computed at $X=+1$ ($X=-1$) are shown in red (blue). In both cases, we took $k_FR^*=0.75$. Data for the attractive case in the gray region are not shown since they were too noisy.
    }
\label{fig:effective_masses}
\end{figure}

The longitudinal and transverse masses of the attractive polaron extracted from our theoretical model at $X=+1$, together with the same quantities for the repulsive polaron at $X=-1$, are shown in Fig.~\ref{fig:effective_masses}. At low momentum, as expected, all effective masses are larger than the bare mass. At higher momenta, the repulsive polaron evolves smoothly into a bare K atom, as its dressing cloud becomes increasingly lighter. This is signaled by a uniform decrease of the effective masses that approach the bare mass. 
In contrast, the scenario is much richer in the attractive case: The effective masses grow rapidly in the region where the attractive polaron is the ground state, then they suddenly drop in the region where the polaron-molecule transition takes place, and they eventually become smaller than the bare mass for even larger momenta. The very uncommon situation of polarons being effectively lighter than the impurities themselves is only possible in the region where the attractive polarons are no longer the ground state (having molecules below them), as described in Sec.~\ref{sec:pol_mol_transition}. 

\section{Thermal averaging over impurity momentum}
\label{app:thermav}

The momentum of an impurity atom can be understood as a sum of two contributions. Besides the deliberate Raman momentum change $\Delta p$, which is always directed along a given axis ($z$-axis), there is a thermal component with random direction. Accordingly, the impurity momentum can be written as
\begin{equation}
    \vec{p} = \Delta p \, \vec{\hat{z}} + p_{\rm th} 
     ( \xi_x \, \vec{\hat{x}}  
     +  \xi_y \, \vec{\hat{y}} + \xi_z \, \vec{\hat{z}}) \, ,
\end{equation}
where the coefficients $\xi_i$ are random numbers following a Gaussian normal distribution with unit standard deviation. The thermal momentum is given by
\begin{equation}
    p_{\rm th} = \sqrt{m_\downarrow k_B T} 
    = \sqrt{\frac{m_\downarrow}{2m_\uparrow} \frac{T}{T_F}} \, p_F \,,
\end{equation}
For our experimental conditions, where $m_\downarrow/m_\uparrow = 41/6$ and typically $T/T_F = 0.1$, we obtain $p _{\rm th} = 0.585\,p_F$.

Our experimental observables, the polaron energy shift $\Delta E_{\rm pol}$ and the spectral linewidth $\gamma_{\rm pol}$, are functions of the momentum $p = |\vec{p}|$. To deal with the thermal broadening, we introduce the normalized momentum distribution function $w(p, \Delta p, p_{\rm th})$, where the momentum change $\Delta p$ and the thermal momentum $p_{\rm th}$ enter as additional parameters. For any $p$-dependent quantity $f(p)$, we can now define the thermal average as
\begin{equation}
    \langle f(\Delta p, p_{\rm th})\rangle \equiv \int_0^\infty {\rm d}p \, w(p, \Delta p, p_{\rm th}) \, f(p) \,.
    \label{eq:prandom}
\end{equation}

\begin{figure}[t]
    \includegraphics[trim=5 5 0 0
    0,clip,width=1\columnwidth]{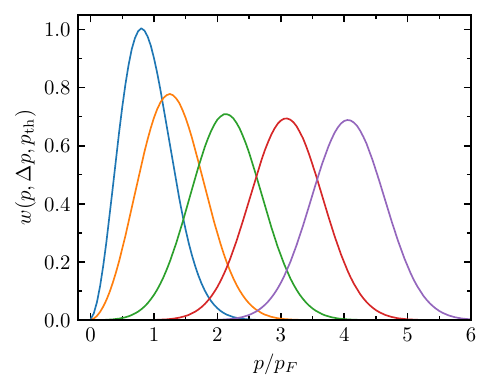}
        \caption{Momentum distribution function $w(p, \Delta p, p_{\rm th})$ for impurity atoms moving through the Fermi sea after a momentum change $\Delta p$. The thermal impurity momentum is $p_{\rm th}/p_F = 0.585$ (corresponding to $T/T_F = 0.1$). From left to right, the momentum change increases from $\Delta p /p_F=0$ (blue curve, pure thermal motion) in steps of $1$ and quickly enters a regime where the transferred momentum dominates the total momentum.}
    \label{fig:pdistrib}
\end{figure}
In Fig.~\ref{fig:pdistrib}, we show examples of the normalized momentum distribution function for increasing values of $\Delta p / p_F$ and fixed $p_{\rm th}/p_F = 0.585$ ($T/T_F = 0.1$). The curves have been computed numerically based on Eq.~\eqref{eq:prandom} using a straightforward Monte-Carlo method.

Without momentum change ($\Delta p =0$, blue curve) one recovers the usual Maxwell-Boltzmann distribution with an average momentum $\bar{p} \equiv \langle p \rangle = \sqrt{8/\pi} \, p_{\rm th} \approx 0.934\,p_F$. This means that any feature at low momentum $p/p_F \lesssim 1$ will be strongly affected and washed out by the thermal averaging, with the consequence of substantially reduced observability in the experiment. This is exactly what our experimental results show in the low-momentum regime (see Fig.~\ref{fig:shiftvskicks} in the main text). In contrast, in the high-momentum regime, where the momentum distribution acquires a Gaussian shape (standard deviation $p_{\rm th}$), thermal averaging has only little effect on the polaron energy curves.

We have also run simulations which take into account the fact that the Fermi sea itself is at a temperature of $T/T_F = 0.1$. Without a thermal average over the impurity momentum, i.e. considering a $T=0$ impurity in a $T>0$ Fermi sea, these simulations led to results that are very similar to the ones discussed above, obtained by thermal-averaging the data with a $T=0$ Fermi sea of lithium atoms. Since the latter approach is conceptually simpler, all theoretical data presented in this work are obtained considering a zero-temperature Fermi sea.
The Fermi polaron at a non-zero temperature has been considered in Refs.~\cite{Mulkerin2019bot,Tajima2018mfp,Hu2022fpa}. 
We finally note that, in most of our illustrations of the momentum dependence of our observables, the experimentally controlled momentum kick $\Delta p$ serves as an independent variable. However, the impurity motion also includes the thermal contribution, which suggests the average momentum $\langle p \rangle$ as an alternative choice. Analyzing the numerical results for the momentum distribution according to Eq.~\eqref{eq:prandom} (see also Fig.~\ref{fig:pdistrib}) and applying a heuristic fit model, we obtain the conversion formula
\begin{equation}
\langle p \rangle = \left( \Delta p^\alpha +  \bar{p}\,^\alpha \right)^{1/\alpha} \ ,  
\label{eq:pconversion}
\end{equation}
which captures the limiting cases of very low and very high momenta ($\Delta p / \bar{p} \rightarrow 0$ and $ \Delta p / \bar{p} \rightarrow \infty$)  and interpolates between them.
We find that already the assumption $\alpha=2$, which corresponds to a simple root-mean-square dependence, provides a very good approximation to the numerical results for $\langle p \rangle$ for given $\Delta p$ and $\bar{p}$.
Our fit to the numerical results yields an optimum value of $\alpha = 2.158$, which leads to an essentially perfect match across the entire momentum range of interest.

\section{Raman acceleration}\label{app:Ramanaccel}

The preparation of the moving impurities consists of accelerating the K~cloud using two hyperfine states that do not interact with the Li cloud, K$|2\rangle$ and K$|3\rangle$. After the impurity acceleration we perform radio-frequency (RF) spectroscopy. Here, we describe how the K cloud is accelerated via a series of light pulses that exploit a two-photon Raman transition. For details on the optical setup and pulse generation, the reader is referred to~\cite{Dobler2026PhD}.

The Raman transition is driven by a pair of counter-propagating laser beams aligned along the weak axis of the cigar-shaped cloud ($z$-direction). Both beams are derived from the same laser source and are red-detuned by $\sim 25$\,GHz from the K D2 transition. The two-photon detuning is set to match the energy splitting between the two hyperfine states K$|2\rangle$ and K$|3\rangle$ with $\Delta\nu_\text{K2,K3}\approx 65.8$\,MHz. A single $\pi$-pulse with a duration of $\sim 2\,\mu$s transfers the atoms from K$|2\rangle$ to K$|3\rangle$ (or vice versa) and imparts two photon momenta to the atoms. Because momentum transfer changes the kinetic energy of the atom, the resonance frequency is shifted in comparison to a transition that does not change the momentum of the atom. The energy shift for an atom with an initial momentum $p_0$ along the Raman beams is expressed as~\cite{Berman1997book, Shkedrov2020ism}
\begin{equation}
    \hbar\delta=\frac{2 p_0\hbar k}{m_\downarrow}+\frac{2(\hbar k)^{\,2}}{m_\downarrow},
    \label{eq:dopplerShift}
\end{equation}
where $\hbar k$ is the single photon momentum, and $m_\downarrow$ denotes the mass of a K atom. For an atom initially at rest, this frequency shift amounts to $33.12$\,kHz in our system.

For final momenta corresponding to integer multiples of $2\,\hbar k$, we apply a sequence of $n$ Raman $\pi$-pulses. Following each pulse, the frequencies of the two Raman beams are interchanged to achieve momentum transfer in the same direction. In addition, the two-photon detuning is adjusted according to Eq.~\eqref{eq:dopplerShift} to account for the changing momentum of the atoms after each pulse. Our scheme requires the impurity cloud to be in the state K$|2\rangle$ after acceleration, which serves as the initial state for the subsequent RF spectroscopy pulse. Therefore, the acceleration scheme begins in K$|3\rangle$ or K$|2\rangle$ for odd $n$ or even $n$, respectively. (After the preparation of the atomic mixture, all K atoms are in the K$|3\rangle$ state and can be transferred to K$|2\rangle$ via an RF pulse.)

\begin{figure}[t]
\includegraphics[width=1\columnwidth]{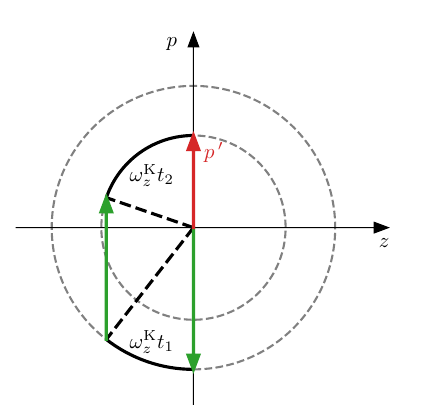}
\caption{\label{fig:two_pulse} Double pulse scheme for imparting momenta in the range $0 \leq |p^{\,\prime}| \leq 4\,\hbar k$: The first Raman pulse (green arrow pointing down) is followed by a delay time $t_1$, during which the cloud's position and momentum evolve in the harmonic trap. The second Raman pulse transfers two photon momenta in the opposite direction. After another delay time $t_2$, the cloud is in the center of the trap with final momentum $p'$.}
\end{figure}

\begin{figure}[t]
\includegraphics[width=1\columnwidth]{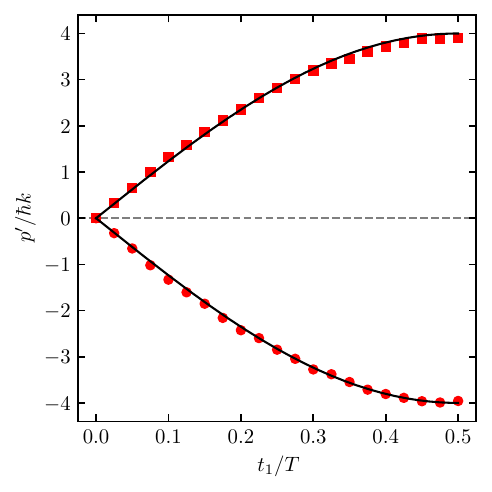}
\caption{\label{fig:two_pulse_data}Measurement of the K momentum after the double pulse scheme. The momentum is shown in units of the photon momentum as a function of the delay time between the two Raman pulses $t_1$. Red circles (squares) are measured with the first Raman pulse pointing in positive (negative) z-direction. The velocity is determined from the position of the atoms after $11\;\text{ms}$ time-of-flight expansion. The black solid line shows the expected velocity according to Eq.~\eqref{eq:pFinal}.}
\end{figure}

To impart momenta smaller than $2\,\hbar k$, we have developed a double-pulse scheme that leverages the harmonic trapping potential. The scheme is depicted in Fig.~\ref{fig:two_pulse}. Two Raman pulses are applied with opposite momentum transfer. After the first and second pulse, the cloud evolves in the harmonic trap for an adjustable delay time $t_1$ and $t_2$. The goal is to prepare the cloud with a specific momentum $p\,'$ in the center of the trap. The cloud reaches the center of the trap after a delay time 
\begin{equation}
    t_2 = \frac{T}{4} - \frac{t_1}{2},
    \label{eq:t2}
\end{equation}
following the seconds pulse, with $T = 2\pi/\omega_z^{\text{K}}
$ being the axial oscillation period. The momentum of the cloud at this point can be expressed as
\begin{equation}
p\,'=4\,\hbar k\sin{\left(\frac{\omega_z^{\text{K}}t_1}{2}\right)}.
\label{eq:pFinal}
\end{equation}
By tuning $t_1$ and adjusting $t_2$ according to Eq.~\eqref{eq:t2}, we can prepare the impurity cloud at any desired momentum in the range $0 \leq |p^{\,\prime}| \leq 4\,\hbar k$. Figure~\ref{fig:two_pulse_data} shows a measurement of the momentum of the K cloud in the non-interacting state after the described protocol, demonstrating our ability to prepare the atomic cloud at any momentum between $-4\,\hbar k$ and $4\,\hbar k$. In our experiments, we use this protocol to realize momenta between $0\,\hbar k$ and $2\,\hbar k$. For $2.5\,\hbar k$  and $3\,\hbar k$, the cloud is accelerated to $0.5\,\hbar k$  and $1\,\hbar k$  using this double-pulse scheme, followed by a third Raman pulse that imparts an additional two-photon recoil.

\section{Spectroscopy of a moving cloud}\label{app:movcloud}
\begin{figure}[t]
\includegraphics[width=1\columnwidth]{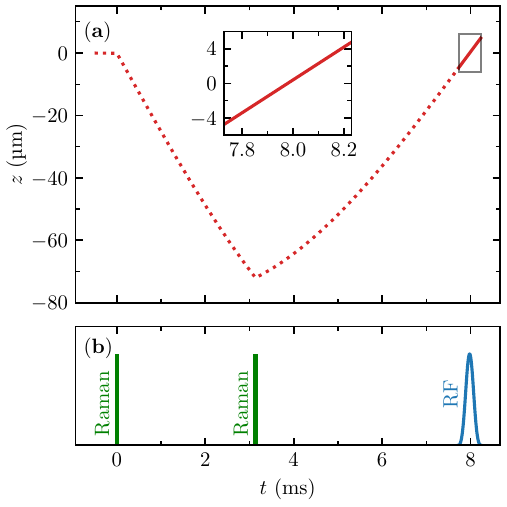}
\caption{\label{fig:position_raman}(a) Axial center-of-mass position of the K cloud during Raman acceleration and RF spectroscopy calculated for a representative measurement at $\Delta p=1.5\,\hbar k$. The dashed red line shows the position in the non-interacting states K$|2\rangle$ and K$|3\rangle$ before the spectroscopy. The solid red line indicates the position of the K cloud during the RF spectroscopy pulse, magnified in the inset. (b) Timing of the two Raman pulses (green) and the RF spectroscopy pulse (blue).}
\end{figure}

An aspect absent from previous experiments on static impurities is the motion of the K cloud during the RF spectroscopy pulse. For the assumption of a quasi-homogeneous environment to remain valid, the distance traveled by the K cloud during the pulse must be much smaller than the size of the Li cloud. Figure~\ref{fig:position_raman} schematically depicts the calculated axial motion of the K cloud during Raman acceleration and RF spectroscopy for a representative measurement at $\Delta p=1.5\,\hbar k$.  The relevant time interval in which the moving impurities are probed by RF spectroscopy is enlarged in the inset.

The duration of the RF pulse is chosen as a compromise between the spectral resolution, the finite lifetime of the polaron, and the motion of the K cloud during the pulse. For small momenta $0 \leq \Delta p \leq 3\,\hbar k$, we use a 0.5\,ms Blackman-shaped pulse with an experimentally determined spectral linewidth $\sigma=1.01(1)$\,kHz. For the largest momentum $\Delta p=3\,\hbar k$, the K cloud moves a distance of $19\,\mu$m during the pulse, which is smaller than the axial width of the K cloud $\sigma_z^K =39\,\mu$m and much smaller than the axial Fermi radius of the Li cloud $R_z^{Li}=160\,\mu$m. To further reduce the spatial variation of the Li density experienced by the K cloud during the motion, we choose the pulse timing such that the pulse is centered around the time at which the cloud crosses the center of the trap. For larger momenta $4 \leq \Delta p \leq 12\hbar k$, the pulse duration is reduced to $0.25$\,ms to account for the faster motion of the cloud. The spectral linewidth of this shorter pulse is  $\sigma=2.07(2)$\,kHz and the maximum distance traveled by the K cloud is $38\,\mu$m for $\Delta p = 12\,\hbar k$.

The intensity of  both RF pulses is adjusted to correspond to a $\pi$-pulse in the absence of Li. In the presence of the Li cloud, the shorter pulse transfers more K atoms because it is spectrally broader. This affects the amplitude and linewidth of the polaron peak in the spectroscopy signal, as seen in Fig.~\ref{fig:polaronpeaks}. The spectral linewidth of the RF pulse is taken into account when fitting the spectra (see App.~\ref{app:dataanal}).

\section{Precise determination of the polaron peak position and linewidth}\label{app:dataanal}

To obtain RF spectra with minimal statistical uncertainty, many repeated measurements over several hours are necessary. Such long acquisition times can lead to slow magnetic field drifts, which limit the precision with which the interaction strength is known. To mitigate these effects, we interleave magnetic field calibrations before and after every two repetitions, discard these repetitions when the bracketing calibrations differ by more than $2\,$mG, and adjust the coil current to restore the target field for the next spectroscopic measurement. In addition, we keep each measurement as short as possible by recording only the polaron peak and modeling the broad background, which is not recorded, with a linear term in the fitting procedure~\cite{Baroni2024mib}. For each fixed momentum and interaction parameter, we obtain the final spectrum by combining 6 to 20 repetitions. The chosen fit function consists of a Voigt profile superimposed on a linear background. This model captures both the Fourier-limited broadening of the RF pulse and the intrinsic linewidth of the polaron peak. The fit function is given by
\begin{equation}
f(\Delta\nu) = y_0 + A V(\Delta\nu - \Delta\nu_{\rm pol}, \sigma, \gamma_{\rm pol}) + m(\Delta\nu - \Delta\nu_{\rm pol}),
\label{eq:1}
\end{equation}
where \( V(\Delta\nu - \Delta\nu_{\rm pol}, \sigma, \gamma_{\rm pol}) \) denotes the Voigt profile, defined as the convolution of a Gaussian (standard deviation \( \sigma \)) and a Lorentzian (halfwidth \( \gamma_{\rm pol} \)), centered at \(\Delta\nu_{\rm pol} \). The parameter \( A \) denotes the amplitude of the peak, \( y_0 \) is a constant background offset, and \( m \) is the slope of the linear background component. We determine the Gaussian width independently by measuring the spectral width of the $0.5\,{\rm ms}$ and $0.25\,{\rm ms}$ Blackman-shaped RF pulses without any interaction with the medium (see App.~\ref{app:movcloud}), and fix it to this value in the fitting procedure.
To obtain robust fit results, we iteratively apply Chauvenet’s criterion~\cite{Taylor2022book} during fitting to identify and remove outliers, minimizing the impact of statistically inconsistent data points. From the Voigt fit, we extract the polaron peak position $\Delta\nu_{\rm pol}$ and linewidth $\gamma_{\rm pol}$.


%


\begin{thebibliography}{99}%
\makeatletter
\providecommand \@ifxundefined [1]{%
 \@ifx{#1\undefined}
}%
\providecommand \@ifnum [1]{%
 \ifnum #1\expandafter \@firstoftwo
 \else \expandafter \@secondoftwo
 \fi
}%
\providecommand \@ifx [1]{%
 \ifx #1\expandafter \@firstoftwo
 \else \expandafter \@secondoftwo
 \fi
}%
\providecommand \natexlab [1]{#1}%
\providecommand \enquote  [1]{``#1''}%
\providecommand \bibnamefont  [1]{#1}%
\providecommand \bibfnamefont [1]{#1}%
\providecommand \citenamefont [1]{#1}%
\providecommand \href@noop [0]{\@secondoftwo}%
\providecommand \href [0]{\begingroup \@sanitize@url \@href}%
\providecommand \@href[1]{\@@startlink{#1}\@@href}%
\providecommand \@@href[1]{\endgroup#1\@@endlink}%
\providecommand \@sanitize@url [0]{\catcode `\\12\catcode `\$12\catcode `\&12\catcode `\#12\catcode `\^12\catcode `\_12\catcode `\%12\relax}%
\providecommand \@@startlink[1]{}%
\providecommand \@@endlink[0]{}%
\providecommand \url  [0]{\begingroup\@sanitize@url \@url }%
\providecommand \@url [1]{\endgroup\@href {#1}{\urlprefix }}%
\providecommand \urlprefix  [0]{URL }%
\providecommand \Eprint [0]{\href }%
\providecommand \doibase [0]{https://doi.org/}%
\providecommand \selectlanguage [0]{\@gobble}%
\providecommand \bibinfo  [0]{\@secondoftwo}%
\providecommand \bibfield  [0]{\@secondoftwo}%
\providecommand \translation [1]{[#1]}%
\providecommand \BibitemOpen [0]{}%
\providecommand \bibitemStop [0]{}%
\providecommand \bibitemNoStop [0]{.\EOS\space}%
\providecommand \EOS [0]{\spacefactor3000\relax}%
\providecommand \BibitemShut  [1]{\csname bibitem#1\endcsname}%
\let\auto@bib@innerbib\@empty
\bibitem [{\citenamefont {Landau}\ and\ \citenamefont {Pekar}(1948)}]{Pekar1948}%
  \BibitemOpen
  \bibfield  {author} {\bibinfo {author} {\bibfnamefont {L.}~\bibnamefont {Landau}}\ and\ \bibinfo {author} {\bibfnamefont {S.}~\bibnamefont {Pekar}},\ }\bibfield  {title} {\bibinfo {title} {{Effective mass of the polaron}},\ }\href {http://www.ujp.bitp.kiev.ua/files/journals/53/si/53SI15p.pdf} {\bibfield  {journal} {\bibinfo  {journal} {J. Exp. Theor. Phys}\ }\textbf {\bibinfo {volume} {423}},\ \bibinfo {pages} {71} (\bibinfo {year} {1948})}\BibitemShut {NoStop}%
\bibitem [{\citenamefont {Baym}\ and\ \citenamefont {Pethick}(1991)}]{BaymPethick1991book}%
  \BibitemOpen
  \bibfield  {author} {\bibinfo {author} {\bibfnamefont {G.}~\bibnamefont {Baym}}\ and\ \bibinfo {author} {\bibfnamefont {C.}~\bibnamefont {Pethick}},\ }\href@noop {} {\emph {\bibinfo {title} {Landau Fermi-Liquid Theory: Concepts and Applications}}}\ (\bibinfo  {publisher} {Wiley-VCH},\ \bibinfo {year} {1991})\BibitemShut {NoStop}%
\bibitem [{\citenamefont {Massignan}\ \emph {et~al.}(2014)\citenamefont {Massignan}, \citenamefont {Zaccanti},\ and\ \citenamefont {Bruun}}]{Massignan2014pdm}%
  \BibitemOpen
  \bibfield  {author} {\bibinfo {author} {\bibfnamefont {P.}~\bibnamefont {Massignan}}, \bibinfo {author} {\bibfnamefont {M.}~\bibnamefont {Zaccanti}},\ and\ \bibinfo {author} {\bibfnamefont {G.~M.}\ \bibnamefont {Bruun}},\ }\bibfield  {title} {\bibinfo {title} {{Polarons, dressed molecules and itinerant ferromagnetism in ultracold Fermi gases}},\ }\href {https://doi.org/10.1088/0034-4885/77/3/034401} {\bibfield  {journal} {\bibinfo  {journal} {Rep. Prog. Phys.}\ }\textbf {\bibinfo {volume} {77}},\ \bibinfo {pages} {034401} (\bibinfo {year} {2014})}\BibitemShut {NoStop}%
\bibitem [{\citenamefont {Schmidt}\ \emph {et~al.}(2018)\citenamefont {Schmidt}, \citenamefont {Knap}, \citenamefont {Ivanov}, \citenamefont {You}, \citenamefont {Cetina},\ and\ \citenamefont {Demler}}]{Schmidt2018umb}%
  \BibitemOpen
  \bibfield  {author} {\bibinfo {author} {\bibfnamefont {R.}~\bibnamefont {Schmidt}}, \bibinfo {author} {\bibfnamefont {M.}~\bibnamefont {Knap}}, \bibinfo {author} {\bibfnamefont {D.~A.}\ \bibnamefont {Ivanov}}, \bibinfo {author} {\bibfnamefont {J.-S.}\ \bibnamefont {You}}, \bibinfo {author} {\bibfnamefont {M.}~\bibnamefont {Cetina}},\ and\ \bibinfo {author} {\bibfnamefont {E.}~\bibnamefont {Demler}},\ }\bibfield  {title} {\bibinfo {title} {Universal many-body response of heavy impurities coupled to a {F}ermi sea: a review of recent progress},\ }\href {https://doi.org/10.1088/1361-6633/aa9593} {\bibfield  {journal} {\bibinfo  {journal} {Rep. Prog. Phys.}\ }\textbf {\bibinfo {volume} {81}},\ \bibinfo {pages} {024401} (\bibinfo {year} {2018})}\BibitemShut {NoStop}%
\bibitem [{\citenamefont {Grimm}\ \emph {et~al.}()\citenamefont {Grimm}, \citenamefont {Inguscio}, \citenamefont {Stringari},\ and\ \citenamefont {Lamporesi}}]{Varenna2022book}%
  \BibitemOpen
  \bibinfo {editor} {\bibfnamefont {R.}~\bibnamefont {Grimm}}, \bibinfo {editor} {\bibfnamefont {M.}~\bibnamefont {Inguscio}}, \bibinfo {editor} {\bibfnamefont {S.}~\bibnamefont {Stringari}},\ and\ \bibinfo {editor} {\bibfnamefont {G.}~\bibnamefont {Lamporesi}},\ eds.,\ \href@noop {} {\emph {\bibinfo {title} {{Quantum Mixtures with Ultra-cold Atoms}}}}\ (\bibinfo  {publisher} {IOS Press, Amsterdam, 2025})\ \bibinfo {note} {{P}roceedings of the International School of Physics ``Enrico Fermi'', Course CCXI, Varenna, 18-23 July 2022}\BibitemShut {NoStop}%
\bibitem [{\citenamefont {Scazza}\ \emph {et~al.}(2022)\citenamefont {Scazza}, \citenamefont {Zaccanti}, \citenamefont {Massignan}, \citenamefont {Parish},\ and\ \citenamefont {Levinsen}}]{Scazza2022rfa}%
  \BibitemOpen
  \bibfield  {author} {\bibinfo {author} {\bibfnamefont {F.}~\bibnamefont {Scazza}}, \bibinfo {author} {\bibfnamefont {M.}~\bibnamefont {Zaccanti}}, \bibinfo {author} {\bibfnamefont {P.}~\bibnamefont {Massignan}}, \bibinfo {author} {\bibfnamefont {M.~M.}\ \bibnamefont {Parish}},\ and\ \bibinfo {author} {\bibfnamefont {J.}~\bibnamefont {Levinsen}},\ }\bibfield  {title} {\bibinfo {title} {{Repulsive Fermi and Bose Polarons in Quantum Gases}},\ }\href {https://doi.org/10.3390/atoms10020055} {\bibfield  {journal} {\bibinfo  {journal} {Atoms}\ }\textbf {\bibinfo {volume} {10}},\ \bibinfo {pages} {55} (\bibinfo {year} {2022})}\BibitemShut {NoStop}%
\bibitem [{\citenamefont {Parish}\ and\ \citenamefont {Levinsen}(2022)}]{Parish2025review}%
  \BibitemOpen
  \bibfield  {author} {\bibinfo {author} {\bibfnamefont {M.~M.}\ \bibnamefont {Parish}}\ and\ \bibinfo {author} {\bibfnamefont {J.}~\bibnamefont {Levinsen}},\ }\href@noop {} {\bibinfo {title} {Fermi polarons and beyond}} (\bibinfo {year} {2022}),\ \bibinfo {note} {in Ref.~\cite{Varenna2022book}}\BibitemShut {NoStop}%
\bibitem [{\citenamefont {Massignan}\ \emph {et~al.}(2026)\citenamefont {Massignan}, \citenamefont {Schmidt}, \citenamefont {Astrakharchik}, \citenamefont {İmamoglu}, \citenamefont {Zwierlein}, \citenamefont {Arlt},\ and\ \citenamefont {Bruun}}]{Massignan2026pia}%
  \BibitemOpen
  \bibfield  {author} {\bibinfo {author} {\bibfnamefont {P.}~\bibnamefont {Massignan}}, \bibinfo {author} {\bibfnamefont {R.}~\bibnamefont {Schmidt}}, \bibinfo {author} {\bibfnamefont {G.~E.}\ \bibnamefont {Astrakharchik}}, \bibinfo {author} {\bibfnamefont {A.}~\bibnamefont {İmamoglu}}, \bibinfo {author} {\bibfnamefont {M.}~\bibnamefont {Zwierlein}}, \bibinfo {author} {\bibfnamefont {J.~J.}\ \bibnamefont {Arlt}},\ and\ \bibinfo {author} {\bibfnamefont {G.~M.}\ \bibnamefont {Bruun}},\ }\bibfield  {title} {\bibinfo {title} {{Polarons in atomic gases and two-dimensional semiconductors}},\ }\href {https://doi.org/10.1103/4nng-bb9z} {\bibfield  {journal} {\bibinfo  {journal} {Rev. Mod. Phys.}\ } (\bibinfo {year} {2026})}\BibitemShut {NoStop}%
\bibitem [{\citenamefont {Schirotzek}\ \emph {et~al.}(2009)\citenamefont {Schirotzek}, \citenamefont {Wu}, \citenamefont {Sommer},\ and\ \citenamefont {Zwierlein}}]{Schirotzek2009oof}%
  \BibitemOpen
  \bibfield  {author} {\bibinfo {author} {\bibfnamefont {A.}~\bibnamefont {Schirotzek}}, \bibinfo {author} {\bibfnamefont {C.-H.}\ \bibnamefont {Wu}}, \bibinfo {author} {\bibfnamefont {A.}~\bibnamefont {Sommer}},\ and\ \bibinfo {author} {\bibfnamefont {M.~W.}\ \bibnamefont {Zwierlein}},\ }\bibfield  {title} {\bibinfo {title} {{Observation of Fermi Polarons in a Tunable Fermi Liquid of Ultracold Atoms}},\ }\href {https://doi.org/10.1103/PhysRevLett.102.230402} {\bibfield  {journal} {\bibinfo  {journal} {Phys. Rev. Lett.}\ }\textbf {\bibinfo {volume} {102}},\ \bibinfo {pages} {230402} (\bibinfo {year} {2009})}\BibitemShut {NoStop}%
\bibitem [{\citenamefont {Nascimb\`ene}\ \emph {et~al.}(2009)\citenamefont {Nascimb\`ene}, \citenamefont {Navon}, \citenamefont {Jiang}, \citenamefont {Tarruell}, \citenamefont {Teichmann}, \citenamefont {McKeever}, \citenamefont {Chevy},\ and\ \citenamefont {Salomon}}]{Nascimbene2009coo}%
  \BibitemOpen
  \bibfield  {author} {\bibinfo {author} {\bibfnamefont {S.}~\bibnamefont {Nascimb\`ene}}, \bibinfo {author} {\bibfnamefont {N.}~\bibnamefont {Navon}}, \bibinfo {author} {\bibfnamefont {K.~J.}\ \bibnamefont {Jiang}}, \bibinfo {author} {\bibfnamefont {L.}~\bibnamefont {Tarruell}}, \bibinfo {author} {\bibfnamefont {M.}~\bibnamefont {Teichmann}}, \bibinfo {author} {\bibfnamefont {J.}~\bibnamefont {McKeever}}, \bibinfo {author} {\bibfnamefont {F.}~\bibnamefont {Chevy}},\ and\ \bibinfo {author} {\bibfnamefont {C.}~\bibnamefont {Salomon}},\ }\bibfield  {title} {\bibinfo {title} {{Collective Oscillations of an Imbalanced Fermi Gas: Axial Compression Modes and Polaron Effective Mass}},\ }\href {https://doi.org/10.1103/PhysRevLett.103.170402} {\bibfield  {journal} {\bibinfo  {journal} {Phys. Rev. Lett.}\ }\textbf {\bibinfo {volume} {103}},\ \bibinfo {pages} {170402} (\bibinfo {year} {2009})}\BibitemShut {NoStop}%
\bibitem [{\citenamefont {Kohstall}\ \emph {et~al.}(2012)\citenamefont {Kohstall}, \citenamefont {Zaccanti}, \citenamefont {Jag}, \citenamefont {Trenkwalder}, \citenamefont {Massignan}, \citenamefont {Bruun}, \citenamefont {Schreck},\ and\ \citenamefont {Grimm}}]{Kohstall2012mac}%
  \BibitemOpen
  \bibfield  {author} {\bibinfo {author} {\bibfnamefont {C.}~\bibnamefont {Kohstall}}, \bibinfo {author} {\bibfnamefont {M.}~\bibnamefont {Zaccanti}}, \bibinfo {author} {\bibfnamefont {M.}~\bibnamefont {Jag}}, \bibinfo {author} {\bibfnamefont {A.}~\bibnamefont {Trenkwalder}}, \bibinfo {author} {\bibfnamefont {P.}~\bibnamefont {Massignan}}, \bibinfo {author} {\bibfnamefont {G.~M.}\ \bibnamefont {Bruun}}, \bibinfo {author} {\bibfnamefont {F.}~\bibnamefont {Schreck}},\ and\ \bibinfo {author} {\bibfnamefont {R.}~\bibnamefont {Grimm}},\ }\bibfield  {title} {\bibinfo {title} {Metastability and coherence of repulsive polarons in a strongly interacting {F}ermi mixture},\ }\href {https://doi.org/10.1038/nature11065} {\bibfield  {journal} {\bibinfo  {journal} {Nature (London)}\ }\textbf {\bibinfo {volume} {485}},\ \bibinfo {pages} {615} (\bibinfo {year} {2012})}\BibitemShut {NoStop}%
\bibitem [{\citenamefont {Koschorreck}\ \emph {et~al.}(2012)\citenamefont {Koschorreck}, \citenamefont {Pertot}, \citenamefont {Vogt}, \citenamefont {Fr\"{o}lich}, \citenamefont {Feld},\ and\ \citenamefont {K\"{o}hl}}]{Koschorreck2012aar}%
  \BibitemOpen
  \bibfield  {author} {\bibinfo {author} {\bibfnamefont {M.}~\bibnamefont {Koschorreck}}, \bibinfo {author} {\bibfnamefont {D.}~\bibnamefont {Pertot}}, \bibinfo {author} {\bibfnamefont {E.}~\bibnamefont {Vogt}}, \bibinfo {author} {\bibfnamefont {B.}~\bibnamefont {Fr\"{o}lich}}, \bibinfo {author} {\bibfnamefont {M.}~\bibnamefont {Feld}},\ and\ \bibinfo {author} {\bibfnamefont {M.}~\bibnamefont {K\"{o}hl}},\ }\bibfield  {title} {\bibinfo {title} {{Attractive and repulsive Fermi polarons in two dimensions}},\ }\href {https://doi.org/10.1038/nature11151} {\bibfield  {journal} {\bibinfo  {journal} {Nature (London)}\ }\textbf {\bibinfo {volume} {485}},\ \bibinfo {pages} {619} (\bibinfo {year} {2012})}\BibitemShut {NoStop}%
\bibitem [{\citenamefont {Scazza}\ \emph {et~al.}(2017)\citenamefont {Scazza}, \citenamefont {Valtolina}, \citenamefont {Massignan}, \citenamefont {Recati}, \citenamefont {Amico}, \citenamefont {Burchianti}, \citenamefont {Fort}, \citenamefont {Inguscio}, \citenamefont {Zaccanti},\ and\ \citenamefont {Roati}}]{Scazza2017rfp}%
  \BibitemOpen
  \bibfield  {author} {\bibinfo {author} {\bibfnamefont {F.}~\bibnamefont {Scazza}}, \bibinfo {author} {\bibfnamefont {G.}~\bibnamefont {Valtolina}}, \bibinfo {author} {\bibfnamefont {P.}~\bibnamefont {Massignan}}, \bibinfo {author} {\bibfnamefont {A.}~\bibnamefont {Recati}}, \bibinfo {author} {\bibfnamefont {A.}~\bibnamefont {Amico}}, \bibinfo {author} {\bibfnamefont {A.}~\bibnamefont {Burchianti}}, \bibinfo {author} {\bibfnamefont {C.}~\bibnamefont {Fort}}, \bibinfo {author} {\bibfnamefont {M.}~\bibnamefont {Inguscio}}, \bibinfo {author} {\bibfnamefont {M.}~\bibnamefont {Zaccanti}},\ and\ \bibinfo {author} {\bibfnamefont {G.}~\bibnamefont {Roati}},\ }\bibfield  {title} {\bibinfo {title} {{Repulsive Fermi Polarons in a Resonant Mixture of Ultracold $^{6}\mathrm{Li}$ Atoms}},\ }\href {https://doi.org/10.1103/PhysRevLett.118.083602} {\bibfield  {journal} {\bibinfo  {journal} {Phys. Rev. Lett.}\ }\textbf {\bibinfo {volume} {118}},\ \bibinfo {pages} {083602} (\bibinfo {year} {2017})}\BibitemShut {NoStop}%
\bibitem [{\citenamefont {Yan}\ \emph {et~al.}(2019)\citenamefont {Yan}, \citenamefont {Patel}, \citenamefont {Mukherjee}, \citenamefont {Fletcher}, \citenamefont {Struck},\ and\ \citenamefont {Zwierlein}}]{Yan2019bau}%
  \BibitemOpen
  \bibfield  {author} {\bibinfo {author} {\bibfnamefont {Z.}~\bibnamefont {Yan}}, \bibinfo {author} {\bibfnamefont {P.~B.}\ \bibnamefont {Patel}}, \bibinfo {author} {\bibfnamefont {B.}~\bibnamefont {Mukherjee}}, \bibinfo {author} {\bibfnamefont {R.~J.}\ \bibnamefont {Fletcher}}, \bibinfo {author} {\bibfnamefont {J.}~\bibnamefont {Struck}},\ and\ \bibinfo {author} {\bibfnamefont {M.~W.}\ \bibnamefont {Zwierlein}},\ }\bibfield  {title} {\bibinfo {title} {{Boiling a Unitary Fermi Liquid}},\ }\href {https://doi.org/10.1103/PhysRevLett.122.093401} {\bibfield  {journal} {\bibinfo  {journal} {Phys. Rev. Lett.}\ }\textbf {\bibinfo {volume} {122}},\ \bibinfo {pages} {093401} (\bibinfo {year} {2019})}\BibitemShut {NoStop}%
\bibitem [{\citenamefont {Baroni}\ \emph {et~al.}(2024)\citenamefont {Baroni}, \citenamefont {Huang}, \citenamefont {Fritsche}, \citenamefont {Dobler}, \citenamefont {Anich}, \citenamefont {Kirilov}, \citenamefont {Grimm}, \citenamefont {Bastarrachea-Magnani}, \citenamefont {Massignan},\ and\ \citenamefont {Bruun}}]{Baroni2024mib}%
  \BibitemOpen
  \bibfield  {author} {\bibinfo {author} {\bibfnamefont {C.}~\bibnamefont {Baroni}}, \bibinfo {author} {\bibfnamefont {B.}~\bibnamefont {Huang}}, \bibinfo {author} {\bibfnamefont {I.}~\bibnamefont {Fritsche}}, \bibinfo {author} {\bibfnamefont {E.}~\bibnamefont {Dobler}}, \bibinfo {author} {\bibfnamefont {G.}~\bibnamefont {Anich}}, \bibinfo {author} {\bibfnamefont {E.}~\bibnamefont {Kirilov}}, \bibinfo {author} {\bibfnamefont {R.}~\bibnamefont {Grimm}}, \bibinfo {author} {\bibfnamefont {M.~A.}\ \bibnamefont {Bastarrachea-Magnani}}, \bibinfo {author} {\bibfnamefont {P.}~\bibnamefont {Massignan}},\ and\ \bibinfo {author} {\bibfnamefont {G.~M.}\ \bibnamefont {Bruun}},\ }\bibfield  {title} {\bibinfo {title} {Mediated interactions between {F}ermi polarons and the role of impurity quantum statistics},\ }\href {https://doi.org/10.1038/s41567-023-02248-4} {\bibfield  {journal} {\bibinfo  {journal} {Nat. Phys.}\ }\textbf {\bibinfo {volume} {20}},\ \bibinfo {pages} {68} (\bibinfo {year} {2024})}\BibitemShut {NoStop}%
\bibitem [{\citenamefont {Hu}\ \emph {et~al.}(2016)\citenamefont {Hu}, \citenamefont {Van~de Graaff}, \citenamefont {Kedar}, \citenamefont {Corson}, \citenamefont {Cornell},\ and\ \citenamefont {Jin}}]{Hu2016bpi}%
  \BibitemOpen
  \bibfield  {author} {\bibinfo {author} {\bibfnamefont {M.-G.}\ \bibnamefont {Hu}}, \bibinfo {author} {\bibfnamefont {M.~J.}\ \bibnamefont {Van~de Graaff}}, \bibinfo {author} {\bibfnamefont {D.}~\bibnamefont {Kedar}}, \bibinfo {author} {\bibfnamefont {J.~P.}\ \bibnamefont {Corson}}, \bibinfo {author} {\bibfnamefont {E.~A.}\ \bibnamefont {Cornell}},\ and\ \bibinfo {author} {\bibfnamefont {D.~S.}\ \bibnamefont {Jin}},\ }\bibfield  {title} {\bibinfo {title} {Bose polarons in the strongly interacting regime},\ }\href {https://doi.org/10.1103/PhysRevLett.117.055301} {\bibfield  {journal} {\bibinfo  {journal} {Phys. Rev. Lett.}\ }\textbf {\bibinfo {volume} {117}},\ \bibinfo {pages} {055301} (\bibinfo {year} {2016})}\BibitemShut {NoStop}%
\bibitem [{\citenamefont {J\o{}rgensen}\ \emph {et~al.}(2016)\citenamefont {J\o{}rgensen}, \citenamefont {Wacker}, \citenamefont {Skalmstang}, \citenamefont {Parish}, \citenamefont {Levinsen}, \citenamefont {Christensen}, \citenamefont {Bruun},\ and\ \citenamefont {Arlt}}]{Jorgensen2016ooa}%
  \BibitemOpen
  \bibfield  {author} {\bibinfo {author} {\bibfnamefont {N.~B.}\ \bibnamefont {J\o{}rgensen}}, \bibinfo {author} {\bibfnamefont {L.}~\bibnamefont {Wacker}}, \bibinfo {author} {\bibfnamefont {K.~T.}\ \bibnamefont {Skalmstang}}, \bibinfo {author} {\bibfnamefont {M.~M.}\ \bibnamefont {Parish}}, \bibinfo {author} {\bibfnamefont {J.}~\bibnamefont {Levinsen}}, \bibinfo {author} {\bibfnamefont {R.~S.}\ \bibnamefont {Christensen}}, \bibinfo {author} {\bibfnamefont {G.~M.}\ \bibnamefont {Bruun}},\ and\ \bibinfo {author} {\bibfnamefont {J.~J.}\ \bibnamefont {Arlt}},\ }\bibfield  {title} {\bibinfo {title} {{Observation of Attractive and Repulsive Polarons in a Bose-Einstein Condensate}},\ }\href {https://doi.org/10.1103/PhysRevLett.117.055302} {\bibfield  {journal} {\bibinfo  {journal} {Phys. Rev. Lett.}\ }\textbf {\bibinfo {volume} {117}},\ \bibinfo {pages} {055302} (\bibinfo {year} {2016})}\BibitemShut {NoStop}%
\bibitem [{\citenamefont {Pe\~na Ardila}\ \emph {et~al.}(2019)\citenamefont {Pe\~na Ardila}, \citenamefont {J\o{}rgensen}, \citenamefont {Pohl}, \citenamefont {Giorgini}, \citenamefont {Bruun},\ and\ \citenamefont {Arlt}}]{Penaardila2019aab}%
  \BibitemOpen
  \bibfield  {author} {\bibinfo {author} {\bibfnamefont {L.~A.}\ \bibnamefont {Pe\~na Ardila}}, \bibinfo {author} {\bibfnamefont {N.~B.}\ \bibnamefont {J\o{}rgensen}}, \bibinfo {author} {\bibfnamefont {T.}~\bibnamefont {Pohl}}, \bibinfo {author} {\bibfnamefont {S.}~\bibnamefont {Giorgini}}, \bibinfo {author} {\bibfnamefont {G.~M.}\ \bibnamefont {Bruun}},\ and\ \bibinfo {author} {\bibfnamefont {J.~J.}\ \bibnamefont {Arlt}},\ }\bibfield  {title} {\bibinfo {title} {Analyzing a {B}ose polaron across resonant interactions},\ }\href {https://doi.org/10.1103/PhysRevA.99.063607} {\bibfield  {journal} {\bibinfo  {journal} {Phys. Rev. A}\ }\textbf {\bibinfo {volume} {99}},\ \bibinfo {pages} {063607} (\bibinfo {year} {2019})}\BibitemShut {NoStop}%
\bibitem [{\citenamefont {Yan}\ \emph {et~al.}(2020)\citenamefont {Yan}, \citenamefont {Ni}, \citenamefont {Robens},\ and\ \citenamefont {Zwierlein}}]{Yan2020bpn}%
  \BibitemOpen
  \bibfield  {author} {\bibinfo {author} {\bibfnamefont {Z.~Z.}\ \bibnamefont {Yan}}, \bibinfo {author} {\bibfnamefont {Y.}~\bibnamefont {Ni}}, \bibinfo {author} {\bibfnamefont {C.}~\bibnamefont {Robens}},\ and\ \bibinfo {author} {\bibfnamefont {M.~W.}\ \bibnamefont {Zwierlein}},\ }\bibfield  {title} {\bibinfo {title} {Bose polarons near quantum criticality},\ }\href {https://doi.org/10.1126/science.aax5850} {\bibfield  {journal} {\bibinfo  {journal} {Science}\ }\textbf {\bibinfo {volume} {368}},\ \bibinfo {pages} {190} (\bibinfo {year} {2020})}\BibitemShut {NoStop}%
\bibitem [{\citenamefont {Skou}\ \emph {et~al.}(2021)\citenamefont {Skou}, \citenamefont {Skov}, \citenamefont {J\o{}rgensen}, \citenamefont {Nielsen}, \citenamefont {Camacho-Guardian}, \citenamefont {Pohl}, \citenamefont {Bruun},\ and\ \citenamefont {Arlt}}]{Skou2021neq}%
  \BibitemOpen
  \bibfield  {author} {\bibinfo {author} {\bibfnamefont {M.~G.}\ \bibnamefont {Skou}}, \bibinfo {author} {\bibfnamefont {T.~G.}\ \bibnamefont {Skov}}, \bibinfo {author} {\bibfnamefont {N.~B.}\ \bibnamefont {J\o{}rgensen}}, \bibinfo {author} {\bibfnamefont {K.~K.}\ \bibnamefont {Nielsen}}, \bibinfo {author} {\bibfnamefont {A.}~\bibnamefont {Camacho-Guardian}}, \bibinfo {author} {\bibfnamefont {T.}~\bibnamefont {Pohl}}, \bibinfo {author} {\bibfnamefont {G.~M.}\ \bibnamefont {Bruun}},\ and\ \bibinfo {author} {\bibfnamefont {J.~J.}\ \bibnamefont {Arlt}},\ }\bibfield  {title} {\bibinfo {title} {Non-equilibrium quantum dynamics and formation of the {B}ose polaron},\ }\href {https://doi.org/10.1038/s41567-021-01184-5} {\bibfield  {journal} {\bibinfo  {journal} {Nat. Phys.}\ }\textbf {\bibinfo {volume} {17}},\ \bibinfo {pages} {731} (\bibinfo {year} {2021})}\BibitemShut {NoStop}%
\bibitem [{\citenamefont {Skou}\ \emph {et~al.}(2022)\citenamefont {Skou}, \citenamefont {Nielsen}, \citenamefont {Skov}, \citenamefont {Morgen}, \citenamefont {J\o{}rgensen}, \citenamefont {Camacho-Guardian}, \citenamefont {Pohl}, \citenamefont {Bruun},\ and\ \citenamefont {Arlt}}]{Skou2022lad}%
  \BibitemOpen
  \bibfield  {author} {\bibinfo {author} {\bibfnamefont {M.~G.}\ \bibnamefont {Skou}}, \bibinfo {author} {\bibfnamefont {K.~K.}\ \bibnamefont {Nielsen}}, \bibinfo {author} {\bibfnamefont {T.~G.}\ \bibnamefont {Skov}}, \bibinfo {author} {\bibfnamefont {A.~M.}\ \bibnamefont {Morgen}}, \bibinfo {author} {\bibfnamefont {N.~B.}\ \bibnamefont {J\o{}rgensen}}, \bibinfo {author} {\bibfnamefont {A.}~\bibnamefont {Camacho-Guardian}}, \bibinfo {author} {\bibfnamefont {T.}~\bibnamefont {Pohl}}, \bibinfo {author} {\bibfnamefont {G.~M.}\ \bibnamefont {Bruun}},\ and\ \bibinfo {author} {\bibfnamefont {J.~J.}\ \bibnamefont {Arlt}},\ }\bibfield  {title} {\bibinfo {title} {{Life and death of the Bose polaron}},\ }\href {https://doi.org/10.1103/PhysRevResearch.4.043093} {\bibfield  {journal} {\bibinfo  {journal} {Phys. Rev. Res.}\ }\textbf {\bibinfo {volume} {4}},\ \bibinfo {pages} {043093} (\bibinfo {year} {2022})}\BibitemShut {NoStop}%
\bibitem [{\citenamefont {Morgen}\ \emph {et~al.}(2025)\citenamefont {Morgen}, \citenamefont {Balling}, \citenamefont {Nielsen}, \citenamefont {Pohl}, \citenamefont {Bruun},\ and\ \citenamefont {Arlt}}]{Morgen2025qbs}%
  \BibitemOpen
  \bibfield  {author} {\bibinfo {author} {\bibfnamefont {A.~M.}\ \bibnamefont {Morgen}}, \bibinfo {author} {\bibfnamefont {S.~S.}\ \bibnamefont {Balling}}, \bibinfo {author} {\bibfnamefont {K.~K.}\ \bibnamefont {Nielsen}}, \bibinfo {author} {\bibfnamefont {T.}~\bibnamefont {Pohl}}, \bibinfo {author} {\bibfnamefont {G.~M.}\ \bibnamefont {Bruun}},\ and\ \bibinfo {author} {\bibfnamefont {J.~J.}\ \bibnamefont {Arlt}},\ }\bibfield  {title} {\bibinfo {title} {Quantum beat spectroscopy of repulsive {B}ose polarons},\ }\href {https://doi.org/10.1103/PhysRevResearch.7.L022002} {\bibfield  {journal} {\bibinfo  {journal} {Phys. Rev. Res.}\ }\textbf {\bibinfo {volume} {7}},\ \bibinfo {pages} {L022002} (\bibinfo {year} {2025})}\BibitemShut {NoStop}%
\bibitem [{\citenamefont {Etrych}\ \emph {et~al.}(2025)\citenamefont {Etrych}, \citenamefont {Martirosyan}, \citenamefont {Cao}, \citenamefont {Ho}, \citenamefont {Hadzibabic},\ and\ \citenamefont {Eigen}}]{Etrych2025uqd}%
  \BibitemOpen
  \bibfield  {author} {\bibinfo {author} {\bibfnamefont {J.}~\bibnamefont {Etrych}}, \bibinfo {author} {\bibfnamefont {G.}~\bibnamefont {Martirosyan}}, \bibinfo {author} {\bibfnamefont {A.}~\bibnamefont {Cao}}, \bibinfo {author} {\bibfnamefont {C.~J.}\ \bibnamefont {Ho}}, \bibinfo {author} {\bibfnamefont {Z.}~\bibnamefont {Hadzibabic}},\ and\ \bibinfo {author} {\bibfnamefont {C.}~\bibnamefont {Eigen}},\ }\bibfield  {title} {\bibinfo {title} {{Universal Quantum Dynamics of Bose Polarons}},\ }\href {https://doi.org/10.1103/PhysRevX.15.021070} {\bibfield  {journal} {\bibinfo  {journal} {Phys. Rev. X}\ }\textbf {\bibinfo {volume} {15}},\ \bibinfo {pages} {021070} (\bibinfo {year} {2025})}\BibitemShut {NoStop}%
\bibitem [{\citenamefont {Henke}\ \emph {et~al.}(2025)\citenamefont {Henke}, \citenamefont {Levinsen}, \citenamefont {Parish}, \citenamefont {Boronat}, \citenamefont {Astrakharchik}, \citenamefont {Moritz},\ and\ \citenamefont {Cabrera}}]{Henke2025ror}%
  \BibitemOpen
  \bibfield  {author} {\bibinfo {author} {\bibfnamefont {R.}~\bibnamefont {Henke}}, \bibinfo {author} {\bibfnamefont {J.}~\bibnamefont {Levinsen}}, \bibinfo {author} {\bibfnamefont {M.~M.}\ \bibnamefont {Parish}}, \bibinfo {author} {\bibfnamefont {J.}~\bibnamefont {Boronat}}, \bibinfo {author} {\bibfnamefont {G.~E.}\ \bibnamefont {Astrakharchik}}, \bibinfo {author} {\bibfnamefont {H.}~\bibnamefont {Moritz}},\ and\ \bibinfo {author} {\bibfnamefont {C.~R.}\ \bibnamefont {Cabrera}},\ }\href@noop {} {\bibinfo {title} {Realization of repulsive polarons in the strongly correlated regime}} (\bibinfo {year} {2025}),\ \Eprint {https://arxiv.org/abs/2511.03569} {arXiv:2511.03569} \BibitemShut {NoStop}%
\bibitem [{\citenamefont {Grusdt}\ \emph {et~al.}(2025)\citenamefont {Grusdt}, \citenamefont {Mostaan}, \citenamefont {Demler},\ and\ \citenamefont {Ardila}}]{Grusdt2025iap}%
  \BibitemOpen
  \bibfield  {author} {\bibinfo {author} {\bibfnamefont {F.}~\bibnamefont {Grusdt}}, \bibinfo {author} {\bibfnamefont {N.}~\bibnamefont {Mostaan}}, \bibinfo {author} {\bibfnamefont {E.}~\bibnamefont {Demler}},\ and\ \bibinfo {author} {\bibfnamefont {L.~A.~P.}\ \bibnamefont {Ardila}},\ }\bibfield  {title} {\bibinfo {title} {Impurities and polarons in bosonic quantum gases: a review on recent progress},\ }\href {https://doi.org/10.1088/1361-6633/add94b} {\bibfield  {journal} {\bibinfo  {journal} {Rep. Progr. Phys.}\ }\textbf {\bibinfo {volume} {88}},\ \bibinfo {pages} {066401} (\bibinfo {year} {2025})}\BibitemShut {NoStop}%
\bibitem [{\citenamefont {Mathy}\ \emph {et~al.}(2012)\citenamefont {Mathy}, \citenamefont {Zvonarev},\ and\ \citenamefont {Demler}}]{Mathy2012qfo}%
  \BibitemOpen
  \bibfield  {author} {\bibinfo {author} {\bibfnamefont {C.~J.~M.}\ \bibnamefont {Mathy}}, \bibinfo {author} {\bibfnamefont {M.~B.}\ \bibnamefont {Zvonarev}},\ and\ \bibinfo {author} {\bibfnamefont {E.}~\bibnamefont {Demler}},\ }\bibfield  {title} {\bibinfo {title} {{Quantum flutter of supersonic particles in one-dimensional quantum liquids}},\ }\href {https://doi.org/10.1038/nphys2455} {\bibfield  {journal} {\bibinfo  {journal} {Nature Physics}\ }\textbf {\bibinfo {volume} {8}},\ \bibinfo {pages} {881} (\bibinfo {year} {2012})}\BibitemShut {NoStop}%
\bibitem [{\citenamefont {Hu}\ and\ \citenamefont {Liu}(2022{\natexlab{a}})}]{Hu2022rso}%
  \BibitemOpen
  \bibfield  {author} {\bibinfo {author} {\bibfnamefont {H.}~\bibnamefont {Hu}}\ and\ \bibinfo {author} {\bibfnamefont {X.-J.}\ \bibnamefont {Liu}},\ }\bibfield  {title} {\bibinfo {title} {{Raman spectroscopy of Fermi polarons}},\ }\href {https://doi.org/10.1103/PhysRevA.106.063306} {\bibfield  {journal} {\bibinfo  {journal} {Phys. Rev. A}\ }\textbf {\bibinfo {volume} {106}},\ \bibinfo {pages} {063306} (\bibinfo {year} {2022}{\natexlab{a}})}\BibitemShut {NoStop}%
\bibitem [{\citenamefont {von Milczewski}\ and\ \citenamefont {Schmidt}(2024)}]{vonMilczewski2024}%
  \BibitemOpen
  \bibfield  {author} {\bibinfo {author} {\bibfnamefont {J.}~\bibnamefont {von Milczewski}}\ and\ \bibinfo {author} {\bibfnamefont {R.}~\bibnamefont {Schmidt}},\ }\bibfield  {title} {\bibinfo {title} {{Momentum-dependent quasiparticle properties of the Fermi polaron from the functional renormalization group}},\ }\href {https://doi.org/10.1103/PhysRevA.110.033309} {\bibfield  {journal} {\bibinfo  {journal} {Phys. Rev. A}\ }\textbf {\bibinfo {volume} {110}},\ \bibinfo {pages} {033309} (\bibinfo {year} {2024})}\BibitemShut {NoStop}%
\bibitem [{\citenamefont {Shi}\ and\ \citenamefont {Cui}(2025)}]{Shi2025vtd}%
  \BibitemOpen
  \bibfield  {author} {\bibinfo {author} {\bibfnamefont {T.}~\bibnamefont {Shi}}\ and\ \bibinfo {author} {\bibfnamefont {X.}~\bibnamefont {Cui}},\ }\href@noop {} {\bibinfo {title} {{Visualizing the dispersions of Fermi polaron and molecule via spin-orbit coupling}}} (\bibinfo {year} {2025}),\ \Eprint {https://arxiv.org/abs/2512.23918} {arXiv:2512.23918} \BibitemShut {NoStop}%
\bibitem [{\citenamefont {Horvath}\ \emph {et~al.}(2026)\citenamefont {Horvath}, \citenamefont {Dhar}, \citenamefont {Wybo}, \citenamefont {Trypogeorgos}, \citenamefont {Guo}, \citenamefont {Zvonarev}, \citenamefont {Knap}, \citenamefont {Landini},\ and\ \citenamefont {N\"agerl}}]{Horvath2026odf}%
  \BibitemOpen
  \bibfield  {author} {\bibinfo {author} {\bibfnamefont {M.}~\bibnamefont {Horvath}}, \bibinfo {author} {\bibfnamefont {S.}~\bibnamefont {Dhar}}, \bibinfo {author} {\bibfnamefont {E.}~\bibnamefont {Wybo}}, \bibinfo {author} {\bibfnamefont {D.}~\bibnamefont {Trypogeorgos}}, \bibinfo {author} {\bibfnamefont {Y.}~\bibnamefont {Guo}}, \bibinfo {author} {\bibfnamefont {M.}~\bibnamefont {Zvonarev}}, \bibinfo {author} {\bibfnamefont {M.}~\bibnamefont {Knap}}, \bibinfo {author} {\bibfnamefont {M.}~\bibnamefont {Landini}},\ and\ \bibinfo {author} {\bibfnamefont {H.-C.}\ \bibnamefont {N\"agerl}},\ }\href@noop {} {\bibinfo {title} {Observing dissipationless flow of an impurity in a strongly repulsive quantum fluid}} (\bibinfo {year} {2026}),\ \Eprint {https://arxiv.org/abs/2602.12320} {arXiv:2602.12320} \BibitemShut {NoStop}%
\bibitem [{\citenamefont {Andrade-Sánchez}\ and\ \citenamefont {Camacho-Guardian}(2026)}]{Andrade2026cbp}%
  \BibitemOpen
  \bibfield  {author} {\bibinfo {author} {\bibfnamefont {G.}~\bibnamefont {Andrade-Sánchez}}\ and\ \bibinfo {author} {\bibfnamefont {A.}~\bibnamefont {Camacho-Guardian}},\ }\href@noop {} {\bibinfo {title} {{Charged Bose polarons at finite momentum}}} (\bibinfo {year} {2026}),\ \Eprint {https://arxiv.org/abs/2605.31255} {arXiv:2605.31255} \BibitemShut {NoStop}%
\bibitem [{\citenamefont {Navon}\ \emph {et~al.}(2010)\citenamefont {Navon}, \citenamefont {Nascimb\`{e}ne}, \citenamefont {Chevy},\ and\ \citenamefont {Salomon}}]{Navon2010teo}%
  \BibitemOpen
  \bibfield  {author} {\bibinfo {author} {\bibfnamefont {N.}~\bibnamefont {Navon}}, \bibinfo {author} {\bibfnamefont {S.}~\bibnamefont {Nascimb\`{e}ne}}, \bibinfo {author} {\bibfnamefont {F.}~\bibnamefont {Chevy}},\ and\ \bibinfo {author} {\bibfnamefont {C.}~\bibnamefont {Salomon}},\ }\bibfield  {title} {\bibinfo {title} {{The Equation of State of a Low-Temperature Fermi Gas with Tunable Interactions}},\ }\href {https://doi.org/10.1126/science.1187582} {\bibfield  {journal} {\bibinfo  {journal} {Science}\ }\textbf {\bibinfo {volume} {328}},\ \bibinfo {pages} {729} (\bibinfo {year} {2010})}\BibitemShut {NoStop}%
\bibitem [{\citenamefont {Chin}\ \emph {et~al.}(2010)\citenamefont {Chin}, \citenamefont {Grimm}, \citenamefont {Julienne},\ and\ \citenamefont {Tiesinga}}]{Chin2010fri}%
  \BibitemOpen
  \bibfield  {author} {\bibinfo {author} {\bibfnamefont {C.}~\bibnamefont {Chin}}, \bibinfo {author} {\bibfnamefont {R.}~\bibnamefont {Grimm}}, \bibinfo {author} {\bibfnamefont {P.~S.}\ \bibnamefont {Julienne}},\ and\ \bibinfo {author} {\bibfnamefont {E.}~\bibnamefont {Tiesinga}},\ }\bibfield  {title} {\bibinfo {title} {Feshbach resonances in ultracold gases},\ }\href {https://doi.org/doi.org/10.1103/RevModPhys.82.1225} {\bibfield  {journal} {\bibinfo  {journal} {Rev. Mod. Phys.}\ }\textbf {\bibinfo {volume} {82}},\ \bibinfo {pages} {1225} (\bibinfo {year} {2010})}\BibitemShut {NoStop}%
\bibitem [{\citenamefont {Fritsche}\ \emph {et~al.}(2021)\citenamefont {Fritsche}, \citenamefont {Baroni}, \citenamefont {Dobler}, \citenamefont {Kirilov}, \citenamefont {Huang}, \citenamefont {Grimm}, \citenamefont {Bruun},\ and\ \citenamefont {Massignan}}]{Fritsche2021sab}%
  \BibitemOpen
  \bibfield  {author} {\bibinfo {author} {\bibfnamefont {I.}~\bibnamefont {Fritsche}}, \bibinfo {author} {\bibfnamefont {C.}~\bibnamefont {Baroni}}, \bibinfo {author} {\bibfnamefont {E.}~\bibnamefont {Dobler}}, \bibinfo {author} {\bibfnamefont {E.}~\bibnamefont {Kirilov}}, \bibinfo {author} {\bibfnamefont {B.}~\bibnamefont {Huang}}, \bibinfo {author} {\bibfnamefont {R.}~\bibnamefont {Grimm}}, \bibinfo {author} {\bibfnamefont {G.~M.}\ \bibnamefont {Bruun}},\ and\ \bibinfo {author} {\bibfnamefont {P.}~\bibnamefont {Massignan}},\ }\bibfield  {title} {\bibinfo {title} {{Stability and breakdown of Fermi polarons in a strongly interacting Fermi-Bose mixture}},\ }\href {https://doi.org/10.1103/PhysRevA.103.053314} {\bibfield  {journal} {\bibinfo  {journal} {Phys. Rev. A}\ }\textbf {\bibinfo {volume} {103}},\ \bibinfo {pages} {053314} (\bibinfo {year} {2021})}\BibitemShut {NoStop}%
\bibitem [{\citenamefont {Grimm}\ and\ \citenamefont {Baroni}(2022)}]{Grimm2025fqm}%
  \BibitemOpen
  \bibfield  {author} {\bibinfo {author} {\bibfnamefont {R.}~\bibnamefont {Grimm}}\ and\ \bibinfo {author} {\bibfnamefont {C.}~\bibnamefont {Baroni}},\ }\href@noop {} {\bibinfo {title} {Fermionic quantum mixtures with tunable interactions}} (\bibinfo {year} {2022}),\ \bibinfo {note} {in Ref.~\cite{Varenna2022book}}\BibitemShut {NoStop}%
\bibitem [{\citenamefont {Dobler}(2026)}]{Dobler2026PhD}%
  \BibitemOpen
  \bibfield  {author} {\bibinfo {author} {\bibfnamefont {E.}~\bibnamefont {Dobler}},\ }\emph {\bibinfo {title} {Ultracold Mixtures of Lithium and Potassium: An Atomic Beam Source, Raman Transitions and Two-Body Scattering}},\ \href {http://www.ultracold.at/theses/2026-dobler.pdf} {Ph.D. thesis},\ \bibinfo  {school} {University of Innsbruck} (\bibinfo {year} {2026})\BibitemShut {NoStop}%
\bibitem [{\citenamefont {Kokkelmans}(2015)}]{Kokkelmans2014fri}%
  \BibitemOpen
  \bibfield  {author} {\bibinfo {author} {\bibfnamefont {S.}~\bibnamefont {Kokkelmans}},\ }\bibfield  {title} {\bibinfo {title} {{Feshbach Resonances in Ultracold Gases}},\ }in\ \href@noop {} {\emph {\bibinfo {booktitle} {Quantum Gas Experiments}}},\ \bibinfo {editor} {edited by\ \bibinfo {editor} {\bibfnamefont {P.}~\bibnamefont {Törma}}\ and\ \bibinfo {editor} {\bibfnamefont {K.}~\bibnamefont {Sengstock}}}\ (\bibinfo  {publisher} {Imperial College Press},\ \bibinfo {year} {2015})\ pp.\ \bibinfo {pages} {63--85}\BibitemShut {NoStop}%
\bibitem [{\citenamefont {Lous}\ \emph {et~al.}(2018)\citenamefont {Lous}, \citenamefont {Fritsche}, \citenamefont {Jag}, \citenamefont {Lehmann}, \citenamefont {Kirilov}, \citenamefont {Huang},\ and\ \citenamefont {Grimm}}]{Lous2018pti}%
  \BibitemOpen
  \bibfield  {author} {\bibinfo {author} {\bibfnamefont {R.~S.}\ \bibnamefont {Lous}}, \bibinfo {author} {\bibfnamefont {I.}~\bibnamefont {Fritsche}}, \bibinfo {author} {\bibfnamefont {M.}~\bibnamefont {Jag}}, \bibinfo {author} {\bibfnamefont {F.}~\bibnamefont {Lehmann}}, \bibinfo {author} {\bibfnamefont {E.}~\bibnamefont {Kirilov}}, \bibinfo {author} {\bibfnamefont {B.}~\bibnamefont {Huang}},\ and\ \bibinfo {author} {\bibfnamefont {R.}~\bibnamefont {Grimm}},\ }\bibfield  {title} {\bibinfo {title} {{Probing the Interface of a Phase-Separated State in a Repulsive Bose-Fermi Mixture}},\ }\href {https://doi.org/10.1103/PhysRevLett.120.243403} {\bibfield  {journal} {\bibinfo  {journal} {Phys. Rev. Lett.}\ }\textbf {\bibinfo {volume} {120}},\ \bibinfo {pages} {243403} (\bibinfo {year} {2018})}\BibitemShut {NoStop}%
\bibitem [{\citenamefont {Petrov}(2004)}]{Petrov2004tbp}%
  \BibitemOpen
  \bibfield  {author} {\bibinfo {author} {\bibfnamefont {D.~S.}\ \bibnamefont {Petrov}},\ }\bibfield  {title} {\bibinfo {title} {{Three-Boson Problem near a Narrow {F}eshbach Resonance}},\ }\href {https://doi.org/10.1103/PhysRevLett.93.143201} {\bibfield  {journal} {\bibinfo  {journal} {Phys. Rev. Lett.}\ }\textbf {\bibinfo {volume} {93}},\ \bibinfo {pages} {143201} (\bibinfo {year} {2004})}\BibitemShut {NoStop}%
\bibitem [{\citenamefont {Vale}\ and\ \citenamefont {Zwierlein}(2021)}]{Vale2021spo}%
  \BibitemOpen
  \bibfield  {author} {\bibinfo {author} {\bibfnamefont {C.~J.}\ \bibnamefont {Vale}}\ and\ \bibinfo {author} {\bibfnamefont {M.}~\bibnamefont {Zwierlein}},\ }\bibfield  {title} {\bibinfo {title} {Spectroscopic probes of quantum gases},\ }\href {https://doi.org/10.1038/s41567-021-01434-6} {\bibfield  {journal} {\bibinfo  {journal} {Nat. Phys.}\ }\textbf {\bibinfo {volume} {17}},\ \bibinfo {pages} {1305} (\bibinfo {year} {2021})}\BibitemShut {NoStop}%
\bibitem [{\citenamefont {Liu}\ \emph {et~al.}(2020{\natexlab{a}})\citenamefont {Liu}, \citenamefont {Shi}, \citenamefont {Levinsen},\ and\ \citenamefont {Parish}}]{Liu2020rfr}%
  \BibitemOpen
  \bibfield  {author} {\bibinfo {author} {\bibfnamefont {W.~E.}\ \bibnamefont {Liu}}, \bibinfo {author} {\bibfnamefont {Z.-Y.}\ \bibnamefont {Shi}}, \bibinfo {author} {\bibfnamefont {J.}~\bibnamefont {Levinsen}},\ and\ \bibinfo {author} {\bibfnamefont {M.~M.}\ \bibnamefont {Parish}},\ }\bibfield  {title} {\bibinfo {title} {{Radio-Frequency Response and Contact of Impurities in a Quantum Gas}},\ }\href {https://doi.org/10.1103/PhysRevLett.125.065301} {\bibfield  {journal} {\bibinfo  {journal} {Phys. Rev. Lett.}\ }\textbf {\bibinfo {volume} {125}},\ \bibinfo {pages} {065301} (\bibinfo {year} {2020}{\natexlab{a}})}\BibitemShut {NoStop}%
\bibitem [{\citenamefont {Törmä}(2016)}]{Torma2016pou}%
  \BibitemOpen
  \bibfield  {author} {\bibinfo {author} {\bibfnamefont {P.}~\bibnamefont {Törmä}},\ }\bibfield  {title} {\bibinfo {title} {{Physics of ultracold Fermi gases revealed by spectroscopies}},\ }\href {https://doi.org/10.1088/0031-8949/91/4/043006} {\bibfield  {journal} {\bibinfo  {journal} {Phys. Scr.}\ }\textbf {\bibinfo {volume} {91}},\ \bibinfo {pages} {043006} (\bibinfo {year} {2016})}\BibitemShut {NoStop}%
\bibitem [{\citenamefont {Horvath}\ \emph {et~al.}(2017)\citenamefont {Horvath}, \citenamefont {Thomas}, \citenamefont {Tiesinga}, \citenamefont {Deb},\ and\ \citenamefont {Kjærgaard}}]{Horvath2017ats}%
  \BibitemOpen
  \bibfield  {author} {\bibinfo {author} {\bibfnamefont {M.~S.}\ \bibnamefont {Horvath}}, \bibinfo {author} {\bibfnamefont {R.}~\bibnamefont {Thomas}}, \bibinfo {author} {\bibfnamefont {E.}~\bibnamefont {Tiesinga}}, \bibinfo {author} {\bibfnamefont {A.~B.}\ \bibnamefont {Deb}},\ and\ \bibinfo {author} {\bibfnamefont {N.}~\bibnamefont {Kjærgaard}},\ }\bibfield  {title} {\bibinfo {title} {{Above-threshold scattering about a Feshbach resonance for ultracold atoms in an optical collider}},\ }\href {https://doi.org/https://doi.org/10.1038/s41467-017-00458-y} {\bibfield  {journal} {\bibinfo  {journal} {Nat. Commun.}\ }\textbf {\bibinfo {volume} {8}},\ \bibinfo {pages} {452} (\bibinfo {year} {2017})}\BibitemShut {NoStop}%
\bibitem [{\citenamefont {Cetina}\ \emph {et~al.}(2015)\citenamefont {Cetina}, \citenamefont {Jag}, \citenamefont {Lous}, \citenamefont {Walraven}, \citenamefont {Grimm}, \citenamefont {Christensen},\ and\ \citenamefont {Bruun}}]{Cetina2015doi}%
  \BibitemOpen
  \bibfield  {author} {\bibinfo {author} {\bibfnamefont {M.}~\bibnamefont {Cetina}}, \bibinfo {author} {\bibfnamefont {M.}~\bibnamefont {Jag}}, \bibinfo {author} {\bibfnamefont {R.~S.}\ \bibnamefont {Lous}}, \bibinfo {author} {\bibfnamefont {J.~T.~M.}\ \bibnamefont {Walraven}}, \bibinfo {author} {\bibfnamefont {R.}~\bibnamefont {Grimm}}, \bibinfo {author} {\bibfnamefont {R.~S.}\ \bibnamefont {Christensen}},\ and\ \bibinfo {author} {\bibfnamefont {G.~M.}\ \bibnamefont {Bruun}},\ }\bibfield  {title} {\bibinfo {title} {Decoherence of {I}mpurities in a {F}ermi {S}ea of {U}ltracold {A}toms},\ }\href {https://doi.org/10.1103/PhysRevLett.115.135302} {\bibfield  {journal} {\bibinfo  {journal} {Phys. Rev. Lett.}\ }\textbf {\bibinfo {volume} {115}},\ \bibinfo {pages} {135302} (\bibinfo {year} {2015})}\BibitemShut {NoStop}%
\bibitem [{\citenamefont {Cetina}\ \emph {et~al.}(2016)\citenamefont {Cetina}, \citenamefont {Jag}, \citenamefont {Lous}, \citenamefont {Fritsche}, \citenamefont {Walraven}, \citenamefont {Grimm}, \citenamefont {Levinsen}, \citenamefont {Parish}, \citenamefont {Schmidt}, \citenamefont {Knap},\ and\ \citenamefont {Demler}}]{Cetina2016umb}%
  \BibitemOpen
  \bibfield  {author} {\bibinfo {author} {\bibfnamefont {M.}~\bibnamefont {Cetina}}, \bibinfo {author} {\bibfnamefont {M.}~\bibnamefont {Jag}}, \bibinfo {author} {\bibfnamefont {R.~S.}\ \bibnamefont {Lous}}, \bibinfo {author} {\bibfnamefont {I.}~\bibnamefont {Fritsche}}, \bibinfo {author} {\bibfnamefont {J.~T.~M.}\ \bibnamefont {Walraven}}, \bibinfo {author} {\bibfnamefont {R.}~\bibnamefont {Grimm}}, \bibinfo {author} {\bibfnamefont {J.}~\bibnamefont {Levinsen}}, \bibinfo {author} {\bibfnamefont {M.~M.}\ \bibnamefont {Parish}}, \bibinfo {author} {\bibfnamefont {R.}~\bibnamefont {Schmidt}}, \bibinfo {author} {\bibfnamefont {M.}~\bibnamefont {Knap}},\ and\ \bibinfo {author} {\bibfnamefont {E.}~\bibnamefont {Demler}},\ }\bibfield  {title} {\bibinfo {title} {{U}ltrafast many-body interferometry of impurities coupled to a {F}ermi sea},\ }\href {https://doi.org/10.1126/science.aaf5134} {\bibfield  {journal} {\bibinfo  {journal} {Science}\ }\textbf {\bibinfo {volume} {354}},\ \bibinfo {pages} {96} (\bibinfo {year}
  {2016})}\BibitemShut {NoStop}%
\bibitem [{\citenamefont {Spiegelhalder}\ \emph {et~al.}(2010)\citenamefont {Spiegelhalder}, \citenamefont {Trenkwalder}, \citenamefont {Naik}, \citenamefont {Kerner}, \citenamefont {Wille}, \citenamefont {Hendl}, \citenamefont {Schreck},\ and\ \citenamefont {Grimm}}]{Spiegelhalder2010aop}%
  \BibitemOpen
  \bibfield  {author} {\bibinfo {author} {\bibfnamefont {F.~M.}\ \bibnamefont {Spiegelhalder}}, \bibinfo {author} {\bibfnamefont {A.}~\bibnamefont {Trenkwalder}}, \bibinfo {author} {\bibfnamefont {D.}~\bibnamefont {Naik}}, \bibinfo {author} {\bibfnamefont {G.}~\bibnamefont {Kerner}}, \bibinfo {author} {\bibfnamefont {E.}~\bibnamefont {Wille}}, \bibinfo {author} {\bibfnamefont {G.}~\bibnamefont {Hendl}}, \bibinfo {author} {\bibfnamefont {F.}~\bibnamefont {Schreck}},\ and\ \bibinfo {author} {\bibfnamefont {R.}~\bibnamefont {Grimm}},\ }\bibfield  {title} {\bibinfo {title} {{All-optical production of a degenerate mixture of $^6$Li and $^{40}$K and creation of heteronuclear molecules}},\ }\href {https://doi.org/10.1103/PhysRevA.81.043637} {\bibfield  {journal} {\bibinfo  {journal} {Phys. Rev. A}\ }\textbf {\bibinfo {volume} {81}},\ \bibinfo {pages} {043637} (\bibinfo {year} {2010})}\BibitemShut {NoStop}%
\bibitem [{\citenamefont {Lous}(2018)}]{Lous2018PhD}%
  \BibitemOpen
  \bibfield  {author} {\bibinfo {author} {\bibfnamefont {R.~S.}\ \bibnamefont {Lous}},\ }\emph {\bibinfo {title} {{Tunable Bose-Fermi and Fermi-Fermi Mixtures of Potassium and Lithium: Phase Separation, Polarons, and Molecules}}},\ \href {http://www.ultracold.at/theses/2018-lous.pdf} {Ph.D. thesis},\ \bibinfo  {school} {University of Innsbruck} (\bibinfo {year} {2018})\BibitemShut {NoStop}%
\bibitem [{\citenamefont {Lous}\ \emph {et~al.}(2017)\citenamefont {Lous}, \citenamefont {Fritsche}, \citenamefont {Jag}, \citenamefont {Huang},\ and\ \citenamefont {Grimm}}]{Lous2017toa}%
  \BibitemOpen
  \bibfield  {author} {\bibinfo {author} {\bibfnamefont {R.~S.}\ \bibnamefont {Lous}}, \bibinfo {author} {\bibfnamefont {I.}~\bibnamefont {Fritsche}}, \bibinfo {author} {\bibfnamefont {M.}~\bibnamefont {Jag}}, \bibinfo {author} {\bibfnamefont {B.}~\bibnamefont {Huang}},\ and\ \bibinfo {author} {\bibfnamefont {R.}~\bibnamefont {Grimm}},\ }\bibfield  {title} {\bibinfo {title} {Thermometry of a deeply degenerate {F}ermi gas with a {B}ose-{E}instein condensate},\ }\href {https://doi.org/10.1103/PhysRevA.95.053627} {\bibfield  {journal} {\bibinfo  {journal} {Phys. Rev. A}\ }\textbf {\bibinfo {volume} {95}},\ \bibinfo {pages} {053627} (\bibinfo {year} {2017})}\BibitemShut {NoStop}%
\bibitem [{\citenamefont {Liu}\ \emph {et~al.}(2020{\natexlab{b}})\citenamefont {Liu}, \citenamefont {Shi}, \citenamefont {Parish},\ and\ \citenamefont {Levinsen}}]{Liu2020tor}%
  \BibitemOpen
  \bibfield  {author} {\bibinfo {author} {\bibfnamefont {W.~E.}\ \bibnamefont {Liu}}, \bibinfo {author} {\bibfnamefont {Z.}~\bibnamefont {Shi}}, \bibinfo {author} {\bibfnamefont {M.~M.}\ \bibnamefont {Parish}},\ and\ \bibinfo {author} {\bibfnamefont {J.}~\bibnamefont {Levinsen}},\ }\bibfield  {title} {\bibinfo {title} {{Theory of Radio-Frequency Spectroscopy of Impurities in Quantum Gases}},\ }\href {https://doi.org/https://doi.org/10.1103/PhysRevA.102.023304} {\bibfield  {journal} {\bibinfo  {journal} {Phys. Rev. A}\ }\textbf {\bibinfo {volume} {102}},\ \bibinfo {pages} {023304} (\bibinfo {year} {2020}{\natexlab{b}})}\BibitemShut {NoStop}%
\bibitem [{\citenamefont {Kasevich}\ \emph {et~al.}(1991)\citenamefont {Kasevich}, \citenamefont {Weiss}, \citenamefont {Riis}, \citenamefont {Moler}, \citenamefont {Kasapi},\ and\ \citenamefont {Chu}}]{Kasevich1991avs}%
  \BibitemOpen
  \bibfield  {author} {\bibinfo {author} {\bibfnamefont {M.}~\bibnamefont {Kasevich}}, \bibinfo {author} {\bibfnamefont {D.~S.}\ \bibnamefont {Weiss}}, \bibinfo {author} {\bibfnamefont {E.}~\bibnamefont {Riis}}, \bibinfo {author} {\bibfnamefont {K.}~\bibnamefont {Moler}}, \bibinfo {author} {\bibfnamefont {S.}~\bibnamefont {Kasapi}},\ and\ \bibinfo {author} {\bibfnamefont {S.}~\bibnamefont {Chu}},\ }\bibfield  {title} {\bibinfo {title} {Atomic velocity selection using stimulated raman transitions},\ }\href {https://doi.org/10.1103/PhysRevLett.66.2297} {\bibfield  {journal} {\bibinfo  {journal} {Phys. Rev. Lett.}\ }\textbf {\bibinfo {volume} {66}},\ \bibinfo {pages} {2297} (\bibinfo {year} {1991})}\BibitemShut {NoStop}%
\bibitem [{\citenamefont {Weitz}\ \emph {et~al.}(1994)\citenamefont {Weitz}, \citenamefont {Young},\ and\ \citenamefont {Chu}}]{Weitz1994aib}%
  \BibitemOpen
  \bibfield  {author} {\bibinfo {author} {\bibfnamefont {M.}~\bibnamefont {Weitz}}, \bibinfo {author} {\bibfnamefont {B.~C.}\ \bibnamefont {Young}},\ and\ \bibinfo {author} {\bibfnamefont {S.}~\bibnamefont {Chu}},\ }\bibfield  {title} {\bibinfo {title} {Atomic interferometer based on adiabatic population transfer},\ }\href {https://doi.org/10.1103/PhysRevLett.73.2563} {\bibfield  {journal} {\bibinfo  {journal} {Phys. Rev. Lett.}\ }\textbf {\bibinfo {volume} {73}},\ \bibinfo {pages} {2563} (\bibinfo {year} {1994})}\BibitemShut {NoStop}%
\bibitem [{\citenamefont {Jaffe}\ \emph {et~al.}(2018)\citenamefont {Jaffe}, \citenamefont {Xu}, \citenamefont {Haslinger}, \citenamefont {M\"uller},\ and\ \citenamefont {Hamilton}}]{Jaffe2018eas}%
  \BibitemOpen
  \bibfield  {author} {\bibinfo {author} {\bibfnamefont {M.}~\bibnamefont {Jaffe}}, \bibinfo {author} {\bibfnamefont {V.}~\bibnamefont {Xu}}, \bibinfo {author} {\bibfnamefont {P.}~\bibnamefont {Haslinger}}, \bibinfo {author} {\bibfnamefont {H.}~\bibnamefont {M\"uller}},\ and\ \bibinfo {author} {\bibfnamefont {P.}~\bibnamefont {Hamilton}},\ }\bibfield  {title} {\bibinfo {title} {Efficient adiabatic spin-dependent kicks in an atom interferometer},\ }\href {https://doi.org/10.1103/PhysRevLett.121.040402} {\bibfield  {journal} {\bibinfo  {journal} {Phys. Rev. Lett.}\ }\textbf {\bibinfo {volume} {121}},\ \bibinfo {pages} {040402} (\bibinfo {year} {2018})}\BibitemShut {NoStop}%
\bibitem [{\citenamefont {McGuirk}\ \emph {et~al.}(2000)\citenamefont {McGuirk}, \citenamefont {Snadden},\ and\ \citenamefont {Kasevich}}]{Mcguirk2000lal}%
  \BibitemOpen
  \bibfield  {author} {\bibinfo {author} {\bibfnamefont {J.~M.}\ \bibnamefont {McGuirk}}, \bibinfo {author} {\bibfnamefont {M.~J.}\ \bibnamefont {Snadden}},\ and\ \bibinfo {author} {\bibfnamefont {M.~A.}\ \bibnamefont {Kasevich}},\ }\bibfield  {title} {\bibinfo {title} {Large area light-pulse atom interferometry},\ }\href {https://doi.org/10.1103/PhysRevLett.85.4498} {\bibfield  {journal} {\bibinfo  {journal} {Phys. Rev. Lett.}\ }\textbf {\bibinfo {volume} {85}},\ \bibinfo {pages} {4498} (\bibinfo {year} {2000})}\BibitemShut {NoStop}%
\bibitem [{not()}]{note6hk}%
  \BibitemOpen
  \href@noop {} {}\bibinfo {note} {A later analysis of our measurements revealed a potential problem for momentum changes involving more than three subsequent Raman transitions. We found indications that some measurements may have been affected by incomplete population transfer. To avoid any contamination of our fit analysis, we generally restrict the fitted momentum range to $\Delta p \le 6\,\hbar k $. In most cases, however, the fits are found to describe also the excluded data at higher momentum very well.}\BibitemShut {Stop}%
\bibitem [{\citenamefont {Massignan}(2012)}]{Massignan2012pad}%
  \BibitemOpen
  \bibfield  {author} {\bibinfo {author} {\bibfnamefont {P.}~\bibnamefont {Massignan}},\ }\bibfield  {title} {\bibinfo {title} {{Polarons and dressed molecules near narrow Feshbach resonances}},\ }\href {https://doi.org/10.1209/0295-5075/98/10012} {\bibfield  {journal} {\bibinfo  {journal} {Europhys. Lett.}\ }\textbf {\bibinfo {volume} {98}},\ \bibinfo {pages} {10012} (\bibinfo {year} {2012})}\BibitemShut {NoStop}%
\bibitem [{\citenamefont {Ness}\ \emph {et~al.}(2020)\citenamefont {Ness}, \citenamefont {Shkedrov}, \citenamefont {Florshaim}, \citenamefont {Diessel}, \citenamefont {von Milczewski}, \citenamefont {Schmidt},\ and\ \citenamefont {Sagi}}]{Ness2020ooa}%
  \BibitemOpen
  \bibfield  {author} {\bibinfo {author} {\bibfnamefont {G.}~\bibnamefont {Ness}}, \bibinfo {author} {\bibfnamefont {C.}~\bibnamefont {Shkedrov}}, \bibinfo {author} {\bibfnamefont {Y.}~\bibnamefont {Florshaim}}, \bibinfo {author} {\bibfnamefont {O.~K.}\ \bibnamefont {Diessel}}, \bibinfo {author} {\bibfnamefont {J.}~\bibnamefont {von Milczewski}}, \bibinfo {author} {\bibfnamefont {R.}~\bibnamefont {Schmidt}},\ and\ \bibinfo {author} {\bibfnamefont {Y.}~\bibnamefont {Sagi}},\ }\bibfield  {title} {\bibinfo {title} {{Observation of a Smooth Polaron-Molecule Transition in a Degenerate Fermi Gas}},\ }\href {https://doi.org/10.1103/PhysRevX.10.041019} {\bibfield  {journal} {\bibinfo  {journal} {Phys. Rev. X}\ }\textbf {\bibinfo {volume} {10}},\ \bibinfo {pages} {041019} (\bibinfo {year} {2020})}\BibitemShut {NoStop}%
\bibitem [{\citenamefont {Bruun}\ \emph {et~al.}(2008)\citenamefont {Bruun}, \citenamefont {Recati}, \citenamefont {Pethick}, \citenamefont {Smith},\ and\ \citenamefont {Stringari}}]{Bruun2008cpo}%
  \BibitemOpen
  \bibfield  {author} {\bibinfo {author} {\bibfnamefont {G.~M.}\ \bibnamefont {Bruun}}, \bibinfo {author} {\bibfnamefont {A.}~\bibnamefont {Recati}}, \bibinfo {author} {\bibfnamefont {C.~J.}\ \bibnamefont {Pethick}}, \bibinfo {author} {\bibfnamefont {H.}~\bibnamefont {Smith}},\ and\ \bibinfo {author} {\bibfnamefont {S.}~\bibnamefont {Stringari}},\ }\bibfield  {title} {\bibinfo {title} {{Collisional Properties of a Polarized Fermi Gas with Resonant Interactions}},\ }\href {https://doi.org/10.1103/PhysRevLett.100.240406} {\bibfield  {journal} {\bibinfo  {journal} {Phys. Rev. Lett.}\ }\textbf {\bibinfo {volume} {100}},\ \bibinfo {pages} {240406} (\bibinfo {year} {2008})}\BibitemShut {NoStop}%
\bibitem [{\citenamefont {Massignan}\ and\ \citenamefont {Bruun}(2011)}]{Massignan2011rpa}%
  \BibitemOpen
  \bibfield  {author} {\bibinfo {author} {\bibfnamefont {P.}~\bibnamefont {Massignan}}\ and\ \bibinfo {author} {\bibfnamefont {G.~M.}\ \bibnamefont {Bruun}},\ }\bibfield  {title} {\bibinfo {title} {{R}epulsive polarons and itinerant ferromagnetismin strongly polarized {F}ermi gases},\ }\href {https://doi.org/10.1140/epjd/e2011-20084-5} {\bibfield  {journal} {\bibinfo  {journal} {Eur. Phys. J D}\ }\textbf {\bibinfo {volume} {65}},\ \bibinfo {pages} {83} (\bibinfo {year} {2011})}\BibitemShut {NoStop}%
\bibitem [{\citenamefont {Prokof'ev}\ and\ \citenamefont {Svistunov}(2008)}]{Prokofev2008fpp}%
  \BibitemOpen
  \bibfield  {author} {\bibinfo {author} {\bibfnamefont {N.}~\bibnamefont {Prokof'ev}}\ and\ \bibinfo {author} {\bibfnamefont {B.}~\bibnamefont {Svistunov}},\ }\bibfield  {title} {\bibinfo {title} {{Fermi-polaron problem: Diagrammatic Monte Carlo method for divergent sign-alternating series}},\ }\href {https://doi.org/10.1103/PhysRevB.77.020408} {\bibfield  {journal} {\bibinfo  {journal} {Phys. Rev. B}\ }\textbf {\bibinfo {volume} {77}},\ \bibinfo {pages} {020408} (\bibinfo {year} {2008})}\BibitemShut {NoStop}%
\bibitem [{\citenamefont {Punk}\ \emph {et~al.}(2009)\citenamefont {Punk}, \citenamefont {Dumitrescu},\ and\ \citenamefont {Zwerger}}]{Punk2009ptm}%
  \BibitemOpen
  \bibfield  {author} {\bibinfo {author} {\bibfnamefont {M.}~\bibnamefont {Punk}}, \bibinfo {author} {\bibfnamefont {P.~T.}\ \bibnamefont {Dumitrescu}},\ and\ \bibinfo {author} {\bibfnamefont {W.}~\bibnamefont {Zwerger}},\ }\bibfield  {title} {\bibinfo {title} {{Polaron-to-molecule transition in a strongly imbalanced Fermi gas}},\ }\href {https://doi.org/10.1103/PhysRevA.80.053605} {\bibfield  {journal} {\bibinfo  {journal} {Phys. Rev. A}\ }\textbf {\bibinfo {volume} {80}},\ \bibinfo {pages} {053605} (\bibinfo {year} {2009})}\BibitemShut {NoStop}%
\bibitem [{\citenamefont {Combescot}\ \emph {et~al.}(2010)\citenamefont {Combescot}, \citenamefont {Giraud},\ and\ \citenamefont {Leyronas}}]{Combescot2010ato}%
  \BibitemOpen
  \bibfield  {author} {\bibinfo {author} {\bibfnamefont {R.}~\bibnamefont {Combescot}}, \bibinfo {author} {\bibfnamefont {S.}~\bibnamefont {Giraud}},\ and\ \bibinfo {author} {\bibfnamefont {X.}~\bibnamefont {Leyronas}},\ }\bibfield  {title} {\bibinfo {title} {{Analytical theory of the dressed bound state in highly polarized Fermi gases}},\ }\href {https://doi.org/10.1209/0295-5075/88/60007} {\bibfield  {journal} {\bibinfo  {journal} {Europhys. Lett.}\ }\textbf {\bibinfo {volume} {88}},\ \bibinfo {pages} {60007} (\bibinfo {year} {2010})}\BibitemShut {NoStop}%
\bibitem [{\citenamefont {Bruun}\ and\ \citenamefont {Massignan}(2010)}]{Bruun2010dop}%
  \BibitemOpen
  \bibfield  {author} {\bibinfo {author} {\bibfnamefont {G.~M.}\ \bibnamefont {Bruun}}\ and\ \bibinfo {author} {\bibfnamefont {P.}~\bibnamefont {Massignan}},\ }\bibfield  {title} {\bibinfo {title} {{Decay of Polarons and Molecules in a Strongly Polarized Fermi Gas}},\ }\href {https://doi.org/10.1103/PhysRevLett.105.020403} {\bibfield  {journal} {\bibinfo  {journal} {Phys. Rev. Lett.}\ }\textbf {\bibinfo {volume} {105}},\ \bibinfo {pages} {020403} (\bibinfo {year} {2010})}\BibitemShut {NoStop}%
\bibitem [{\citenamefont {Schmidt}\ and\ \citenamefont {Enss}(2011)}]{Schmidt2011esa}%
  \BibitemOpen
  \bibfield  {author} {\bibinfo {author} {\bibfnamefont {R.}~\bibnamefont {Schmidt}}\ and\ \bibinfo {author} {\bibfnamefont {T.}~\bibnamefont {Enss}},\ }\bibfield  {title} {\bibinfo {title} {{Excitation spectra and rf response near the polaron-to-molecule transition from the functional renormalization group}},\ }\href {https://doi.org/10.1103/PhysRevA.83.063620} {\bibfield  {journal} {\bibinfo  {journal} {Phys. Rev. A}\ }\textbf {\bibinfo {volume} {83}},\ \bibinfo {pages} {063620} (\bibinfo {year} {2011})}\BibitemShut {NoStop}%
\bibitem [{\citenamefont {Trefzger}\ and\ \citenamefont {Castin}(2012)}]{Trefzger2012iia}%
  \BibitemOpen
  \bibfield  {author} {\bibinfo {author} {\bibfnamefont {C.}~\bibnamefont {Trefzger}}\ and\ \bibinfo {author} {\bibfnamefont {Y.}~\bibnamefont {Castin}},\ }\bibfield  {title} {\bibinfo {title} {{I}mpurity in a {F}ermi sea on a narrow {F}eshbach resonance: {A} variational study of the polaronic and dimeronic branches},\ }\href {https://doi.org/10.1103/PhysRevA.85.053612} {\bibfield  {journal} {\bibinfo  {journal} {Phys. Rev. A}\ }\textbf {\bibinfo {volume} {85}},\ \bibinfo {pages} {053612} (\bibinfo {year} {2012})}\BibitemShut {NoStop}%
\bibitem [{\citenamefont {Cui}(2020)}]{Cui2020fpr}%
  \BibitemOpen
  \bibfield  {author} {\bibinfo {author} {\bibfnamefont {X.}~\bibnamefont {Cui}},\ }\bibfield  {title} {\bibinfo {title} {{Fermi polaron revisited: Polaron-molecule transition and coexistence}},\ }\href {https://doi.org/10.1103/PhysRevA.102.061301} {\bibfield  {journal} {\bibinfo  {journal} {Phys. Rev. A}\ }\textbf {\bibinfo {volume} {102}},\ \bibinfo {pages} {061301(R)} (\bibinfo {year} {2020})}\BibitemShut {NoStop}%
\bibitem [{\citenamefont {Peng}\ \emph {et~al.}(2021)\citenamefont {Peng}, \citenamefont {Liu}, \citenamefont {Zhang},\ and\ \citenamefont {Cui}}]{Peng2021not}%
  \BibitemOpen
  \bibfield  {author} {\bibinfo {author} {\bibfnamefont {C.}~\bibnamefont {Peng}}, \bibinfo {author} {\bibfnamefont {R.}~\bibnamefont {Liu}}, \bibinfo {author} {\bibfnamefont {W.}~\bibnamefont {Zhang}},\ and\ \bibinfo {author} {\bibfnamefont {X.}~\bibnamefont {Cui}},\ }\bibfield  {title} {\bibinfo {title} {{Nature of the polaron-molecule transition in Fermi polarons}},\ }\href {https://doi.org/10.1103/PhysRevA.103.063312} {\bibfield  {journal} {\bibinfo  {journal} {Phys. Rev. A}\ }\textbf {\bibinfo {volume} {103}},\ \bibinfo {pages} {063312} (\bibinfo {year} {2021})}\BibitemShut {NoStop}%
\bibitem [{\citenamefont {Parish}\ \emph {et~al.}(2021)\citenamefont {Parish}, \citenamefont {Adlong}, \citenamefont {Liu},\ and\ \citenamefont {Levinsen}}]{Parish2021tso}%
  \BibitemOpen
  \bibfield  {author} {\bibinfo {author} {\bibfnamefont {M.~M.}\ \bibnamefont {Parish}}, \bibinfo {author} {\bibfnamefont {H.~S.}\ \bibnamefont {Adlong}}, \bibinfo {author} {\bibfnamefont {W.~E.}\ \bibnamefont {Liu}},\ and\ \bibinfo {author} {\bibfnamefont {J.}~\bibnamefont {Levinsen}},\ }\bibfield  {title} {\bibinfo {title} {Thermodynamic signatures of the polaron-molecule transition in a {F}ermi gas},\ }\href {https://doi.org/10.1103/PhysRevA.103.023312} {\bibfield  {journal} {\bibinfo  {journal} {Phys. Rev. A}\ }\textbf {\bibinfo {volume} {103}},\ \bibinfo {pages} {023312} (\bibinfo {year} {2021})}\BibitemShut {NoStop}%
\bibitem [{\citenamefont {Repp}\ \emph {et~al.}(2013)\citenamefont {Repp}, \citenamefont {Pires}, \citenamefont {Ulmanis}, \citenamefont {Heck}, \citenamefont {Kuhnle}, \citenamefont {Weidem\"uller},\ and\ \citenamefont {Tiemann}}]{Repp2013ooi}%
  \BibitemOpen
  \bibfield  {author} {\bibinfo {author} {\bibfnamefont {M.}~\bibnamefont {Repp}}, \bibinfo {author} {\bibfnamefont {R.}~\bibnamefont {Pires}}, \bibinfo {author} {\bibfnamefont {J.}~\bibnamefont {Ulmanis}}, \bibinfo {author} {\bibfnamefont {R.}~\bibnamefont {Heck}}, \bibinfo {author} {\bibfnamefont {E.~D.}\ \bibnamefont {Kuhnle}}, \bibinfo {author} {\bibfnamefont {M.}~\bibnamefont {Weidem\"uller}},\ and\ \bibinfo {author} {\bibfnamefont {E.}~\bibnamefont {Tiemann}},\ }\bibfield  {title} {\bibinfo {title} {{Observation of interspecies $^6$Li-$^133$Cs Feshbach resonances}},\ }\href {https://doi.org/10.1103/PhysRevA.87.010701} {\bibfield  {journal} {\bibinfo  {journal} {Phys. Rev. A}\ }\textbf {\bibinfo {volume} {87}},\ \bibinfo {pages} {010701(R)} (\bibinfo {year} {2013})}\BibitemShut {NoStop}%
\bibitem [{\citenamefont {Tung}\ \emph {et~al.}(2013)\citenamefont {Tung}, \citenamefont {Parker}, \citenamefont {Johansen}, \citenamefont {Chin}, \citenamefont {Wang},\ and\ \citenamefont {Julienne}}]{Tung2013umo}%
  \BibitemOpen
  \bibfield  {author} {\bibinfo {author} {\bibfnamefont {S.-K.}\ \bibnamefont {Tung}}, \bibinfo {author} {\bibfnamefont {C.}~\bibnamefont {Parker}}, \bibinfo {author} {\bibfnamefont {J.}~\bibnamefont {Johansen}}, \bibinfo {author} {\bibfnamefont {C.}~\bibnamefont {Chin}}, \bibinfo {author} {\bibfnamefont {Y.}~\bibnamefont {Wang}},\ and\ \bibinfo {author} {\bibfnamefont {P.~S.}\ \bibnamefont {Julienne}},\ }\bibfield  {title} {\bibinfo {title} {{Ultracold mixtures of atomic ${}^{6}$Li and ${}^{133}$Cs with tunable interactions}},\ }\href {https://doi.org/10.1103/PhysRevA.87.010702} {\bibfield  {journal} {\bibinfo  {journal} {Phys. Rev. A}\ }\textbf {\bibinfo {volume} {87}},\ \bibinfo {pages} {010702(R)} (\bibinfo {year} {2013})}\BibitemShut {NoStop}%
\bibitem [{\citenamefont {Sch\"afer}\ \emph {et~al.}(2022)\citenamefont {Sch\"afer}, \citenamefont {Mizukami},\ and\ \citenamefont {Takahashi}}]{Schafer2022fro}%
  \BibitemOpen
  \bibfield  {author} {\bibinfo {author} {\bibfnamefont {F.}~\bibnamefont {Sch\"afer}}, \bibinfo {author} {\bibfnamefont {N.}~\bibnamefont {Mizukami}},\ and\ \bibinfo {author} {\bibfnamefont {Y.}~\bibnamefont {Takahashi}},\ }\bibfield  {title} {\bibinfo {title} {{Feshbach resonances of large-mass-imbalance Er-Li mixtures}},\ }\href {https://doi.org/10.1103/PhysRevA.105.012816} {\bibfield  {journal} {\bibinfo  {journal} {Phys. Rev. A}\ }\textbf {\bibinfo {volume} {105}},\ \bibinfo {pages} {012816} (\bibinfo {year} {2022})}\BibitemShut {NoStop}%
\bibitem [{\citenamefont {Xie}\ \emph {et~al.}(2025)\citenamefont {Xie}, \citenamefont {Li}, \citenamefont {Zhou}, \citenamefont {Luo}, \citenamefont {Wang}, \citenamefont {Nie}, \citenamefont {Shen}, \citenamefont {Chen}, \citenamefont {Yao},\ and\ \citenamefont {Pan}}]{Xie2025fso}%
  \BibitemOpen
  \bibfield  {author} {\bibinfo {author} {\bibfnamefont {K.}~\bibnamefont {Xie}}, \bibinfo {author} {\bibfnamefont {X.}~\bibnamefont {Li}}, \bibinfo {author} {\bibfnamefont {Y.-Y.}\ \bibnamefont {Zhou}}, \bibinfo {author} {\bibfnamefont {J.-H.}\ \bibnamefont {Luo}}, \bibinfo {author} {\bibfnamefont {S.}~\bibnamefont {Wang}}, \bibinfo {author} {\bibfnamefont {Y.-Z.}\ \bibnamefont {Nie}}, \bibinfo {author} {\bibfnamefont {H.-C.}\ \bibnamefont {Shen}}, \bibinfo {author} {\bibfnamefont {Y.-A.}\ \bibnamefont {Chen}}, \bibinfo {author} {\bibfnamefont {X.-C.}\ \bibnamefont {Yao}},\ and\ \bibinfo {author} {\bibfnamefont {J.-W.}\ \bibnamefont {Pan}},\ }\bibfield  {title} {\bibinfo {title} {{Feshbach spectroscopy of ultracold mixtures of $^{6}\mathrm{Li}$ and $^{164}\mathrm{Dy}$ atoms}},\ }\href {https://doi.org/10.1103/PhysRevA.111.023327} {\bibfield  {journal} {\bibinfo  {journal} {Phys. Rev. A}\ }\textbf {\bibinfo {volume} {111}},\ \bibinfo {pages} {023327} (\bibinfo {year} {2025})}\BibitemShut {NoStop}%
\bibitem [{\citenamefont {Lippi}(2024)}]{Lippi2024PhD}%
  \BibitemOpen
  \bibfield  {author} {\bibinfo {author} {\bibfnamefont {E.}~\bibnamefont {Lippi}},\ }\emph {\bibinfo {title} {$^{133}$Cs Atoms in a $^6$Li Fermi Sea for Exploring Polaron Physics in the Heavy Impurity Limit}},\ \href {https://archiv.ub.uni-heidelberg.de/volltextserver/35527/1/Dissertation_EleonoraLippi.pdf} {Ph.D. thesis},\ \bibinfo  {school} {University of Heidelberg} (\bibinfo {year} {2024})\BibitemShut {NoStop}%
\bibitem [{\citenamefont {Anderson}(1967)}]{Anderson1967ici}%
  \BibitemOpen
  \bibfield  {author} {\bibinfo {author} {\bibfnamefont {P.~W.}\ \bibnamefont {Anderson}},\ }\bibfield  {title} {\bibinfo {title} {{Infrared Catastrophe in Fermi Gases with Local Scattering Potentials}},\ }\href {https://doi.org/10.1103/PhysRevLett.18.1049} {\bibfield  {journal} {\bibinfo  {journal} {Phys. Rev. Lett.}\ }\textbf {\bibinfo {volume} {18}},\ \bibinfo {pages} {1049} (\bibinfo {year} {1967})}\BibitemShut {NoStop}%
\bibitem [{\citenamefont {Knap}\ \emph {et~al.}(2012)\citenamefont {Knap}, \citenamefont {Shashi}, \citenamefont {Nishida}, \citenamefont {Imambekov}, \citenamefont {Abanin},\ and\ \citenamefont {Demler}}]{Knap2012tdi}%
  \BibitemOpen
  \bibfield  {author} {\bibinfo {author} {\bibfnamefont {M.}~\bibnamefont {Knap}}, \bibinfo {author} {\bibfnamefont {A.}~\bibnamefont {Shashi}}, \bibinfo {author} {\bibfnamefont {Y.}~\bibnamefont {Nishida}}, \bibinfo {author} {\bibfnamefont {A.}~\bibnamefont {Imambekov}}, \bibinfo {author} {\bibfnamefont {D.~A.}\ \bibnamefont {Abanin}},\ and\ \bibinfo {author} {\bibfnamefont {E.}~\bibnamefont {Demler}},\ }\bibfield  {title} {\bibinfo {title} {Time-dependent impurity in ultracold fermions: Orthogonality catastrophe and beyond},\ }\href {https://doi.org/10.1103/PhysRevX.2.041020} {\bibfield  {journal} {\bibinfo  {journal} {Phys. Rev. X}\ }\textbf {\bibinfo {volume} {2}},\ \bibinfo {pages} {041020} (\bibinfo {year} {2012})}\BibitemShut {NoStop}%
\bibitem [{\citenamefont {Kartavtsev}\ and\ \citenamefont {Malykh}(2007)}]{Kartavtsev2007let}%
  \BibitemOpen
  \bibfield  {author} {\bibinfo {author} {\bibfnamefont {O.~I.}\ \bibnamefont {Kartavtsev}}\ and\ \bibinfo {author} {\bibfnamefont {A.~V.}\ \bibnamefont {Malykh}},\ }\bibfield  {title} {\bibinfo {title} {Low-energy three-body dynamics in binary quantum gases},\ }\href {https://doi.org/10.1088/0953-4075/40/7/011} {\bibfield  {journal} {\bibinfo  {journal} {J. Phys. B: At. Mol. Opt. Phys.}\ }\textbf {\bibinfo {volume} {40}},\ \bibinfo {pages} {1429} (\bibinfo {year} {2007})}\BibitemShut {NoStop}%
\bibitem [{\citenamefont {Greene}\ \emph {et~al.}(2017)\citenamefont {Greene}, \citenamefont {Giannakeas},\ and\ \citenamefont {P\'erez-R\'{\i}os}}]{Greene2017ufb}%
  \BibitemOpen
  \bibfield  {author} {\bibinfo {author} {\bibfnamefont {C.~H.}\ \bibnamefont {Greene}}, \bibinfo {author} {\bibfnamefont {P.}~\bibnamefont {Giannakeas}},\ and\ \bibinfo {author} {\bibfnamefont {J.}~\bibnamefont {P\'erez-R\'{\i}os}},\ }\bibfield  {title} {\bibinfo {title} {Universal few-body physics and cluster formation},\ }\href {https://doi.org/10.1103/RevModPhys.89.035006} {\bibfield  {journal} {\bibinfo  {journal} {Rev. Mod. Phys.}\ }\textbf {\bibinfo {volume} {89}},\ \bibinfo {pages} {035006} (\bibinfo {year} {2017})}\BibitemShut {NoStop}%
\bibitem [{\citenamefont {Naidon}\ and\ \citenamefont {Endo}(2017)}]{Naidon2017epa}%
  \BibitemOpen
  \bibfield  {author} {\bibinfo {author} {\bibfnamefont {P.}~\bibnamefont {Naidon}}\ and\ \bibinfo {author} {\bibfnamefont {S.}~\bibnamefont {Endo}},\ }\bibfield  {title} {\bibinfo {title} {Efimov physics: a review},\ }\href {https://doi.org/10.1088/1361-6633/aa50e8} {\bibfield  {journal} {\bibinfo  {journal} {Rep. Prog. Phys.}\ }\textbf {\bibinfo {volume} {80}},\ \bibinfo {pages} {056001} (\bibinfo {year} {2017})}\BibitemShut {NoStop}%
\bibitem [{\citenamefont {Christianen}\ \emph {et~al.}(2022)\citenamefont {Christianen}, \citenamefont {Cirac},\ and\ \citenamefont {Schmidt}}]{Christianen2022bpa}%
  \BibitemOpen
  \bibfield  {author} {\bibinfo {author} {\bibfnamefont {A.}~\bibnamefont {Christianen}}, \bibinfo {author} {\bibfnamefont {J.~I.}\ \bibnamefont {Cirac}},\ and\ \bibinfo {author} {\bibfnamefont {R.}~\bibnamefont {Schmidt}},\ }\bibfield  {title} {\bibinfo {title} {{Bose polaron and the Efimov effect: A Gaussian-state approach}},\ }\href {https://doi.org/10.1103/PhysRevA.105.053302} {\bibfield  {journal} {\bibinfo  {journal} {Phys. Rev. A}\ }\textbf {\bibinfo {volume} {105}},\ \bibinfo {pages} {053302} (\bibinfo {year} {2022})}\BibitemShut {NoStop}%
\bibitem [{\citenamefont {Nishimura}\ \emph {et~al.}(2021)\citenamefont {Nishimura}, \citenamefont {Nakano}, \citenamefont {Iida}, \citenamefont {Tajima}, \citenamefont {Miyakawa},\ and\ \citenamefont {Yabu}}]{Nishimura2021gso}%
  \BibitemOpen
  \bibfield  {author} {\bibinfo {author} {\bibfnamefont {K.}~\bibnamefont {Nishimura}}, \bibinfo {author} {\bibfnamefont {E.}~\bibnamefont {Nakano}}, \bibinfo {author} {\bibfnamefont {K.}~\bibnamefont {Iida}}, \bibinfo {author} {\bibfnamefont {H.}~\bibnamefont {Tajima}}, \bibinfo {author} {\bibfnamefont {T.}~\bibnamefont {Miyakawa}},\ and\ \bibinfo {author} {\bibfnamefont {H.}~\bibnamefont {Yabu}},\ }\bibfield  {title} {\bibinfo {title} {Ground state of the polaron in an ultracold dipolar fermi gas},\ }\href {https://doi.org/10.1103/PhysRevA.103.033324} {\bibfield  {journal} {\bibinfo  {journal} {Phys. Rev. A}\ }\textbf {\bibinfo {volume} {103}},\ \bibinfo {pages} {033324} (\bibinfo {year} {2021})}\BibitemShut {NoStop}%
\bibitem [{\citenamefont {Kain}\ and\ \citenamefont {Ling}(2014)}]{Kain2014pia}%
  \BibitemOpen
  \bibfield  {author} {\bibinfo {author} {\bibfnamefont {B.}~\bibnamefont {Kain}}\ and\ \bibinfo {author} {\bibfnamefont {H.~Y.}\ \bibnamefont {Ling}},\ }\bibfield  {title} {\bibinfo {title} {Polarons in a dipolar condensate},\ }\href {https://doi.org/10.1103/PhysRevA.89.023612} {\bibfield  {journal} {\bibinfo  {journal} {Phys. Rev. A}\ }\textbf {\bibinfo {volume} {89}},\ \bibinfo {pages} {023612} (\bibinfo {year} {2014})}\BibitemShut {NoStop}%
\bibitem [{\citenamefont {Enss}(2025)}]{Enss2025qti}%
  \BibitemOpen
  \bibfield  {author} {\bibinfo {author} {\bibfnamefont {T.}~\bibnamefont {Enss}},\ }\bibfield  {title} {\bibinfo {title} {Quantum transport in strongly correlated {Fermi} gases},\ }\href {https://doi.org/10.5802/crphys.237} {\bibfield  {journal} {\bibinfo  {journal} {Comptes Rendus. Physique}\ }\textbf {\bibinfo {volume} {26}},\ \bibinfo {pages} {217} (\bibinfo {year} {2025})}\BibitemShut {NoStop}%
\bibitem [{\citenamefont {Astrakharchik}\ and\ \citenamefont {Pitaevskii}(2004)}]{Astrakharchik2004moa}%
  \BibitemOpen
  \bibfield  {author} {\bibinfo {author} {\bibfnamefont {G.~E.}\ \bibnamefont {Astrakharchik}}\ and\ \bibinfo {author} {\bibfnamefont {L.~P.}\ \bibnamefont {Pitaevskii}},\ }\bibfield  {title} {\bibinfo {title} {{Motion of a heavy impurity through a Bose-Einstein condensate}},\ }\href {https://doi.org/10.1103/PhysRevA.70.013608} {\bibfield  {journal} {\bibinfo  {journal} {Phys. Rev. A}\ }\textbf {\bibinfo {volume} {70}},\ \bibinfo {pages} {013608} (\bibinfo {year} {2004})}\BibitemShut {NoStop}%
\bibitem [{\citenamefont {Castin}\ \emph {et~al.}(2015)\citenamefont {Castin}, \citenamefont {Ferrier-Barbut},\ and\ \citenamefont {Salomon}}]{Castin2015lvc}%
  \BibitemOpen
  \bibfield  {author} {\bibinfo {author} {\bibfnamefont {Y.}~\bibnamefont {Castin}}, \bibinfo {author} {\bibfnamefont {I.}~\bibnamefont {Ferrier-Barbut}},\ and\ \bibinfo {author} {\bibfnamefont {C.}~\bibnamefont {Salomon}},\ }\bibfield  {title} {\bibinfo {title} {La vitesse critique de {Landau} d'une particule dans un superfluide de fermions},\ }\href {https://doi.org/10.1016/j.crhy.2015.01.005} {\bibfield  {journal} {\bibinfo  {journal} {Comptes Rendus. Physique}\ }\textbf {\bibinfo {volume} {16}},\ \bibinfo {pages} {241} (\bibinfo {year} {2015})}\BibitemShut {NoStop}%
\bibitem [{\citenamefont {Koepsell}\ \emph {et~al.}(2019)\citenamefont {Koepsell}, \citenamefont {Vijayan}, \citenamefont {Sompet}, \citenamefont {Grusdt}, \citenamefont {Hilker}, \citenamefont {Demler}, \citenamefont {Salomon}, \citenamefont {Bloch},\ and\ \citenamefont {Gross}}]{Koepsell2019imp}%
  \BibitemOpen
  \bibfield  {author} {\bibinfo {author} {\bibfnamefont {J.}~\bibnamefont {Koepsell}}, \bibinfo {author} {\bibfnamefont {J.}~\bibnamefont {Vijayan}}, \bibinfo {author} {\bibfnamefont {P.}~\bibnamefont {Sompet}}, \bibinfo {author} {\bibfnamefont {F.}~\bibnamefont {Grusdt}}, \bibinfo {author} {\bibfnamefont {T.~A.}\ \bibnamefont {Hilker}}, \bibinfo {author} {\bibfnamefont {E.}~\bibnamefont {Demler}}, \bibinfo {author} {\bibfnamefont {G.}~\bibnamefont {Salomon}}, \bibinfo {author} {\bibfnamefont {I.}~\bibnamefont {Bloch}},\ and\ \bibinfo {author} {\bibfnamefont {C.}~\bibnamefont {Gross}},\ }\bibfield  {title} {\bibinfo {title} {{Imaging magnetic polarons in the doped Fermi-Hubbard model}},\ }\href {https://doi.org/10.1038/s41586-019-1463-1} {\bibfield  {journal} {\bibinfo  {journal} {Nature}\ }\textbf {\bibinfo {volume} {572}},\ \bibinfo {pages} {358} (\bibinfo {year} {2019})}\BibitemShut {NoStop}%
\bibitem [{\citenamefont {Lin}\ \emph {et~al.}(2011)\citenamefont {Lin}, \citenamefont {Jimenez-Garcia},\ and\ \citenamefont {Spielman}}]{Lin2011soc}%
  \BibitemOpen
  \bibfield  {author} {\bibinfo {author} {\bibfnamefont {Y.-J.}\ \bibnamefont {Lin}}, \bibinfo {author} {\bibfnamefont {K.}~\bibnamefont {Jimenez-Garcia}},\ and\ \bibinfo {author} {\bibfnamefont {I.~B.}\ \bibnamefont {Spielman}},\ }\bibfield  {title} {\bibinfo {title} {{Spin-orbit-coupled Bose-Einstein condensates}},\ }\href {https://doi.org/10.1038/nature09887} {\bibfield  {journal} {\bibinfo  {journal} {Nature (London)}\ }\textbf {\bibinfo {volume} {471}},\ \bibinfo {pages} {83} (\bibinfo {year} {2011})}\BibitemShut {NoStop}%
\bibitem [{\citenamefont {Cheuk}\ \emph {et~al.}(2012)\citenamefont {Cheuk}, \citenamefont {Sommer}, \citenamefont {Hadzibabic}, \citenamefont {Yefsah}, \citenamefont {Bakr},\ and\ \citenamefont {Zwierlein}}]{Cheuk2012sis}%
  \BibitemOpen
  \bibfield  {author} {\bibinfo {author} {\bibfnamefont {L.~W.}\ \bibnamefont {Cheuk}}, \bibinfo {author} {\bibfnamefont {A.~T.}\ \bibnamefont {Sommer}}, \bibinfo {author} {\bibfnamefont {Z.}~\bibnamefont {Hadzibabic}}, \bibinfo {author} {\bibfnamefont {T.}~\bibnamefont {Yefsah}}, \bibinfo {author} {\bibfnamefont {W.~S.}\ \bibnamefont {Bakr}},\ and\ \bibinfo {author} {\bibfnamefont {M.~W.}\ \bibnamefont {Zwierlein}},\ }\bibfield  {title} {\bibinfo {title} {{Spin-Injection Spectroscopy of a Spin-Orbit Coupled Fermi Gas}},\ }\href {https://doi.org/10.1103/PhysRevLett.109.095302} {\bibfield  {journal} {\bibinfo  {journal} {Phys. Rev. Lett.}\ }\textbf {\bibinfo {volume} {109}},\ \bibinfo {pages} {095302} (\bibinfo {year} {2012})}\BibitemShut {NoStop}%
\bibitem [{\citenamefont {Wang}\ \emph {et~al.}(2012)\citenamefont {Wang}, \citenamefont {Yu}, \citenamefont {Fu}, \citenamefont {Miao}, \citenamefont {Huang}, \citenamefont {Chai}, \citenamefont {Zhai},\ and\ \citenamefont {Zhang}}]{Wang2012spc}%
  \BibitemOpen
  \bibfield  {author} {\bibinfo {author} {\bibfnamefont {P.}~\bibnamefont {Wang}}, \bibinfo {author} {\bibfnamefont {Z.-Q.}\ \bibnamefont {Yu}}, \bibinfo {author} {\bibfnamefont {Z.}~\bibnamefont {Fu}}, \bibinfo {author} {\bibfnamefont {J.}~\bibnamefont {Miao}}, \bibinfo {author} {\bibfnamefont {L.}~\bibnamefont {Huang}}, \bibinfo {author} {\bibfnamefont {S.}~\bibnamefont {Chai}}, \bibinfo {author} {\bibfnamefont {H.}~\bibnamefont {Zhai}},\ and\ \bibinfo {author} {\bibfnamefont {J.}~\bibnamefont {Zhang}},\ }\bibfield  {title} {\bibinfo {title} {{Spin-Orbit Coupled Degenerate Fermi Gases}},\ }\href {https://doi.org/10.1103/PhysRevLett.109.095301} {\bibfield  {journal} {\bibinfo  {journal} {Phys. Rev. Lett.}\ }\textbf {\bibinfo {volume} {109}},\ \bibinfo {pages} {095301} (\bibinfo {year} {2012})}\BibitemShut {NoStop}%
\bibitem [{dat()}]{data-repository}%
  \BibitemOpen
  \href {https://doi.org/10.48323/jv1s5-t5234} {}\bibinfo {note} {\url{https://doi.org/10.48323/jv1s5-t5234}}\BibitemShut {NoStop}%
\bibitem [{\citenamefont {Bruun}\ \emph {et~al.}(2005)\citenamefont {Bruun}, \citenamefont {Jackson},\ and\ \citenamefont {Kolomeitsev}}]{Bruun2005msa}%
  \BibitemOpen
  \bibfield  {author} {\bibinfo {author} {\bibfnamefont {G.~M.}\ \bibnamefont {Bruun}}, \bibinfo {author} {\bibfnamefont {A.~D.}\ \bibnamefont {Jackson}},\ and\ \bibinfo {author} {\bibfnamefont {E.~E.}\ \bibnamefont {Kolomeitsev}},\ }\bibfield  {title} {\bibinfo {title} {{Multichannel scattering and Feshbach resonances: Effective theory, phenomenology, and many-body effects}},\ }\href {https://doi.org/10.1103/PhysRevA.71.052713} {\bibfield  {journal} {\bibinfo  {journal} {Phys. Rev. A}\ }\textbf {\bibinfo {volume} {71}},\ \bibinfo {pages} {052713} (\bibinfo {year} {2005})}\BibitemShut {NoStop}%
\bibitem [{\citenamefont {Trefzger}\ and\ \citenamefont {Castin}(2013)}]{Trefzger2013edr}%
  \BibitemOpen
  \bibfield  {author} {\bibinfo {author} {\bibfnamefont {C.}~\bibnamefont {Trefzger}}\ and\ \bibinfo {author} {\bibfnamefont {Y.}~\bibnamefont {Castin}},\ }\bibfield  {title} {\bibinfo {title} {{Energy, decay rate, and effective masses for a moving polaron in a Fermi sea: Explicit results in the weakly attractive limit}},\ }\href {https://doi.org/10.1209/0295-5075/104/50005} {\bibfield  {journal} {\bibinfo  {journal} {Europhys. Lett.}\ }\textbf {\bibinfo {volume} {104}},\ \bibinfo {pages} {50005} (\bibinfo {year} {2013})}\BibitemShut {NoStop}%
\bibitem [{\citenamefont {Hu}\ \emph {et~al.}(2018)\citenamefont {Hu}, \citenamefont {Mulkerin}, \citenamefont {Wang},\ and\ \citenamefont {Liu}}]{Hu2018}%
  \BibitemOpen
  \bibfield  {author} {\bibinfo {author} {\bibfnamefont {H.}~\bibnamefont {Hu}}, \bibinfo {author} {\bibfnamefont {B.~C.}\ \bibnamefont {Mulkerin}}, \bibinfo {author} {\bibfnamefont {J.}~\bibnamefont {Wang}},\ and\ \bibinfo {author} {\bibfnamefont {X.-J.}\ \bibnamefont {Liu}},\ }\bibfield  {title} {\bibinfo {title} {{Attractive Fermi polarons at nonzero temperatures with a finite impurity concentration}},\ }\href {https://doi.org/10.1103/PhysRevA.98.013626} {\bibfield  {journal} {\bibinfo  {journal} {Phys. Rev. A}\ }\textbf {\bibinfo {volume} {98}},\ \bibinfo {pages} {013626} (\bibinfo {year} {2018})}\BibitemShut {NoStop}%
\bibitem [{\citenamefont {Zaccanti}(2022)}]{Zaccanti2025mif}%
  \BibitemOpen
  \bibfield  {author} {\bibinfo {author} {\bibfnamefont {M.}~\bibnamefont {Zaccanti}},\ }\href@noop {} {\bibinfo {title} {Mass-imbalanced {F}ermi mixtures with resonant interactions}} (\bibinfo {year} {2022}),\ \bibinfo {note} {in Ref.~\cite{Varenna2022book}}\BibitemShut {NoStop}%
\bibitem [{\citenamefont {Ashcroft}\ and\ \citenamefont {Mermin}(1976)}]{Ashcroft1976ssp}%
  \BibitemOpen
  \bibfield  {author} {\bibinfo {author} {\bibfnamefont {N.}~\bibnamefont {Ashcroft}}\ and\ \bibinfo {author} {\bibfnamefont {N.}~\bibnamefont {Mermin}},\ }\href {https://books.google.it/books?id=oXIfAQAAMAAJ} {\emph {\bibinfo {title} {{Solid State Physics}}}},\ HRW international editions\ (\bibinfo  {publisher} {{Saunders College Publishing}},\ \bibinfo {year} {1976})\BibitemShut {NoStop}%
\bibitem [{\citenamefont {Mulkerin}\ \emph {et~al.}(2019)\citenamefont {Mulkerin}, \citenamefont {Liu},\ and\ \citenamefont {Hu}}]{Mulkerin2019bot}%
  \BibitemOpen
  \bibfield  {author} {\bibinfo {author} {\bibfnamefont {B.~C.}\ \bibnamefont {Mulkerin}}, \bibinfo {author} {\bibfnamefont {X.-J.}\ \bibnamefont {Liu}},\ and\ \bibinfo {author} {\bibfnamefont {H.}~\bibnamefont {Hu}},\ }\bibfield  {title} {\bibinfo {title} {{Breakdown of the Fermi polaron description near Fermi degeneracy at unitarity}},\ }\href {https://doi.org/https://doi.org/10.1016/j.aop.2019.04.018} {\bibfield  {journal} {\bibinfo  {journal} {Ann. Phys.}\ }\textbf {\bibinfo {volume} {407}},\ \bibinfo {pages} {29} (\bibinfo {year} {2019})}\BibitemShut {NoStop}%
\bibitem [{\citenamefont {Tajima}\ and\ \citenamefont {Uchino}(2018)}]{Tajima2018mfp}%
  \BibitemOpen
  \bibfield  {author} {\bibinfo {author} {\bibfnamefont {H.}~\bibnamefont {Tajima}}\ and\ \bibinfo {author} {\bibfnamefont {S.}~\bibnamefont {Uchino}},\ }\bibfield  {title} {\bibinfo {title} {{Many Fermi polarons at nonzero temperature}},\ }\href {https://doi.org/10.1088/1367-2630/aad1e7} {\bibfield  {journal} {\bibinfo  {journal} {New J. Phys.}\ }\textbf {\bibinfo {volume} {20}},\ \bibinfo {pages} {073048} (\bibinfo {year} {2018})}\BibitemShut {NoStop}%
\bibitem [{\citenamefont {Hu}\ and\ \citenamefont {Liu}(2022{\natexlab{b}})}]{Hu2022fpa}%
  \BibitemOpen
  \bibfield  {author} {\bibinfo {author} {\bibfnamefont {H.}~\bibnamefont {Hu}}\ and\ \bibinfo {author} {\bibfnamefont {X.-J.}\ \bibnamefont {Liu}},\ }\bibfield  {title} {\bibinfo {title} {{Fermi polarons at finite temperature: Spectral function and rf spectroscopy}},\ }\href {https://doi.org/10.1103/PhysRevA.105.043303} {\bibfield  {journal} {\bibinfo  {journal} {Phys. Rev. A}\ }\textbf {\bibinfo {volume} {105}},\ \bibinfo {pages} {043303} (\bibinfo {year} {2022}{\natexlab{b}})}\BibitemShut {NoStop}%
\bibitem [{\citenamefont {Berman}(1997)}]{Berman1997book}%
  \BibitemOpen
  \bibfield  {author} {\bibinfo {author} {\bibfnamefont {P.~R.}\ \bibnamefont {Berman}},\ }\href@noop {} {\emph {\bibinfo {title} {{Atom Interferomety}}}}\ (\bibinfo  {publisher} {Academic Press},\ \bibinfo {year} {1997})\BibitemShut {NoStop}%
\bibitem [{\citenamefont {Shkedrov}\ \emph {et~al.}(2020)\citenamefont {Shkedrov}, \citenamefont {Ness}, \citenamefont {Florshaim},\ and\ \citenamefont {Sagi}}]{Shkedrov2020ism}%
  \BibitemOpen
  \bibfield  {author} {\bibinfo {author} {\bibfnamefont {C.}~\bibnamefont {Shkedrov}}, \bibinfo {author} {\bibfnamefont {G.}~\bibnamefont {Ness}}, \bibinfo {author} {\bibfnamefont {Y.}~\bibnamefont {Florshaim}},\ and\ \bibinfo {author} {\bibfnamefont {Y.}~\bibnamefont {Sagi}},\ }\bibfield  {title} {\bibinfo {title} {{In situ momentum-distribution measurement of a quantum degenerate Fermi gas using Raman spectroscopy}},\ }\href {https://doi.org/10.1103/PhysRevA.101.013609} {\bibfield  {journal} {\bibinfo  {journal} {Phys. Rev. A}\ }\textbf {\bibinfo {volume} {101}},\ \bibinfo {pages} {013609} (\bibinfo {year} {2020})}\BibitemShut {NoStop}%
\bibitem [{\citenamefont {Taylor}(2022)}]{Taylor2022book}%
  \BibitemOpen
  \bibfield  {author} {\bibinfo {author} {\bibfnamefont {J.~R.}\ \bibnamefont {Taylor}},\ }\href@noop {} {\emph {\bibinfo {title} {{An Introduction to Error Analysis: The Study of Uncertainties in Physical Measurements, 3rd Edition}}}}\ (\bibinfo  {publisher} {University Science Books},\ \bibinfo {year} {2022})\BibitemShut {NoStop}%
\end{thebibliography}
\end{document}